\documentclass[12pt]{article}\usepackage[]{graphicx}\usepackage[]{xcolor}
\makeatletter
\def\maxwidth{ %
  \ifdim\Gin@nat@width>\linewidth
    \linewidth
  \else
    \Gin@nat@width
  \fi
}
\makeatother

\definecolor{fgcolor}{rgb}{0.345, 0.345, 0.345}
\newcommand{\hlkwb}[1]{\textcolor[rgb]{0.69,0.353,0.396}{#1}}%

\usepackage{framed}
\makeatletter
 {\par\unskip\endMakeFramed%
 \at@end@of@kframe}
\makeatother

\definecolor{shadecolor}{rgb}{.97, .97, .97}
\definecolor{messagecolor}{rgb}{0, 0, 0}
\definecolor{warningcolor}{rgb}{1, 0, 1}
\definecolor{errorcolor}{rgb}{1, 0, 0}
\newenvironment{knitrout}{}{} 

\usepackage{alltt}
\usepackage{lmodern}
\PassOptionsToPackage{unicode}{hyperref}
\PassOptionsToPackage{hyphens}{url}
\PassOptionsToPackage{dvipsnames,svgnames,x11names}{xcolor}
\usepackage{longtable,booktabs,array}
\usepackage{calc}
\usepackage{amsmath}
\usepackage{enumerate}
\usepackage[authoryear,numbers]{natbib}
\usepackage{graphicx}
\usepackage[english]{babel}
\usepackage[colorlinks=true, allcolors=blue]{hyperref}
\usepackage{caption}
\usepackage{subcaption}
\usepackage{adjustbox}
\usepackage{placeins}
\usepackage{amssymb}
\usepackage{hyperref}
\usepackage{float}
\floatstyle{ruled}
\newfloat{codelisting}{h}{lop}
\floatname{codelisting}{Listing}
\usepackage{bookmark}
\usepackage[ruled,vlined]{algorithm2e}
\usepackage[utf8]{inputenc}
\usepackage{hyphenat}
\usepackage{url}
\usepackage{textcomp}
\usepackage{tikz}
\usetikzlibrary{shapes.geometric, arrows, fit}
\usetikzlibrary{automata,positioning}
\usetikzlibrary{shapes.multipart}
\tikzset{
    process/.style = {rectangle, rounded corners, minimum width=2cm,minimum height=1cm, text centered, text width=6cm, draw=black, fill=orange!30},
    process2/.style = {rectangle, rounded corners, minimum width=2cm,minimum height=1cm, text centered, text width=6cm, draw=black, fill=blue!30},
    decision/.style = {diamond, minimum width=2cm, minimum height=1cm, text centered, draw=black, fill=green!30},
    decision2/.style = {diamond, minimum width=2cm, minimum height=1cm, text centered, draw=black, fill=blue!30},
    arrow/.style = {thick,->,>=stealth},
    myfit/.style={draw,dashed,black, inner xsep=10pt, inner ysep=15pt, rounded corners=5pt},
    mytitle/.style={draw,black, fill=orange!50, inner sep=5pt, right, xshift=10pt, align=center, execute at begin node=\setlength{\baselineskip}{2em}}
    }
\newcommand{\blind}{1}
\usetikzlibrary{positioning}
\usetikzlibrary {arrows.meta}
\usetikzlibrary{shapes.geometric}
\usetikzlibrary{calc}
\newcommand\authoraddress{ \vspace{-1mm} \\ \normalsize}

\newcommand\Unit{U}
\newcommand\unit{u}
\newcommand\seq[2]{{#1}\!:\!{#2}}
\newcommand\species{\mathbbm{k}}
\usepackage{bbm}
\newcommand\native{\mathbbm{n}}
\newcommand\invasive{\mathbbm{i}}
\newcommand\Normal{\mathrm{N}}
\IfFileExists{upquote.sty}{\usepackage{upquote}}{}
\begin{document}
\def\spacingset#1{\renewcommand{\baselinestretch}%
{#1}\small\normalsize} \spacingset{1}

\if1\blind
{
  \title{\bf Mechanistic models for panel data: Analysis of ecological experiments with four interacting species}
\author{Bo Yang \authoraddress Department of Biostatistics, University of Michigan \\
        Jesse Wheeler  \authoraddress Department of Statistics, University of Michigan \\
	Meghan A. Duffy  \authoraddress Department of Ecology and Evolutionary Biology, University of Michigan\\
        Aaron A. King  \authoraddress Department of Ecology and Evolutionary Biology \&  \authoraddress Center for the Study of Complex Systems, University of Michigan\\
        Edward L. Ionides \authoraddress Department of Statistics, University of Michigan}
  \maketitle
} \fi

\if0\blind
{
  \bigskip
  \bigskip
  \bigskip
  \begin{center}
    {\LARGE\bf Mechanistic models for panel data: Analysis of ecological experiments with four interacting species}
\end{center}
  \medskip
} \fi


\begin{abstract} 
\noindent In an ecological context, panel data arise when time series measurements are made on a collection of ecological processes.
Each process may correspond to a spatial location for field data, or to an experimental ecosystem in a designed experiment.
Statistical models for ecological panel data should capture the high levels of nonlinearity, stochasticity, and measurement uncertainty inherent in ecological systems.
Furthermore, the system dynamics may depend on unobservable variables.
This study applies iterated particle filtering techniques to explore new possibilities for likelihood-based statistical analysis of these complex systems.
We analyze data from a mesocosm experiment in which two species of the freshwater planktonic crustacean genus, {\it Daphnia}, coexist with an alga and a fungal parasite.
Time series data were collected on replicated mesocosms under six treatment conditions.
Iterated filtering enables maximization of the likelihood for scientifically motivated nonlinear partially observed Markov process models, providing access to standard likelihood-based methods for parameter estimation, confidence intervals, hypothesis testing, model selection and diagnostics.
This toolbox allows scientists to propose and evaluate scientifically motivated stochastic dynamic models for panel data, constrained only by the requirement to write code to simulate from the model and to specify a measurement distribution describing how the system state is observed.
\end{abstract}

\noindent%
\vfill

\newpage
\spacingset{1.8} 
\section{Introduction}
\label{sec:intro}

Biological dynamic systems often contain elements or processes that are difficult to observe, even under controlled experimental conditions.
These unobservable, or latent, components are often crucial to understanding system dynamics, or are of scientific interest themselves.
Additionally, both the measurements obtained from the systems in question and their underlying dynamic processes often exhibit substantial stochasticity.
The combination of these two factors complicates data interpretation and analysis in ecological studies.
Given these inherent properties of ecological systems, it is essential to use models that can account for key latent mechanisms and the stochastic nature of the system when conducting inference \citep{young98,bjornstad01}.

Partially observed Markov process (POMP) models, also known as hidden Markov models or state space models \citep{breto09,auger-methe21}, comprise a class of models suited to the description of such systems, which describe the latent process using a Markovian representation.
 A POMP model is mechanistic if it is based on scientific principles, which is usually equivalent to providing a causal description of the system that can explain the consequence of interventions.
  Statistical inferences on mechanistic models can provide insight into key latent mechanisms.
  Further, statistical tests between rival mechanistic hypotheses can inform qualitative understanding about suitable model structures.
  Thus, mechanistic models are invaluable tools for addressing challenges in ecology \citep{mouquet15}.

  Collections of time series on related but disjoint systems are called panel data, also known as longitudinal data.
  Each component system is called a unit, and the units may be sites in a field study (e.g., plots, lakes) or independent mesocosms or microcosms in a controlled laboratory setting.
  Studying the entire panel may allow precision in statistical conclusions that is unattainable from a single unit.
  Alternatively, scientists may be interested in specific differences between the units, requiring the investigation of the entire panel of data.
  For either of these goals, investigation of panel data leads to the challenging statistical task of fitting a collection of related nonlinear or non-Gaussian vector-valued POMP models, called a PanelPOMP model \citep{breto20,breto25}.
  In this article, we demonstrate how recent methodological advances can facilitate data analysis for such models in a multi-species system.
  We use data from a controlled experiment involving the population dynamics of two freshwater plankton species ({\it Daphnia dentifera} and {\it Daphnia lumholtzi}) together with an alga, \textit{Ankistrodesmus falcatus}, as food and a fungal parasite, \textit{Metschnikowia bicuspidata}, that can infect both {\it Daphnia} species.
  These data were collected by \citet{searle16} to investigate how the competition between the North American native {\it D.~dentifera} and the invasive {\it D.~lumholtzi} is modified by their differing susceptibility to the parasite.
  Advances in data collection make this experiment representative of a growing class of situations where panel data is available on a complex multi-species system.

  Our model builds upon the ordinary differential equation model of \citet{searle16}.
  We include dynamic stochasticity, leading to a stochastic differential equation model which provides an improved statistical fit to the data while also assisting identification of model misspecification \citep{king15}.
  We carry out likelihood-based inference using panel iterated filtering \citep{breto20} which is a PanelPOMP extension of basic iterated filtering methodology \citep{ionides15,king16} for POMP models.
  Iterated filtering methods have a plug-and-play property \citep{he10} that they require a simulator for the dynamic model but do not need the ability to evaluate transition densities.
  The plug-and-play property permits practical consideration of a broad class of PanelPOMP models, assisting the creative task of developing scientifically plausible models that simultaneously have good statistical fit.
  POMP models have long been advocated as an appropriate framework for ecological systems \citep{schnute94,buckland04}.
  A major practical limitation has been the computational difficulty in carrying out inference.
  When Laplace approximations are applicable, the Template Model Builder approach is available \citep{kristensen16}.
Iterated filtering techniques apply to a more general class of POMP models, but the high-dimensional nature of PanelPOMP models adds additional methodological challenges.  
  We are not aware of any previous scientific studies where nonlinear, non-Gaussian PanelPOMP models (with some parameters shared between units and some parameters specific to each unit) have been calibrated to multi-species panel data via maximum likelihood.

  A central question when using model-based statistical inference to investigate ecological dynamics is to discover which biological details are necessary to include in a model to provide an adequate statistical explanation for the data.
  Our approach to answering this question uses standard tools of likelihood-based inference.
  We compare the model likelihood to statistical benchmarks, such as linear or autoregressive models, seeking to continue improving a mechanistic model until it captures the predictability of the data with at least comparable skill to a simple non-mechanistic analysis \citep{wheeler24,li24}.
  In doing so, we ensure that the proposed dynamic model provides a plausible mechanistic description of the dynamic system, supported by evidence that the model also provides a suitable quantitative description of data that arise from the system.

  To do this, we study residuals and conditional log-likelihood anomalies for each data point to identify relative weaknesses and strengths of different models.
  We calculate Akaike's Information Criterion (AIC) to compare the overall fit of rival mechanistic models \citep{aic74}.
We compute profile likelihood confidence intervals, making appropriate adjustment for Monte Carlo uncertainty \citep{ionides17,ning21}.
All these techniques are predicated on an ability to evaluate and maximize the log-likelihood function.
This ability was unlocked by panel iterated filtering \citep{breto20} and we demonstrate empirically that this is further strengthened by a new marginalized panel iterated filtering algorithm.
A contemporaneous investigation of marginalized panel iterated filtering from a theory and methodology perspective is provided by \cite{wheeler25}.
These algorithms are discussed in more detail in Section~\ref{sec:meth}, after description of the experimental system in Section~\ref{sec:studysystem} and presentation of our models and their fitted parameter values in Section~\ref{sec:mecmod}.
Section~\ref{sec:res} presents additional results showing that our stochastic model, that describes depletion of the algal food resource together with age-structured population dynamics, provides a better statistical explanation of the data than the previous deterministic model by \citet{searle16} that supposed the dynamics are driven by a switch of reproductive strategies from the asexual production of juveniles to the sexual production of long-lasting dormant eggs.
Our findings have consequences for the interpretation of this experiment and the planning of future experiments.
Section~\ref{sec:dis} is a concluding discussion.

\section{A {\it Daphnia} mesocosm experiment}
\label{sec:studysystem}

\textit{D.~dentifera} is native to North American stratified lakes, whereas \textit{D.~lumholtzi} is an invasive species originating from Africa, Asia, and Australia.
First observed in North America in the 1990s, \textit{D.~lumholtzi} has rapidly expanded across the United States, impacting native species and ecosystem dynamics \citep{havel93,searle16salinization}.
This leads to ecological interest in understanding competition between \textit{D.~dentifera} and \textit{D.~lumholtzi}, two species with high evolutionary divergence having their most recent common ancestor around 145 million years ago \citep{kotov11,cornetti19}.
Here, we investigate how this competition is affected by a parasite that infects both species.
\citet{searle16} hypothesized that the presence of a fast-growing invasive species, which has a high susceptibility to the native parasite {\it M.~bicuspidata}, could amplify the parasite in the native population.

Female \textit{Daphnia} reproduce asexually in favorable situations, parthenogenetically producing female offspring equipped for parthenogenetic reproduction.
Maturation of juveniles to adults takes approximately five to ten days, depending on environmental conditions.
In response to adverse environmental cues, \textit{Daphnia} produce males, and females reproduce sexually to produce durable resting eggs called ephippia \citep{radzikowski18,searle16salinization}.
The ephippia can remain viable for many years, awaiting the return of favorable conditions.

{\it Daphnia} become infected by the parasite by ingesting free-floating spores as they filter the water to extract food.
The spores pass through the gut wall to infect the body cavity of the {\it Daphnia}, where the fungus grows until the body cavity is full of spores and the host dies, releasing the spores into the water.
Infected \textit{Daphnia} die within two to three weeks \citep{ebert05}, whereas healthy \textit{Daphnia} can survive for many months.
Infection leads to sterility and reduced feeding, and an opaque body cavity which can be readily distinguished from the transparent cavity of uninfected {\it Daphnia}.

\citet{searle16} conducted an experiment involving a collection of mesocosms, each consisting of 15 liter of growth medium in a bucket, initialized with  2.5 $\times$ $10^8$ cells of green alga, {\it A.~falcatus}, and with one or two species of {\it Daphnia}.
Six treaments were carried out, comprised of three levels of the host ({\it D.\ dentifera} or {\it D. lumholtzi} or both) factored with two levels of the parasite (present or absent).
The experimental units are replicates of the treatments; the design consisted of 8, 9 or 10 units for each treatment.
The initial {\it Daphnia} population consisted of 45 adult females, which in two-host treatments consisted of 35 native \textit{Daphnia} and 10 invasives.
After a four day acclimatization period, spores of the fungal parasite, {\it M.~bicuspidata}, were added at a concentration of 25 per \text{mL} if the mesocosm was assigned to the parasite exposure treatment.
Every five days, a well-mixed sample of 1 liter was removed, and the species (native or invasive), infection status, sex, and age (juvenile or adult) were identified for each \textit{Daphnia} using a microscope.
A negligible number of infected juveniles were observed so henceforth we model infection only in adults.
Algal and nutrient levels were supplemented twice per week, with medium added as necessary to maintain the volume.
The mesocosms were kept inside, at 23.3$^\circ C$ and a 16-hour light, 8-hour dark cycle.
Population densities and infection prevalence (if appropriate) were quantified every five days, beginning a week after the experiment started.
Regular stirring of mesocosms ensured resuspension of algae and parasites.
The experiment concluded after 52 days, totaling 10 sampling sessions.

\begin{figure}[ht]
\centering
\begin{knitrout}
\definecolor{shadecolor}{rgb}{0.969, 0.969, 0.969}\color{fgcolor}
\includegraphics[width=\maxwidth]{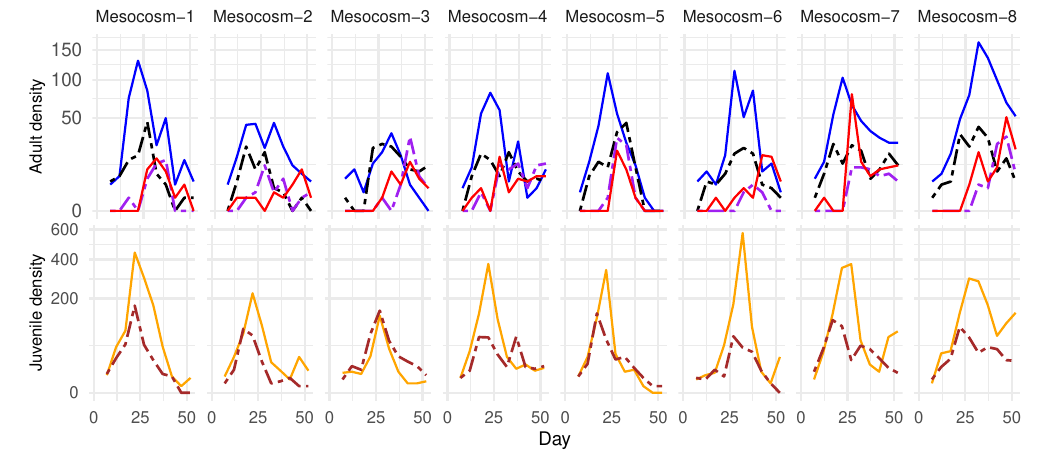} 
\end{knitrout}
\caption{\label{fig:data_vis}
The experimental data.
Density (Individuals$/$Liter) of \textit{D.~dentifera} (solid lines) and \textit{D.~lumholtzi} (dashed lines).
The top panel shows adult susceptibles (\textit{D.~dentifera}, blue; \textit{D.~lumholtzi}, black) and infecteds (\textit{D.~dentifera}, red; \textit{D.~lumholtzi}, purple).
The bottom panel shows juvenile susceptibles (\textit{D.~dentifera}, orange; \textit{D.~lumholtzi}, brown).
There were negligible infected juveniles.
Columns are buckets corresponding to replications with same treatment setting.
}
\end{figure}

Here, we focus on the most complex experimental treatment, namely, the four species treatment with both host species, the alga and the parasite.
The simpler treatments are presented in Supplementary Sections~S4,~S5 and~S6.
Data from the four species treatment are displayed in Fig.~\ref{fig:data_vis}.
The females and juveniles show a population peak around day 25, typically followed by a collapse and a mild resurgence around day 50.
The males arise predominantly later in the experiment as shown in Table S12.
The hypothesis that motivated the model of \citet{searle16} is that the arrival of the males, and the subsequent switch to the generation of ephippia by sexually reproduction, explains the decay of the other populations.
This hypothesis is related to resource depletion, since shortage of food and high population density are both triggers for a change in reproductive strategy \citep{radzikowski18,searle16salinization}.
However, resource depletion affects parthenogenetic reproduction as well influencing a switch to sexual reproduction.
Therefore, we investigate the hypothesis that explicit modeling of resource depletion can help to explain the observed dynamics, via the introduction of a latent food compartment.
We also hypothesize that the inclusion of age structure, via a juvenile stage, could help to explain the population dynamics even though very few juveniles show disease.
Our paradigm is to assess models that represent known or hypothesized biological mechanisms in terms of their ability to explain observed data, quantified by likelihood penalized by model complexity.
Our hypothesized model is supported so far as it is the best model for the available data according to this criterion.

\section{The Susceptible-Infected-Removed-Juvenile-Parasite-Food 2-species model (SIRJPF2)}
\label{sec:mecmod}

\begin{figure}[H]
\begin{center}
\resizebox{7cm}{!}{
\begin{tikzpicture}[
  square/.style={rectangle, draw=black, minimum width=0.5cm, minimum height=0.5cm, rounded corners=.1cm, fill=blue!8},
  rhombus/.style={diamond, draw=black, minimum width=0.1cm, minimum height=0.1cm, fill=purple!8,aspect = 1},
  travel/.style={circle, draw=black, minimum width=0.5cm, minimum height=0.5cm, fill=green!8},
  report/.style={shape=regular polygon, regular polygon sides=8, draw, fill=red!8,minimum size=0.6cm,inner sep=0cm},
  bendy/.style={bend left=10},
  bendy2/.style={bend left=100},
  bendy3/.style={bend left=-100},
  >/.style={shorten >=0.25mm},
  >/.tip={Stealth[length=1.5mm,width=1.5mm]}
]
\tikzset{>={}};

\node (Sn) at (5.5,0) [square] {$S^\native$};
\node (In) at (5.5,-1.5) [square] {$I^\native$};
\node (Si) at (-0.5,0) [square] {$S^\invasive$};
\node (Ii) at (-0.5,-1.5) [square] {$I^\invasive$};
\node (Jn) at (5.5,1.5) [square] {$J^\native$};
\node (Ji) at (-0.5,1.5) [square] {$J^\invasive$};
\node (R) at (2.5,0)  [rhombus] {$R$};
\node (P) at (2.5,-1.5)[travel] {$P$};
\node (F) at (2.5,1.5) [report] {$F$};

\draw [->, bendy] (Sn) to  (In);
\draw [->, bendy] (Si) to  (Ii);
\draw [->, bendy] (In) to  (Sn);
\draw [->, bendy] (Ii) to  (Si);
\draw [->, bendy] (Jn) to  (Sn);
\draw [->, bendy] (Ji) to  (Si);
\draw [->, bendy] (Sn) to  (Jn);
\draw [->, bendy] (Si) to  (Ji);

\draw [->] (F) --  (R);
\draw [->] (P) --  (R);
\draw [->] (Sn) -- (R);
\draw [->] (In) -- (R);
\draw [->] (Si) -- (R);
\draw [->] (Ii) -- (R);
\draw [->] (Jn) -- (R);
\draw [->] (Ji) -- (R);
\draw [->] (F) -- (Sn);
\draw [->] (F) -- (Si);
\draw [->] (F) -- (Jn);
\draw [->] (F) -- (Ji);
\draw [->] (F) -- (In);
\draw [->] (F) -- (Ii);
\draw (F) edge[loop above] (F);

\draw [->, bendy] (P) to (In);
\draw [->, bendy] (P) to (Ii);
\draw [->, bendy] (In) to (P);
\draw [->, bendy] (Ii) to (P);

\end{tikzpicture}
}
\end{center}
\vspace{-5mm}
\caption{A flow diagram for the SIRJPF2 model.
The two species of \textit{Daphnia} interact with each other indirectly through $F$ (Food) and $P$ (Parasite) compartments.
Transition to the $R$ state represents death or removal of any species.
}
\label{fig:flow_SIRJPF2}
\end{figure}
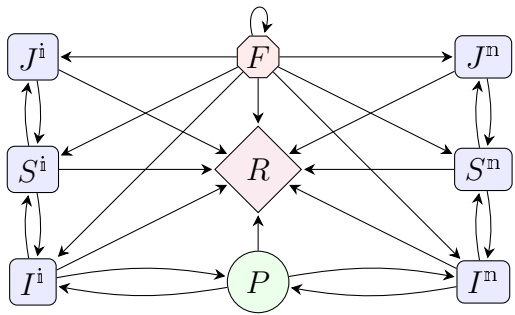

Fig.~\ref{fig:flow_SIRJPF2} shows a flow diagram for a two-species model where the \textit{Daphnia} are susceptible adult (S), infected adult (I), or juvenile (J), with a superscript $\invasive$ for invasive and $\native$ for native.
We also keep track of the aquatic parasite spore density (P) and the food density (F).
Removal (i.e., death, consumption or destructive sampling) for any species is described by transition to class R.
Similar to many epidemiological models, this description of the dynamic system extends the basic SIR model \citep{weiss13}.
Mathematically, the model is described by equations~(\ref{eq:model:first}--\ref{eq:model:last}), a system of stochastic differential equations which we interpret in the It\^{o} sense \citep{oksendal98}.
Nonlinearity arises through product terms in the infection and consumption specifications.
We use Gaussian noise, though multiplicative gamma noise can be employed to enforce non-negativity of rates \citep{bhadra11}.
A version of SIRJPF2 with gamma noise, called SIRJPF2-Gamma, is presented in Section~S9, though we found that a simpler Gaussian noise model is sufficient.

The nonlinearity and intrinsic stochasticity of the SIRPF2 model are characteristic features of dynamics in biological systems \citep{coulson04,wilkinson18}.
In the following equations, the superscript $\species$ describes the species of \textit{Daphnia}, taking value ${\native}$ for native (\textit{D.~dentifera}) and ${\invasive}$ for invasive  (\textit{D.~lumholtzi}).
The subscript $u$ ranges in $\seq{1}{U}$, and most parameters are written with potential dependence on $u$ to enable investigation of factors explaining differences between replicates.
We will later establish that this dependence is not needed; random dynamic noise is sufficient to explain the observed differences between units.
The experimentally determined sampling rate, $\delta = 0.013$ per liter per day, and food replenishment rate, $\mu = 0.37 \times 10^5$ cells per liter per day, are fixed parameter values that are shared between units based on the experimental protocol.
{\small
\begin{align}
    \label{eq:model:first}
    dS^{\species}_{\unit}(t) &= \lambda^{\species}_{J,{\unit}} \,  J^{\species}_{\unit}(t) \, dt -
    \big\{
    \theta^{\species}_{S,{\unit}} + p^{\species}_{\unit} \, f^{\species}_{S,{\unit}} \, P_{\unit}(t) + \delta
    \big\}
    \, S^{\species}_{\unit}(t)\, dt + S^{\species}_{\unit}(t)\, d\zeta^{\species}_{S,{\unit}},\\
    dI^{\species}_{\unit}(t) &= p^{\species}_{\unit} \, f^{\species}_{S,{\unit}}\,  S^{\species}_{\unit}(t) \, P_{\unit}(t)\, dt -
    \big\{
    \theta^{\species}_{I,{\unit}} + \delta
    \big\}
    \, I^{\species}_{\unit}(t)\, dt + I^{\species}_{\unit}(t) \, d\zeta^{\species}_{I,{\unit}},\\
    dJ^{\species}_{\unit}(t) &= r^{\species}_{\unit} \, f^{\species}_{S,{\unit}} \, F_{\unit}(t) \, S^{\species}_{\unit}(t)\, dt -
    \big\{
    \theta^{\species}_{J,{\unit}}  + \delta + \lambda^{\species}_{J,{\unit}}
    \big\} 
    \, J^{\species}_{\unit}(t)\, dt + J^{\species}_{\unit}(t)\, d\zeta^{\species}_{J,{\unit}}, \\
    dP_{\unit}(t) &= \hspace{-1mm} \sum_{\species \in \{\native, \invasive\}} \hspace{-1mm} \left(\beta^{\species}_{\unit} \, \theta^{\species}_{I,{\unit}} \,  I^{\species}_{\unit}(t) - f^{\species}_{S,{\unit}} \big\{ S^{\species}_{\unit}(t) + \xi_{\unit} I^{\species}_{\unit}(t) \big\} P_{\unit}(t)\right) \! dt - \theta_{P,{\unit}} \, P_{\unit}(t)\, dt + P_{\unit}(t)\, d\zeta_{P,{\unit}},\\
    dF_{\unit}(t) &= - \hspace{-1mm} \sum_{\species \in \{\native, \invasive\}} f^{\species}_{S,{\unit}} \, F_{\unit}(t) \left(S^{\species}_{\unit}(t)+\xi_{J,{\unit}} \, J^{\species}_{\unit}(t) +\xi_{\unit} \, I^{\species}_{\unit}(t)\right)dt  + \mu \, dt + F_{\unit}(t)\, d\zeta_{F,{\unit}},\\
d\zeta^{\species}_{S,{\unit}} &\sim \Normal \big[0, (\sigma^{\species}_{S,{\unit}})^{2}\, dt\big],
\quad
d\zeta^{\species}_{I,{\unit}} \sim \Normal \big[0, (\sigma^{\species}_{I,{\unit}})^{2}\, dt\big],
\quad
d\zeta^{\species}_{J,{\unit}} \sim \Normal \big[0, (\sigma^{\species}_{J,{\unit}})^{2}\, dt\big],\\
d\zeta_{F,{\unit}} &\sim \Normal \big[0, \sigma_{F,{\unit}}^{2}\, dt\big],
\quad
d\zeta_{P,{\unit}} \sim \Normal \big[0, \sigma_{P,{\unit}}^{2}\, dt\big].
\label{eq:model:last}
\end{align}}
Here, $\Normal[\mu,\sigma^2]$ is the normal distribution with mean $\mu$ and variance $\sigma^2$.
Intraspecific competition among hosts is conceptualized via limitation of food and environment resources, with $r^{\species}_{\unit}$ denoting the reproductive rate for species $\species \in \{\native,\invasive\}$.
Host density growth, influenced by algal density, is posited to be linear, indicating a direct relationship between food availability and \textit{Daphnia} reproduction.
Mortality rates of susceptible hosts of species $\species$ are denoted as $\theta^{\species}_{S,{\unit}}$.
The model represents the dynamics of infection through \textit{M.~bicuspidata} ingestion by susceptible hosts.
Specifically, susceptible hosts become infected at a rate depending on the density of parasite spores, $P_{\unit}(t)$ and the host filtration rate, $f^{\species}_{S,{\unit}}$.
This filtration rate is expected to be smaller for infected and juvenile individuals, modeled by rate factors $\xi_{\unit}$ and $\xi_{J,{\unit}}$ which are modeled as equal for both host species.

Infected hosts of species $\species$ die at rate $\theta^{\species}_{I,{\unit}}$, at which point $\beta^{\species}_{\unit}$ parasite spores are released.
We do not measure the spores in this experiment, and so only the product $\beta^{\species}_{\unit}\, p^{\species}_{\unit}$ is identified but not the separate values of $\beta^{\species}_{\unit}$ and $p^{\species}_{\unit}$.
We therefore fix $\beta^{\species}_{\unit}=3\times 10^3$ for $\species\in\{\native,\invasive\}$ and for all units \citep{penczykowski14}, though \citet{searle16} estimated a slightly smaller value.
Maturation and death of juvenile individuals are modeled to occur at rate  $\lambda^{\species}_{J,{\unit}}$ and  $\theta^{\species}_{J,{\unit}}$ respectively.

The model's effectiveness at statistically describing the variation in \textit{Daphnia} population dynamics depends critically on its ability to account for overdispersion, a phenomenon which is overlooked in the common practice of using binomial and Poisson variation to add stochasticity to deterministic models \citep{breto11}.
For the process model, this overdispersion is implemented by the noise parameters, $\sigma^{\species}_{S,{\unit}}$, $\sigma^{\species}_{I,{\unit}}$,  $\sigma^{\species}_{J,{\unit}}$, $\sigma^{}_{F,\unit}$, $\sigma^{}_{P,{\unit}}$.
For the measurement model, we obtain overdispersion by using a negative binomial model for the observed counts.
\begin{align}
N^{\species}_{S,{\unit},n} &\sim \operatorname{NBinomial}\big(
  S^{\species}_{\unit}(t_n), \tau^{\species}_{S,{\unit}}
\big), \quad
N^{\species}_{I,{\unit},n} \sim \operatorname{NBinomial}\big(
  I^{\species}_{\unit}(t_n), \tau^{\species}_{I,{\unit}}
\big)
\end{align}
where $\operatorname{NBinomial}(\mu,\tau)$ is the negative binomial distribution with mean $\mu$ and variance $\mu+ \frac{\mu^2}{\tau}$.
Since the sampled volume is $1$ liter, the densities $S^{\species}_{\unit}(t_n)$ and $I^{\species}_{\unit}(t_n)$ are on the same scale as the respective counts, $N^{\species}_{S,{\unit},n}$ and  $N^{\species}_{I,{\unit},n}$.
We do not include the data on juveniles during the model fitting process, to keep our approach comparable to the investigation by \citet{searle16}.
Therefore, these data are available for an out-of-sample model validation.

Simpler treatments with a single \textit{Dapnia} species (SIRJPF) and without the parasite (SRJF and SRJF2) arise as special cases of SIRJPF2, with appropriate states and parameters set to zero.
These experiment treatments and corresponding models are presented in Section~S4, S5 and S6.
By isolating the food variable in the absence of parasites and/or inter-species competition among {\it Daphnia}, these results provide additional evidence on how food resource availability or scarcity impacts the population dynamics of native and invasive species.

\section{Methodology for PanelPOMP models}
\label{sec:meth}

The applications of POMP models span various fields, including epidemiology \citep{mietchen2024,fox22,wen2024}, ecology \citep{auger-methe21,marino19,blackwood13}, and finance \citep{breto14}, among others.
In epidemiological studies, POMP models are instrumental in estimating disease transmission rates and predicting future outbreaks by modeling the spread of infectious diseases when only a fraction of cases are observed \citep{subramanian21}.
Similarly, POMP models help in understanding animal population dynamics and migration patterns based on limited or sporadic sighting data \citep{auger-methe21}.
Widely used latent state estimation and parameter inference methods for nonlinear non-Gaussian POMP models build on the particle filter, also known as sequential Monte Carlo \citep{arulampalam02}.
Particle filters involve simulating a large number of potential realizations of the latent process and updating these simulations as new observations become available, thereby approximating the conditional distribution of the state given the data.
Various plug-and-play inference methods including iterated filtering, particle Markov chain Monte Carlo, and approximate Bayesian computing (ABC) are available using software such as pomp \citep{king16} or NIMBLE \citep{de17}.

Algorithms and software for vector-valued POMP models could in principle be applied to PanelPOMP models, but in practice the high dimensionality of panel data requires modified algorithms.
The panel iterated filter (PIF) algorithm of \citet{breto20} enables plug-and-play likelihood-based inference for general PanelPOMP models.
We start by setting up some general notation that we use to describe PIF, a variant of PIF called the marginalized panel iterated filter (MPIF), and our specific model for the mesocosm experiment.

Time series data are collected on $\Unit$ units, labeled $1,2,\dots,\Unit$.
Unit $\unit$ is observed at $N_u$ measurement times, denoted by $t_{\unit,1},t_{\unit,2},\dots,t_{\unit,N_\unit}$ which we write as $t_{\unit,1:N_\unit}$.
For our example, $N_{\unit}=N=10$, and $t_{\unit,n}=t_n=5n+2$ days, though the algorithms and software used do not need regular observation times.
The observation on unit $\unit$ at time $t_{\unit,n}$ is denoted by $y^*_{\unit,n}$,
and in our example this corresponds to a vector of measured densities of {\it Daphnia} classified by species, life stage, and infection status.
The data are modeled by a stochastic process $\{Y_{\unit,n}, \unit\in \seq{1}{\Unit}, n\in \seq{1}{N_\unit}\}$.
We suppose there is an unobserved continuous-time Markov process $\{ X_\unit(t), t_{u,0}\le t\le t_{N_u}\}$ such that $Y_{u,n}$ is a noisy measurement of $X_u(t_{u,n})$.
In our example,
\begin{eqnarray}
X_{\unit}(t)&=&\big(
  S^{\native}_{\unit}(t), I^{\native}_{\unit}(t),  J^{\native}_{\unit}(t),
  S^{\invasive}_{\unit}(t), I^{\invasive}_{\unit}(t),  J^{\invasive}_{\unit}(t),
  P_{\unit}(t), F_{\unit}(t)
\big),
\\
Y_{\unit,n}&=&\big(
  N^{\native}_{S,\unit,n},  N^{\native}_{I,\unit,n},
  N^{\invasive}_{S,\unit,n},  N^{\invasive}_{I,\unit,n}
\big).
\end{eqnarray}
The time at which the latent process for unit $u$ is initialized is $t_{u,0}$, and in our specific example $t_{u,0}=t_0=0$.
We write $X_{\unit,n}=X_{\unit}(t_{\unit,n})$, and we note that the collection of these discrete-time latent state values is sufficient to describe the model for the data.
Nevertheless, in many situations (including this case study) it is scientifically helpful to construct the latent process in continuous time.

The PanelPOMP model depends on a parameter vector, $\theta$, which we write as $\theta=(\phi,\psi_{1:\Unit})$ where $\phi$ is the vector of parameters shared between all units and $\psi_\unit$ is a vector of parameters specific to unit $\unit$.
We can then define a general PanelPOMP model in terms of the transition density $f_{X_{\unit,n}|X_{\unit,n-1}}(x_{\unit,n}|x_{\unit,n-1};\phi,\psi_{\unit})$, the initial density $f_{X_{\unit,0}}(x_{\unit,0};\phi,\psi_{\unit})$, and the measurement density $f_{Y_{\unit,n}|X_{\unit,n}}(y_{\unit,n}|x_{\unit,n};\phi,\psi_{\unit})$ of each unit $\unit$ in the panel.
For a model such as ours, the transition density is defined implicitly by the model equations.
The plug-and-play property means that we never require explicit evaluation of this density, only the ability to obtain simulated trajectories satisfying equations~(\ref{eq:model:first}--\ref{eq:model:last}).
We do that using Euler's method \citep{kloeden99}.

A defining feature of a PanelPOMP is that the latent processes are modeled as independent across units, so the data generating models for each unit are linked only via the shared parameter vector, $\phi$.
The likelihood function can therefore be written as
\begin{equation}
\mathcal{L}(\theta) = \prod_{\unit=1}^{U} \int f_{Y_{\unit,1:N}|X_{\unit,1:N}}(y^*_{\unit,1:N}|x_{\unit,1:N};\phi,\psi_{\unit}) \, f_{X_{\unit,1:N}}(x_{\unit,1:N};\phi,\psi_\unit)\, dx_{\unit,1:N}.
\end{equation}
Here, we are supposing the existence of conditional probability densities; these may be interpreted as probability mass functions if the latent process and/or the measurement model are discrete-valued.
The likelihood function cannot generally be evaluated explicitly.
However, techniques that let us approximately evaluate and maximize this likelihood, while making proper accommodation for the approximation error, let us employ likelihood-based inference tools for parameter estimation, confidence intervals, diagnosis of model misspecification, and model selection.
Our subsequent investigation demonstrates all these aspects of data analysis.

\begin{figure}[h!]
\centering
\resizebox{\textwidth}{!}{%
\begin{tikzpicture}[
    node distance=5cm
  ]

\node (N) at (0,0) [decision, yshift=-0.5cm] {\texttt{n=1:N}};
\node (initialize) [process, left of=N,xshift= 0cm] {Initialize particle \texttt{j}, unit \texttt{u},\\
time $\mathtt{t_0}$, states \& parameters};
\node[fit=(initialize),myfit] (initialize1) {};
\node[mytitle] at (initialize1.north west) {\texttt{j=1:J}};

\node (U) [decision, left of=initialize1, xshift= 0cm] {\texttt{u=1:U}};
\node (M) [decision2, left of=U, xshift= 2.5cm] {\texttt{m=1:M}};
\node (S) [mytitle, above of=M, xshift=1cm] {\textbf{Input:} PanelPOMP model\\ and data};

\draw[arrow,black] (initialize1) -- (N);
\draw[arrow,black] (U) -- (initialize1);
\draw[arrow,black] (M) -- (U);
\draw[arrow,black] ($(S.south)!.5!(S.south west)$) -- (M);    
\node (predict) [process, right of=N,xshift=0cm,yshift=0.1cm] {Simulate stochastic dynamics,\\ unit $\mathtt{u}$, time $\mathtt{t_{n-1}}$ to $\mathtt{t_{n}}$};
\node (perturb) [process2, above of=predict,yshift=-3cm] {Perturb parameters $\mathtt{\phi}$, $\mathtt{\psi_u}$};
\node (weight) [process, below of=predict,yshift=2.8cm] {Calculate weights for data,\\ unit $\mathtt{u}$, time $\mathtt{t_{n}}$};
\draw[arrow,black] (perturb) -- (predict);
\draw[arrow,black] (predict) -- (weight);
\node[fit=(perturb)(predict)(weight),myfit] (ppw) {};
\node[mytitle] at (ppw.north west) {\texttt{j=1:J}};
\draw[arrow,black] (N) -- (ppw);

\node (resampleParams) [process2, right of=predict,xshift=2.5cm,yshift=-1.1cm] {Resample parameters};
\node (resampleState) [process, above of=resampleParams,yshift=-3cm] {Resample states};
\node[fit=(resampleParams)(resampleState),myfit] (resampleProcess) {};
\node[mytitle] at (resampleProcess.north west) {\texttt{j=1:J}};
\draw[arrow,black] (ppw) -- (resampleProcess);

\node (resampleProcess_south1) [below of=resampleProcess,yshift= 0.75cm] {};
\node (resampleProcess_south2) [below of=resampleProcess,yshift= 0.25cm] {};
\node (resampleProcess_south3) [below of=resampleProcess,yshift=-0.25cm] {};
\node (N_south) [below of=N,yshift=0.75cm] {};
\node (U_south) [below of=U,yshift=0.25cm] {};
\node (M_south) [below of=M,yshift=-0.25cm] {};

\draw[arrow,black] (resampleProcess.south) -- (resampleProcess_south1.south) -- (N_south.south) -- (N.south);
\draw[arrow,black] (resampleProcess.south) -- (resampleProcess_south2.south) -- (U_south.south) -- (U.south);
\draw[arrow,black] (resampleProcess.south) -- (resampleProcess_south3.south) -- (M_south.south) -- (M.south);

\node (O) [mytitle, below of=resampleProcess, yshift=-2.5cm,xshift=-0.36cm] {\textbf{Output: }Monte Carlo Maximum\\
 Likelihood Parameter Estimate};
\draw[arrow,black] (resampleProcess.south) -- (O.north) ;
\end{tikzpicture}
}
\caption{
A flow diagram for panel iterated filtering.
If the blue shaded blocks are omitted, this becomes a standard particle filter for each unit. Full pseudocode is in Section~S2. In the PIF algorithm of \citet{breto20}, the parameter resampling step involves re-selecting the unit-specific parameters corresponding to particle $j$ for units $\tilde u \neq u$ as well as $\psi_u$.
For the marginalized PIF algorithm, only $\phi$ and $\psi_u$ are reselected in this step.
}
\label{fig:pif}
\end{figure}
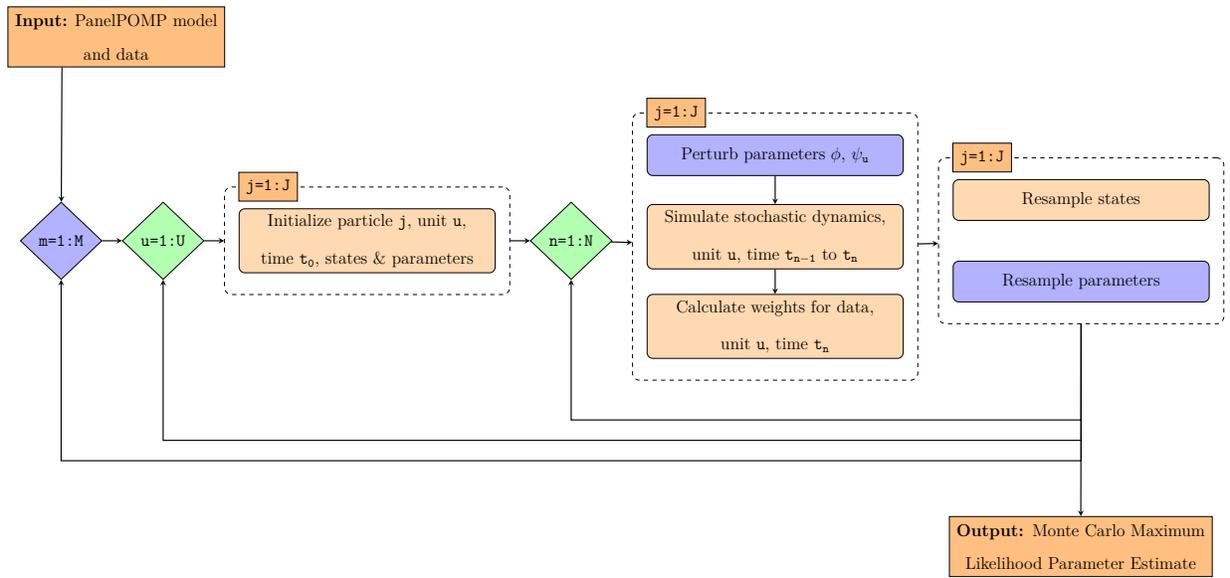

Particle filters use a Monte Carlo swarm of latent state values, known as particles, to represent the conditional distribution of the latent variables given the data.
Particles are updated according to the dynamics of the latent process and then resampled according to their likelihood given the data.
This resampling is key to the particle filter's efficacy, since it allows the algorithm to focus its computational effort on the most probable states of the latent process, as indicated by the observed data.
Iterated particle filters maximize the likelihood for a POMP model by associating a parameter value with each particle.
An analogy with Darwinian evolution explains heuristically how this happens.
The parameter values are the heritable DNA belonging to each particle.
The parameter perturbations are analogous to mutations, and the resampling is analogous to natural selection which preferentially favors the fittest particles, i.e., those with parameter values most consistent with the data.
\citet{kitagawa98} noticed that this favorable property of a particle filter, with perturbed parameters, can be empirically shown to move the population of these parameter values toward a region of increasing likelihood.
\citet{ionides06} and \citet{ionides15} found ways to iterate this process that lead to concentration around the MLE.
The panel extension of iterated filtering by \citet{breto20} applies the same principle to PanelPOMP models.
A flow diagram for a panel iterated filtering (PIF) algorithm is shown in Fig.~\ref{fig:pif}, and full details are given in Section~S2.
At each iteration, $m=\seq{1}{M}$, the magnitude of the perturbations is reduced, and we use geometric ``cooling'' with a reduction factor $\rho<1$.
By providing likelihood-based inference for a general class of nonlinear, non-Gaussian PanelPOMP models, PIF has the potential to facilitate discoveries in a wide range of scientific domains.
This article provides a case study for such an application in the field of ecology.
Parameter estimates and corresponding confidence intervals were obtained using PIF.
A sufficiently large number of particles is needed to maintain particle diversity and achieve robust exploration of the parameter space.
However, computational cost increases with the number of particles.
We found that for this model and data, $J=500$ particles and cooling parameter $\rho = 0.7^{1/50} = 0.993$ gave sufficient Monte Carlo accuracy and computational feasibility.
The likelihood maximization via the PIF algorithm was conducted in three successive stages of $M=150$, $M=150$, and $M=250$ iterations, respectively.
This tempering approach has commonly been found useful for iterated filtering applications \citep{bhadra11}. 
After completing each stage, we selected the top 25\% of parameter values ranked by their corresponding log-likelihoods and used these as the initial parameter swarm for the subsequent stage.
Numerical experimentation under this setting provided log-likelihood estimates with standard error $0.35$ and adequately supported parameter inference while keeping computations within the capabilities of one 36-core Linux machine.

In PanelPOMP models, it is necessary to decide which parameters should be shared across units, versus which should be unit-specific.
This decision can be guided both by scientific considerations and by empirical evidence.
As the number of units in the PanelPOMP model increases, the number of parameters increases linearly with the number of unit-specific parameters.
Therefore, balancing model fit and parsimony becomes essential to avoid overly complex models.
Here, we use AIC, defined to be twice the number of estimated parameters minus twice the maximized log-likelihood.
AIC compensates for over-fitting by seeking to optimize out-of-sample forecast skill for a log-likelihood scoring rule \citep{aic74}.
Thus, AIC penalizes complexity only so far as it affects generalization error for prediction; parsimony has additional scientific benefits, but the scientific value of model simplicity cannot be determined by a purely statistical criterion.
Various other information criteria have been proposed and used in ecological data analysis, in addition to likelihood ratio hypothesis tests \citep{johnson04}.
However, AIC has earned widespread use due to its clear theoretical motivation and its applicability for evaluating collections of non-nested hypotheses, including comparing mechanistic models to non-mechanistic benchmark models.

Profile likelihood provides a useful approach for constructing confidence intervals in nonlinear stochastic systems \citep{simpson23}.
Calculation of the profiles can validate that sufficient computational effort was conducted in order for the Monte Carlo optimization to converge: by increasing Monte Carlo replications, consistent convergence to the same maximum of the profile likelihood curve confirms adequate numerical effort for optimization.
To obtain confidence intervals, we correct for the Monte Carlo uncertainty in the profile likelihood evaluations by using the Monte Carlo Adjusted Profile (MCAP) method of \citet{ionides17}.

\begin{table}[t]
\renewcommand{\arraystretch}{0.5}
\centering
\resizebox{\textwidth}{!}{
\begin{tabular}{||c|c |c|c|c||}
\hline
Parameter & Definition &Unit & Value & CI\\[0.5ex]
\hline
\hline
$S^{k}$ & Susceptible host density for species $k$  & $\mathrm{individual} \cdot L ^{-1}$                 & Variable                & \\
$I^{k}$ & Infected host density for species $k$     & $ \mathrm{individual} \cdot L ^{-1}$                & Variable                & \\
$J^{k}$ & Juvenile host density for species $k$     & $ \mathrm{individual} \cdot L ^{-1}$                & Variable                & \\
$F$   & Alga density                              & $10^6 \cdot \mathrm{cell} \cdot L ^{-1}$             & Variable                & \\
$P$   & Spore density                             & $10^3 \cdot \mathrm{spore} \cdot L ^{-1}$            & Variable                & \\
$r^{\native}$ & Birth rate of native juvenile                         & $\mathrm{individual} \cdot 10^{-6}\cdot \mathrm{cell}^{-1}$ & $4.08\cdot 10$  & ($2.64\cdot 10$,$1.94\cdot 10^{2}$)\\
$r^{\invasive}$ & Birth rate of invasive juvenile                       & $\mathrm{individual} \cdot 10^{-6}\cdot \mathrm{cell}^{-1}$ & $2.15\cdot 10^{5}$  & ($1.74\cdot 10^{2}$, $\infty$)\\
$f_{S^{\native}}$ & Native susceptible adult host filtering rate   &  $\mathrm{L} \cdot \mathrm{individual}^{-1} \cdot \mathrm{day}^{-1}$   & $1.10\cdot 10^{-3}$  & ($4.21\cdot 10^{-4}$,$1.64\cdot 10^{-3}$)\\
$f_{S^{\invasive}}$ & Invasive susceptible adult host filtering rate &  $\mathrm{L} \cdot \mathrm{individual}^{-1} \cdot \mathrm{day}^{-1}$   & $2.42\cdot 10^{-7}$  & (0,$9.47\cdot 10^{-4}$)\\
$p^{\native}$          & Number of native infections per spore             &     $10^{-3} \cdot \mathrm{individual} \cdot \mathrm{spore}^{-1} $ & $2.72\cdot 10^{-1}$  & ($1.54\cdot 10^{-1}$,$7.28\cdot 10^{-1}$)\\
$p^{\invasive}$          & Number of invasive infections per spore           &     $10^{-3} \cdot \mathrm{individual} \cdot \mathrm{spore}^{-1} $ & $1.34\cdot 10^{3}$  & ($1.13$,$\infty$)\\
$p^{\native} \cdot f_{S^{\native}}$          & Effective infection rate of native adult hosts per‐spore      &     $10^{-3} \cdot \mathrm{L} \cdot \mathrm{spore}^{-1} \cdot \mathrm{day}^{-1}$  & $2.99\cdot 10^{-4}$  & ($2.13\cdot 10^{-4}$,$4.61\cdot 10^{-4}$)\\
$p^{\invasive} \cdot f_{S^{\invasive}}$          & Effective infection rate of invasive adult hosts per‐spore      &     $10^{-3} \cdot \mathrm{L} \cdot \mathrm{spore}^{-1} \cdot \mathrm{day}^{-1}$ & $3.25\cdot 10^{-4}$  & ($2.44\cdot 10^{-4}$,$4.87\cdot 10^{-4}$)\\
$r^{\native} \cdot f_{S^{\native}}$          & Effective birth rate of native juvenile            &     $10^{-6} \cdot \mathrm{L} \cdot \mathrm{cell}^{-1} \cdot \mathrm{day}^{-1} $ & $4.49\cdot 10^{-2}$  & ($3.83\cdot 10^{-2}$,$8.53\cdot 10^{-2}$)\\
$r^{\invasive} \cdot f_{S^{\invasive}}$          & Effective birth rate of invasive juvenile            &     $10^{-6} \cdot \mathrm{L} \cdot \mathrm{cell}^{-1} \cdot \mathrm{day}^{-1} $ & $5.21\cdot 10^{-2}$  & ($4.47\cdot 10^{-2}$,$6.91\cdot 10^{-2}$)\\
$\theta^{\native}_{S}$ & Native susceptible adult host mortality rate     &     $ \mathrm{day}^{-1}$     & $8.48\cdot 10^{-4}$  & (0,$1.68\cdot 10^{-1}$)\\
$\theta^{\invasive}_{S}$ & Invasive susceptible adult host mortality rate   &     $  \mathrm{day}^{-1}$     & $2.52\cdot 10^{-3}$  & (0,$5.83\cdot 10^{-3}$)\\
$\theta^{\native}_{I}$ & Native infected adult host mortality rate        &     $  \mathrm{day}^{-1}$     & $5.84\cdot 10^{-1}$  & ($2.75\cdot 10^{-1}$,$8.84\cdot 10^{-1}$)\\
$\theta^{\invasive}_{I}$ & Invasive infected adult host mortality rate      &     $  \mathrm{day}^{-1}$     & $3.85\cdot 10^{-1}$  & ($2.21\cdot 10^{-1}$,$5.92\cdot 10^{-1}$)\\
$\theta^{\native}_{J}$ & Native juvenile mortality rate        &     $ \mathrm{day}^{-1}$     & $1.87\cdot 10^{-5}$  & (0,$1.66\cdot 10^{-2}$)\\
$\theta^{\invasive}_{J}$ & Invasive juvenile mortality rate      &     $  \mathrm{day}^{-1}$     & $5.62\cdot 10^{-4}$  & (0,$1.79\cdot 10^{-2}$)\\
$\theta_{P}$   & Spore degradation rate                           &     $\mathrm{day}^{-1}$     & $9.48\cdot 10^{-4}$   & (0,$7.05\cdot 10^{-2}$)\\
$\lambda^{\native}_{J}$ & Maturation rate of native juvenile                  &     $ \mathrm{day}^{-1} $                  & $1.00\cdot 10^{-1}$  & \\
$\lambda^{\invasive}_{J}$ & Maturation rate of invasive juvenile                &     $ \mathrm{day}^{-1} $                  & $1.00\cdot 10^{-1}$  & \\
$\xi$           & Ratio of infected host filtering rate       &     $\mathrm{Unitless}$                    & $2.22\cdot 10$         & ($1.01\cdot 10$,$7.10\cdot 10$)\\
$\xi_J$         & Ratio of juvenile individual filtering rate &     $\mathrm{Unitless}$                    & $1.00$ & \\
$\beta^{\native}$       & Spores produced per infected native individual     &  $10^3 \cdot \mathrm{spores} \cdot \mathrm{individual}^{-1} \cdot \mathrm{day}^{-1}$  & $3.00\cdot 10$ & \\
$\beta^{\invasive}$       & Spores produced per infected invasive individual   &  $10^3 \cdot \mathrm{spores} \cdot \mathrm{individual}^{-1} \cdot \mathrm{day}^{-1}$  & $3.00\cdot 10$ & \\
$\mu$           & Alga refilling rate                                &  $10^6 \cdot \mathrm{cell} \cdot \mathrm{L} ^{-1} \cdot \mathrm{day}^{-1}$                 & $3.70\cdot 10^{-1}$ & \\
$\delta$        & Sampling rate                                      &  $\mathrm{day}^{-1}$                                                                         & $1.30\cdot 10^{-2}$ & \\
$\sigma^{\native}_{S}$ & Standard deviation of Brownian motion of susceptible native adult         &   $\sqrt{\mathrm{individual} \cdot \mathrm{day}^{-1}}$  & 0  & \\
$\sigma^{\invasive}_{S}$ & Standard deviation of Brownian motion of susceptible invasive adult         &   $\sqrt{\mathrm{individual} \cdot \mathrm{day}^{-1}}$  & 0  & \\
$\sigma^{\native}_{I}$ & Standard deviation of Brownian motion of infected native adult         &   $\sqrt{\mathrm{individual} \cdot \mathrm{day}^{-1}}$  & $2.93\cdot 10^{-4}$  & (0,$\infty$)\\
$\sigma^{\invasive}_{I}$ & Standard deviation of Brownian motion of infected invasive adult       &   $\sqrt{\mathrm{individual} \cdot \mathrm{day}^{-1}}$  & $1.73\cdot 10^{-7}$  & (0,$\infty$)\\
$\sigma^{\native}_{J}$ & Standard deviation of Brownian motion of native juvenile               &   $\sqrt{\mathrm{individual} \cdot \mathrm{day}^{-1}}$  & $2.84\cdot 10^{-1}$  & ($1.50\cdot 10^{-1}$,$4.80\cdot 10^{-1}$)\\
$\sigma^{\invasive}_{J}$ & Standard deviation of Brownian motion of invasive juvenile             &   $\sqrt{\mathrm{individual} \cdot \mathrm{day}^{-1}}$  & $3.02\cdot 10^{-1}$  & ($1.35\cdot 10^{-1}$,$4.20\cdot 10^{-1}$)\\
$\sigma_{F}$   & Standard deviation of Brownian motion of alga                          &   $\sqrt{\mathrm{individual} \cdot \mathrm{day}^{-1}}$  & $1.44\cdot 10^{-1}$  & ($5.41\cdot 10^{-2}$,$2.33\cdot 10^{-1}$)\\
$\sigma_{P}$   & Standard deviation of Brownian motion of parasite                      &   $\sqrt{\mathrm{individual} \cdot \mathrm{day}^{-1}}$  & $2.71\cdot 10^{-1}$  & ($4.41\cdot 10^{-2}$,$4.20\cdot 10^{-1}$)\\
$\tau^{\native}_{S}$      &  Measurement dispersion for susceptible native adult       &$\mathrm{Unitless}$           & $4.10$  & ($2.92$,$7.65$)\\
$\tau^{\invasive}_{S}$      &  Measurement dispersion for susceptible invasive adult       &$\mathrm{Unitless}$            & $5.26$  & ($2.66$,$8.60$)\\
$\tau^{\native}_{I}$      &  Measurement dispersion for infected native adult        &$\mathrm{Unitless}$            & $9.02\cdot 10^{-1}$  & ($6.69\cdot 10^{-1}$,$1.70$)\\
$\tau^{\invasive}_{I}$      &  Measurement dispersion for infected invasive adult        &$\mathrm{Unitless}$            & $1.39$  & ($6.77\cdot 10^{-1}$,$2.27$)\\
\hline
\end{tabular}}
\caption{
Variables and parameter definitions and estimates.
The superscript $\native$ and $\invasive$ represents native (\textit{D.~dentifera}) and invasive (\textit{D.~lumholtzi}) respectively.
}
\label{tab:estimates}
\renewcommand{\arraystretch}{1}
\end{table}

\section{Results}
\label{sec:res}

\begin{table}[t]
\renewcommand{\arraystretch}{0.5}
\centering
\small
\begin{center}
\small
\begin{tabular}{||c |c |c |c||}
 \hline
Model & Parameters & Max log-likelihood & AIC \\ [0.5ex]
 \hline\hline
SIRJPF2 & 26 & -880.56 & 1809.12 \\
SIRPF2 & 20 & -891.80 & 1823.60 \\
SIRJPF2-Gamma & 26 & -892.03 & 1836.06 \\
Negative Binomial (Cubic) & 24 & -932.61 & 1913.21 \\
Negative Binomial (Quadratic) & 20 & -943.79 & 1927.59 \\
Negative Binomial (Linear) & 16 & -1009.25 & 2050.50 \\
Searle et al. (2016) & 19 & -2483.49 & 5004.97 \\[0.5ex]
\hline
\end{tabular}
\end{center}
\caption{
Comparison of model fit for mechanistic models and non-mechanistic benchmarks.}
\label{Table:Model_comparison}
\renewcommand{\arraystretch}{1}
\end{table}

Parameter estimates and confidence intervals for the SIRJPF2 model are in Table~\ref{tab:estimates}, which also includes a brief description of the parameter mechanism and the corresponding measurement units.
The log-likelihood, AIC and number of parameters for this model is compared against alternatives in Table~\ref{Table:Model_comparison}.
Differences in AIC values between models are intepretable only between alternative models for the same data, so we fit statistical benchmark models to calibrate these results.
Specifically, we fit negative binomial regression models with linear, quadratic, and cubic dependence on time.
Comparing a mechanistic model to a simple statistical benchmark model is a useful test of model specification, since it reveals quickly whether the mechanistic model is constructed in a way that is statistically compatible with the data \citep{wheeler24,li24}.
The negative binomial models with polynomial (quadratic or cubic) terms fit the dynamics better than a linear trend.
Although these more flexible polynomial expansions do not represent latent biological processes directly, they better capture the non-linearlity of the system, leading to a higher log-likelihood and a lower AIC than the linear model.

Table~\ref{Table:Model_comparison} includes the log-likelihood and AIC values for the fitted model and the deterministic ODE model of \citet{searle16}, which served as a starting point for our model.
Although the mechanistic model of \citet{searle16} was successfully used to provide scientific insight into the ecological dynamics of the experiment, we found it does not provide a competitive statistical fit to the data compared to simple alternatives.
We also consider several variants of the SIRJPF2 model described here, namely a model that excludes the juvenile state (SIRPF2, described in Supplement~S7.3), and a version of the SIRJPF2 model where the Gaussian noise has been replaced with Gamma noise (SIRJPF2-Gamma). 
The SIRPF2 model ignores the process of maturation of juvenile and instead highlights the infection process of the \textit{Daphnia} adult populations.
While this model does have an improved statistical fit compared to the deterministic ODE model of \citet{searle16} and other benchmark models, it has weaker performance than the model that accounts for the effect of juveniles on population dynamics.

Our SIRJPF2 model quantitatively describes the observed data better than any of the alternatives, while simultaneously providing a plausible mechanistic description of the experimental system.
Having assessed the fit of our model, we have some support for interpreting the estimated parameters of the model.
It is hard to rule out the possibility that some biological details in the model are misspecified in ways that may lead to confounding with the modeled mechanisms and therefore bias in the causal interpretation of estimated parameters \citep{wheeler24}.
With that caveat acknowledged, we proceed to interpret the model and parameters at face value.

There are some substantial differences in the observed trajectories between experimental replicates (Figs.~\ref{fig:data_vis}, S5, S6, S14, S19, S20).
This could result from small differences in the initialization of the replicates, or it could simply be a consequence of the natural variation in any biological system.
In the context of our model, these hypotheses correspond to unit-specific parameters and stochastic dynamics, respectively.
Here, the evidence suggests that stochastic dynamics are sufficient to explain the variation in the data.
Table~S1 in the supplement presents a comparison of model fit and complexity across various configurations of unit-specific parameters within the PanelPOMP framework for the SIRJPF2 model. These distinct configurations represent varying hypotheses explaining the differences across replicates.
The models assessed include those with unit-specific parameters for combinations of mortality rates ($\theta^\native_{S},\theta^\invasive_{S}$,$\theta^\native_{I},\theta^\invasive_{I}$), filter rates ($f^\native_{S},f^\invasive_{S}$), growth efficiency factors ($r^{\native},r^{\invasive}$), and spore infection rates ($p^{\native},p^{\invasive}$).

When parameters are shared between units, we obtain a substantial reduction in the number of parameters to be estimated.
For field data, one may expect the need for unit-specific parameters describing structural differences between units.
For example, in a PanelPOMP analysis of pre-vaccine polio transmission dynamics for state-level data in USA, \citet{breto20} used unit-specific parameters to describe variations in the seasonality of transmission across units.
Modeling decisions concerning shared and unit-specific parameters are testable hypotheses that are of scientific interest in the quest for a parsimonious understanding of the dynamics, and for understanding sources of variation.

In our selected model, the feeding rate of the invasive {\it Daphnia} species ($f^{\invasive}_{S}$) is not statistically significantly different from that of the native {\it Daphnia} species ($f^{\native}_{S}$).
Also, the invasive species has a lower rate of infection per consumed spore ($p^{\native} < p^{\invasive}$), consistent with Fig.~1C of \citet{searle16}.
The native parameters—juvenile growth $r^{\native}$, feeding rate $f^{\native}_{S}$ and infection probability $p^{\native}$ show full identifiability as shown in Fig.~S1, whereas the corresponding invasive parameters $r^{\invasive},\,f^{\invasive}_{S},\,p^{\invasive}$ are only bounded on one side. 
This limited identifiability is attributable to
(i) the lower density of \textit{D. lumholtzi} which may increase stochastic effects;
(ii) model complexity, since the parameters become identifiable in host-only experiments with simplified dynamics (Fig.~S8);
(iii) the product of these parameters governs the juvenile growth and infection process, which allows individual invasive coefficients to vary so long as their products remain fixed.
To investigate this last explanation, Table~\ref{tab:estimates} shows relatively precise identifiability of the effective infection rate ($p^{\native}f^{\native}_{S}$, $p^{\invasive}f^{\invasive}_{S}$), and the effective juvenile birth rate ($r^{\native}f^{\native}_{S}$, $r^{\invasive}f^{\invasive}_{S}$).
At the same time, the identifiability of the measurement error parameter ($\tau^{\native}_{S},\,\tau^{\invasive}_{S},\,\tau^{\native}_{I},\,\tau^{\invasive}_{I}$) explains variability that would otherwise inflate weakly identified process-noise terms.
Our results indicate that stochasticity in the juvenile stage ($\sigma^{\native}_{J}$ and $\sigma^{\invasive}_{J}$), food ($\sigma_{F}$) and parasite ($\sigma_{P}$) dynamics are important model features, needed to adequately describe the data (see the corresponding confidence intervals in Table~\ref{tab:estimates} and the profile likelihood plots in Fig.~S1).
That suggests the juvenile stage, food and parasite dynamics are particularly sensitive to unmodeled phenomena, such as lower fitness resulting from resource depletion, since unmodeled phenomena can be explained by the model as stochastic uncertainty.
Noise in other stages is small, and often insignificantly different from zero, as shown by profile likelihood plots for $\sigma^{\native}_{I}$, $\sigma^{\invasive}_{I}$.
Since the profile for $\sigma^{\native}_{S}$, $\sigma^{\invasive}_{S}$ is flat and insignificantly different from zero, we fixed the $\sigma^{\native}_{S}$ and $\sigma^{\invasive}_{S}$ to be zero as constants.
The proposed model is flexible enough to describe the predictability and uncertainty inherent in the system in various ways, and the data have come down in favor of $\sigma^{\native}_{J}$ and $\sigma^{\invasive}_{J}$.
This indicates that the model might be recalibrated after fixing some or all of the other noise intensities to zero, but there is no pressing need to do so.
The noise term is helpful when describing other dynamics, for example, dynamics with two species of \textit{Daphnia} and no parasite as shown in Supplement~S5.
Natural death rates, ($\theta^{\native}_{S}$, $\theta^{\invasive}_{S}$, $\theta^{\native}_{J}$ and $\theta^{\invasive}_{J}$), are poorly identified in the presence of the parasite since most death comes through the infection process.
By contrast, these parameters are better identified in the treatments without the parasite (Fig.~S15).

Alternative  SIRJPF2 model specifications were fitted with various parameters taken to be unit-specific.
The log-likelihood estimates and corresponding AIC values for these models are presented in Supplementary Table~S1.
These results suggest that introducing unit-specific parameters increased model complexity without significantly improving model performance.
Similar results hold for the other treatments (Tables~S4, S5, S7, S10, S11).
Consequently, we choose to select the model with all shared parameters, as it offers a more parsimonious and robust representation of the system dynamics without substantially compromising the explanatory power.
And the confidence intervals are obtained by the MCAP are shown in Fig.~S1 for the SIRJPF2 model, and Figs.~S7, S8, S15, S21, S22 for other treatments.

Our estimated value of the filtration rate factor for infected {\it Daphnia} is $\xi > 1$, which is superficially incompatible with previous empirical observations that infection leads to reduced filtration, especially as disease progresses \citep{searle16,penczykowski14}.
One way to reconcile these results is the possibility that \textit{Daphnia} which filter faster are exposed to more spores and are consequently more likely to become infected.
There are six genotypes in each experimental unit, and we only observe population-level dynamics rather than individual-level data.
When modeling these dynamics, our parameters of interest actually represent population-level means.
In other words, given this specific combination of genotypes, $\xi > 1$ indicates that at the population level, the average filtration rate of infected \textit{Daphnia} is higher than that of susceptible \textit{Daphnia}.
However, the direction can be reversed at the individual level.
$\xi > 1$ could also be a result of reverse causation due to unmodeled heterogeneity: \textit{Daphnia} with higher filtering rate are exposed to more pathogen spores and so filtering rate can causally affect infection status in addition to any causal effect of infection status on filtering rate.
A subsequent investigation extending the model and measurements to include \textit{Daphna} size and genotype could help to clarify this.

As shown in Table~\ref{tab:estimates}, the estimated values of $p^{\native}$ and $p^{\invasive}$ differ significantly, suggesting that the parasite influences the proportion of individuals becoming infected within each species in distinct scale.
This difference, in turn, affects the interactions between the two species.

\section{Model diagnostics}
\begin{figure}[t]
\centering
\begin{knitrout}
\definecolor{shadecolor}{rgb}{0.969, 0.969, 0.969}\color{fgcolor}
\includegraphics[width=\textwidth]{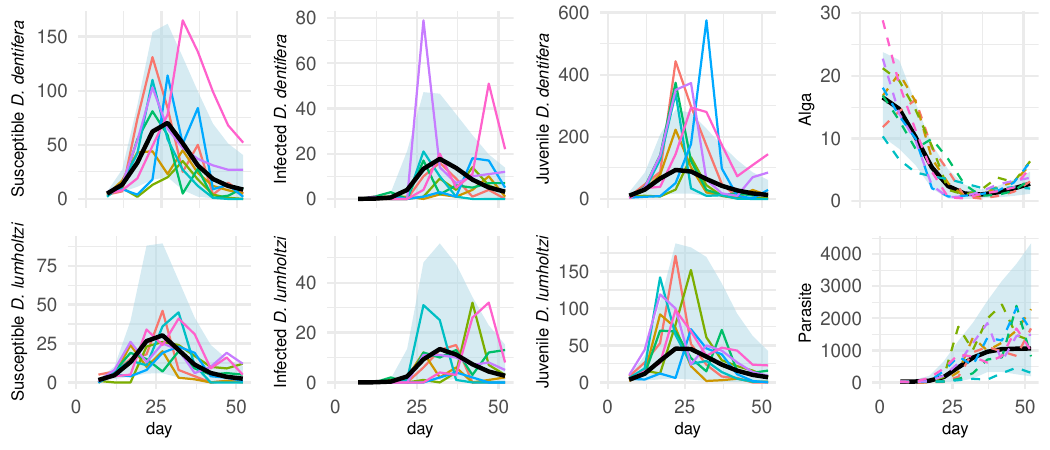} 
\end{knitrout}
\caption{\label{fig:sim_plot}
Simulated \textit{Daphnia} densities (Individuals$/$Liter), alga density ($10^6\cdot$cells$/$Liter) and parasite density ($10^3\cdot$spores$/$Liter), over time (days), in the mixed species parasitized dynamics with parameters from Table~\ref{Table:Model_comparison}. 
Blue bands are $95\%$ pointwise confidence intervals; black lines show the mean; solid color lines are the experimental data.
The parasite and food densities were not observed, and the dashed colored lines are simulations.
}
\end{figure}

Our best-fitting model includes explicit modeling of juveniles and their maturation, a feature not present in the model of \citet{searle16}. 
The extra delay in the adult dynamics introduced by the juvenile stage appears to be necessary for the model to provide a strong statistical fit to the data.
In the original study, \citet{searle16} collected juvenile data but did not include it as part of the data analysis. 
To make our analysis comparable to \citet{searle16}, we fitted to the same data, and therefore did not include a measurement model for juveniles.
Instead, the juvenile data are available to use as model validation, which is done by comparing simulations from the fitted model to the unmodeled data in Fig.~\ref{fig:sim_plot}.

Knowing that the average body length of juvenile \textit{Daphnia} is shorter than the average body size of adult \textit{Daphnia} for both species, we expect that juvenile individuals will have a lower filtering rate.
Yet, without fitting to juvenile data, it is unclear whether that question is identifiable from the model at hand.
To explore this assumption, we compared the model using the value $\xi_{J,{\unit}} = 1$ (juveniles and adults filter at the same rate) against the extreme value $\xi_{J,{\unit}} = 0.001$ (juveniles filter much slower than adults).

Likelihood residual analysis is an effective tool for comparing alternative models by assessing differences in the conditional log-likelihood for each data point \citep{wheeler24,li24}.
We maximized the log-likelihood subject to the fixed values of $\xi_{J,{\unit}}$, and we compared the conditional log-likelihood for each observation category at each time point for each fixed $\xi_{J,{\unit}}$ value, as shown in Fig.~S3(a).
These results show that setting $\xi_{J,{\unit}} = 1$ leads to only a slightly higher maximized log-likelihood ($-881.13$) than $\xi_{J,{\unit}} = 0.001$ ($-882.24$), indicating that the parameter is weakly identified.
Moreover, plotting the conditional log-likelihood for each state at each time, we observe that the log-likelihood components for susceptible and infected \textit{D. dentifera} and \textit{D. lumholtzi} are similar between the two parameter settings.
These results suggest that both parameter values provide indistinguishable explanations of the observed data, which is a stronger conclusion than obtaining a similar total log-likelihood.

The choice of $\xi_{J,{\unit}}$ nevertheless affects the interpretation of the fitted model: the maximum likelihood estimates constrained to different values of $\xi_{J,{\unit}}$ lead to distinct behavior of juvenile \textit{Daphnia} in simulations, as shown in Fig.~S3(b).
If $\xi_{J,{\unit}}$ is set to be lower than is biologically plausible, simulated trajectories from the fitted model have far fewer juveniles than was observed, which is shown in Fig.~S3(b) as well.
This indicates model misspecification which could bias the estimates of $r^{\species}_{\unit}, p^{\species}_{\unit}$ and $ f^{\species}_{S,\unit}$.
When a weakly identified parameter has consequences for the biological interpretation of the model, it is appropriate to fix that parameter at a biologically plausible value, and we therefore fixed $\xi_{J,{\unit}} = 1$ for subsequent analysis, reasoning that larger juveniles should have filtering rates of the same order of magnitude as adults.

In addition to comparing model fit using AIC, we further validate the calibrated SIRJPF2 model via a simulation study.
We generate 1000 simulations from the model and compare these simulations to the data (Fig.~\ref{fig:sim_plot}).
Unit level simulation plots are shown in Fig.~S2.
The simulation bands presented in Fig.~\ref{fig:sim_plot} closely align with the empirical data curves from the actual experiment, underscoring the model's capability to replicate the observed dynamics.
Notably, the model captures not only the trends but also the variability of the experimental data.
Simulations from other models considered are included in the supplementary Figs.~S9--S12, S16, S17, S23--S26.

The consistency of simulated data with the empirical data provides a qualitative indicator of the fit of the model that complements a comparison of log-likelihood values.
Our simulations from  the fitted SIRJPF2 model align with the observed data for the susceptible and infected species of \textit{Daphnia}, and from the out-of fit juvenile data.
This encourages us to explore the dynamics of latent states in the dynamic system by simulating the unobserved densities of \textit{M.~bicuspidata} and \textit{A.~falcatus} (see the third column of Fig.~\ref{fig:sim_plot}).
The model predicts that the alga density decreases dramatically from its starting value, and then starts to rebound, due to resupply, once the \textit{Daphnia} population has crashed.
The parasite density is predicted to rise sharply during the population growth phase and then plateaus at a high level.

As mentioned in Section~\ref{sec:mecmod}, high population density of \textit{Daphnia} and limited food resources will prompt the production of male \textit{Daphnia} leading to sexual reproduction and the generation of diapausing eggs.
That biological feature plays a large role in the model of \cite{searle16}, though Table~\ref{Table:Model_comparison} shows that our proposed mechanism for the role of resource depletion provides a better statistical explanation of the analyzed data.
Our modeling and inference approach could be extended to include data on males and ephippial females, collected by \cite{searle16} but not included in either their model or ours.

\section{Discussion}
\label{sec:dis}

Data science is having a transformative impact on ecological science \citep{mouquet15} and yet too few studies undertake the task of reconciling ecological theory with time series data \citep{peacor22}.
We have shown that our panel data methodology can allow full-information plug-and-play likelihood-based inference for the complex partially observed nonlinear stochastic dynamic models required to carry out this reconciliation.
If the previous shortage of such data analyses is evidence for the previous lack of effective methods, PanelPOMP methodology should facilitate many previously intractable scientific investigations.
Further advances in methodology, such as automatic differentiation for particle filters \citep{tan24}, will continue to increase the numerical tractability of PanelPOMP models.

Despite the complexity of our SIRJPF2 model, there are various ways in which further biological mechanisms could be incorporated.
Our general plug-and-play framework can be further extended, and information criteria can be used to assess the importance of the proposed extension.
For example, we could have incorporated the phenomenon that \textit{Daphnia} with advanced infection have reduced feeding rate, or we could have allowed for the possibility that recently infected \textit{Daphnia} can continue to reproduce.
Additional residual analysis can search for patterns suggesting model improvements, following a similar approach to the residual analysis carried out here.

Although we did not calibrate the model using juvenile data, the addition of latent $J^{\species}_{\unit}$ states leads to an improved statistical fit.
This could arise because the juvenile stage provides a delay in the {\em Daphnia} population growth dynamics, and because juveniles compete with adults for available food resources.
Juveniles rarely exhibit disease, perhaps because of the latency in the parasite's development within its host, but they nevertheless play a significant role in the infection dynamics.
While we have shown the relevance of age-structured dynamics for understanding the data on adults, future work could identify the role of juveniles in the system dynamics more precisely.
Adding juvenile data to the measurement model in the fitting procedure would enable higher statistical resolution on these dynamics.
Our general POMP modeling and inference framework allows for the possibility of additional measurements or alternative model hypotheses.

Fitting a mechanistic model can provide a coherent explanation of system dynamics, but one must remain cautious about the possibility of confounding factors: mechanisms that are omitted from the model and which provide alternative (and perhaps scientifically pertinent) explanations of the observed phenomena \citep{li24,wheeler24}.
We have an external test of our data by showing that it fits the juvenile data even though it was not fit to them, but this validation does not guarantee that all conclusions from the fitted model are correct.
For example, while our results explain the decrease of \textit{Daphnia} density by resource depletion, data on the exact food level were not available.
The mechanism, although logical within the model’s framework, remains an unvalidated assumption in the theory rather than being directly supported by data.
Subsequent experimentation could substantiate or refute this prediction, and our data analysis methodology can assist such investigations.

Deterministic dynamic models continue to be widely used for ecological theory, despite the fact that inclusion of stochasticity can both improve model fit and enable diagnosis of model shortcomings \citep{king15}.
We demonstrate that including stochasticity can also lead to more parsimonious understanding of the system, since, in our case study, the dynamic stochasticity obviates a need for mesocosm-specific parameters to explain the difference between the experimental replicates.

This study has focused on the task of fitting models to understand the biological dynamics.
This is closely related to the task of forecasting: fitting by maximum likelihood is equivalent to assessing one-step forecasting skill, and the AIC criterion is designed to evaluate out-of-sample predictive skill \citep{aic74}.
Within this framework, the goals of model-based biological understanding and forecasting are therefore closely aligned.

We have analyzed data collected in a controlled environment, raising the question of the applicability of our methods to field ecosystems. 
The applicability of our approach is contingent on having panel data of sufficient length and quality to inform interactions between species.
Such panel datasets are available, for example, on pathogen-host ecosystems \citep{martinez-bakker15} and fisheries \citep{naiman02}.
The flexibility of the PanelPOMP model class provides many opportunities for developing new models that integrate ecological theory with panel time series data.

\section{Supplementary Materials}
The Supplementary Material includes detailed Panel Iterated Filtering algorithm and the benchmark model specifications.
It also provides profile likelihoods, residual analyses, and diagnostics for the SIRJPF2, SIRJPF, and SRJF2 models, together with models for the additional dynamics discussed in the manuscript.

\section{Data Availability Statement}
Code and data for the analysis is available at the Github repository: https://github.com/Megumi-ybb/Daphnia-ms or https://doi.org/10.5281/zenodo.15595669.

\section{Disclosure Statement}
No potential conflict of interest was reported by the authors.

\end{document}


\def\spacingset#1{\renewcommand{\baselinestretch}%
{#1}\small\normalsize} \spacingset{1}

\if1\blind
{
  \title{\bf Supplement to ``Mechanistic models for panel data: Analysis of ecological experiments with four interacting species''}
\author{Bo Yang$^1$, Jesse Wheeler$^2$, Meghan A. Duffy$^3$, Aaron A. King$^{3,4}$ \\ and Edward L. Ionides$^2$\\
\authoraddress University of Michigan, Departments of $^1$Biostatistics, $^2$Statistics,
\\
\authoraddress $^3$Ecology \& Evolutionary Biology, $^4$Complex Systems}
  \maketitle
} \fi

\if0\blind
{
  \bigskip
  \bigskip
  \bigskip
  \begin{center}
    {\LARGE\bf Supplement to "Mechanistic models for panel data: analysis of ecological experiments with four interacting species"}
\end{center}
  \medskip
} \fi

\tableofcontents
\newpage
\section{Monte Carlo Adjusted Profile}

Monte Carlo Adjusted Profile (MCAP) methods \citep{ionides17,ning21} provide confidence intervals for situations where the likelihood is evaluated and maximized by Monte Carlo algorithms.
A smoothed estimate of the profile likelihood is used to reduce Monte Carlo error, quantify this error, and adjust the confidence intervals accordingly to maintain their coverage.

\parskip 3mm

MCAP is particularly useful for high-dimensional situations, such as arise in panel data analysis, since it becomes practically impossible to apply sufficient computational effort to make Monte Carlo error negligible.
MCAP has previously been demonstrated for panel iterated filtering by \citet{breto20}.

\parskip 3mm

To investigate identifiability, we carried out an additional exploration to obtain estimates, profiles and confidence intervals for the composite parameters,  $r^{\species}f^{\species}_{S}$ and $r^{\species}f^{\species}_{S}$, as well as the individual parameters.
We could have reparameterized to apply MCAP to these composite parameters exactly as we did for the individual parameters.
However, for this additional analysis we made a simplification by re-using the calculations for the original profiles.
For every point on every profile, computed for the individual parameters, we calculated the value of the composite parameters.
We obtained the highest log-likelihoods within a grid of values for the composite parameters and applied MCAP to these values.
We have found previously that these ``poor man's profiles'' provide a reasonable approximation to the true profile, while avoiding the need to work with a reparameterized version of the model.
We calculated these composite parameters only for the SIRJPF2 model, and the results are shown together with the individual parameters, in Table~1 and Figure~S-1.

\section{The Panel Iterated Filtering Algorithm}

\begin{algorithm}[h]
\caption{Panel iterated filter (PIF)}
\label{alg:pif}
\DontPrintSemicolon
\KwIn{PanelPOMP model, as defined in Section~3 of the main text\;
  Number of particles, $J$, and filter iterations, $M$\;
  Cooling factor $\rho$\;
  Starting parameter vector swarm, \( \Theta^{(0)}_{j} = (\Phi^{(0)}_{j},\Psi^{(0)}_{1:U,j}) \),
  for $j = \seq{1}{J}$\;
  Perturbation variance, $V^\Theta_{u,n}$, for $u=\seq{1}{U}$, $n=\seq{1}{N_u}$\;
  Logical variable determining marginalization, $\mathrm{MARGINALIZE}$
  }
\KwOut{Parameter swarm, $\Theta^{(M)}_{1:J}$, approximating the MLE}
\For{\( m = 1 \) \KwTo \( M \)}{
  $\Theta^{F,m}_{0,j} = \Theta^{(m-1)}_{j}$ for $j = \seq{1}{J}$\;
  \For{\( u = 1 \) \KwTo \( U \)}{
    $(\Phi^{F,m}_{u,0,j},\Psi^{F,m}_{u,0,j}) \sim \mathcal{N}(\Theta^{F,m}_{u-1,j},\rho^{2m}V^\Theta_{u,0})$ for $j = \seq{1}{J}$\;
    \( X_{u,0,j}^{F,m} \sim f_{X_{u,0}}(\, \cdot \params \Phi^{F,m}_{u,0,j},\Psi^{F,m}_{u,0,j}) \) for $j = \seq{1}{J}$\;
    \For{\( n = 1 \) \KwTo \( N_u \)}{
       $\Phi^{P,m}_{u,n,j} \sim \mathcal{N}\big(\Phi^{F,m}_{u,n-1,j}, \rho^{2m}V^\Phi_{u,n}\big)$ for $j = \seq{1}{J}$\;
       $\Psi^{P,m}_{u,n,j} \sim \mathcal{N}\big(\Psi^{F,m}_{u,n-1,j}, \rho^{2m}V^\Psi_{u,n}\big)$ for $j = \seq{1}{J}$\;
       \( X_{u,n,j}^{P,m} \sim f_{X_{u,n} | X_{u,n-1}}\big(\, \cdot \mid X_{u,n-1,j}^{F,m} \params \Phi^{P,m}_{u,n,j},\Psi^{P.m}_{u,n,j}\big) \) for \( j = \seq{1}{J} \)\;
       $w_{u,n,j}^m = f_{Y_{u,n} | X_{u,n}}\big(y^*_{u,n} \mid X^{F,m}_{u,n,j} \params \Phi^{P,m}_{u,n,j},\Psi^{P.m}_{u,n,j}\big)$ for \( j = \seq{1}{J} \)\;
       $k_j = i$ with probability proportional to $ w_{u,n,i}^m$ for $i,j = \seq{1}{J}$\;
       $X^{F,m}_{u,n,j} = X^{P,m}_{u,n,k_j}$ for $j = \seq{1}{J}$\;
       $\Phi^{F,m}_{u,n,j} = \Phi^{P,m}_{u,n,k_j}$ for $j = \seq{1}{J}$\;
       $\Psi^{F,m}_{u,n,j} = \Psi^{P,m}_{u,n,k_j}$ for $j = \seq{1}{J}$\;
              \uIf{$\mathrm{MARGINALIZE}$}{$\Psi^{F,m}_{\tilde u,n,j} = \Psi^{P,m}_{\tilde u,n,j}$ for all $\tilde u \neq u$, $j=\seq{1}{J}$}
       \Else{$\Psi^{F,m}_{\tilde u,n,j} = \Psi^{P,m}_{\tilde u,n,k_j}$ for all $\tilde u \neq u$, $j=\seq{1}{J}$}
       }
  $\Theta^{F,m}_{u,j} = \big( \Phi^{F,m}_{u,N_u,j},  \Psi^{F,m}_{1:U,N_u,j} \big)$
  }
  $\Theta^{(m)}_{j} = \Theta^{F,m}_{U,j}$;
}
\end{algorithm}

Algorithm~\ref{alg:pif} presents pseudocode for PIF.
Here, $\Phi^{F,m}_{u,n,j}$ is the $j$th filter particle searching for the MLE of the shared parameter $\phi$ in the $m$th filter iteration, at time point $n$ and unit $u$.
 $\Psi^{F,m}_{u,n,j}$ is the corresponding quantity searching for the MLE of the unit-specific parameter, $\psi_u$.
Cooling factor $\rho$ determines how aggressively the parameter perturbations are dampened as the number of iterations increases.
By gradually reducing the size of these parameter random walks, the algorithm refines the parameter estimates and converges more tightly to the maximum likelihood point.

When $\mathrm{MARGINALIZE}=\mathrm{TRUE}$, the estimate of the unit-specific parameter $\psi_u$ belonging to particle $j$ is unchanged when filtering throuth a unit $\tilde u\neq u$. 
The unmarginalized PIF was proposed by \cite{breto20} and was provided with theoretical convergence results that are not yet available for MPIF.
However, our results demonstrate that MPIF has superior empirical performance on the ecological model considered here.
Theoretical support for MPIF will be published elsewhere \citep{wheeler25}.

To briefly review, the model is specified by the initial condition, $ f_{X_{u,0}}(\, \cdot \, \params \phi,\psi_u)$, the latent state transition density, $f_{X_{u,n} | X_{u,n-1}}\big(\, \cdot \mid \,  \params \phi,\psi_u \big)$, and the measurement density, $f_{Y_{u,n} | X_{u,n}}\big(\, \cdot \mid \cdot \params \phi,\psi_u\big)$, for $u = \seq{1}{U}$ and $n=\seq{1}{N_u}$.
The full parameter vector is written as $\theta=(\phi,\psi_{1:U})$.
For notational convenience, we define $u=0$ to be an empty panel with $N_0=0$.
Algorithm~\ref{alg:pif} supposes Gaussian perturbations of the parameters, which is a common practical choice but may require re-parameterizing to avoid boundaries.
For example, non-negative parameters are typically log-transformed.
Additional generality is provided by the pseudocode in \cite{breto25} that is implemented in the \texttt{mif2} function of the R package \texttt{panelPomp}.

\clearpage

\section{SIRJPF2 Model}
\subsection{Results}
\subsubsection{MCAP Results}
\begin{figure}[H]
    \centering
    \includegraphics[width=\linewidth]{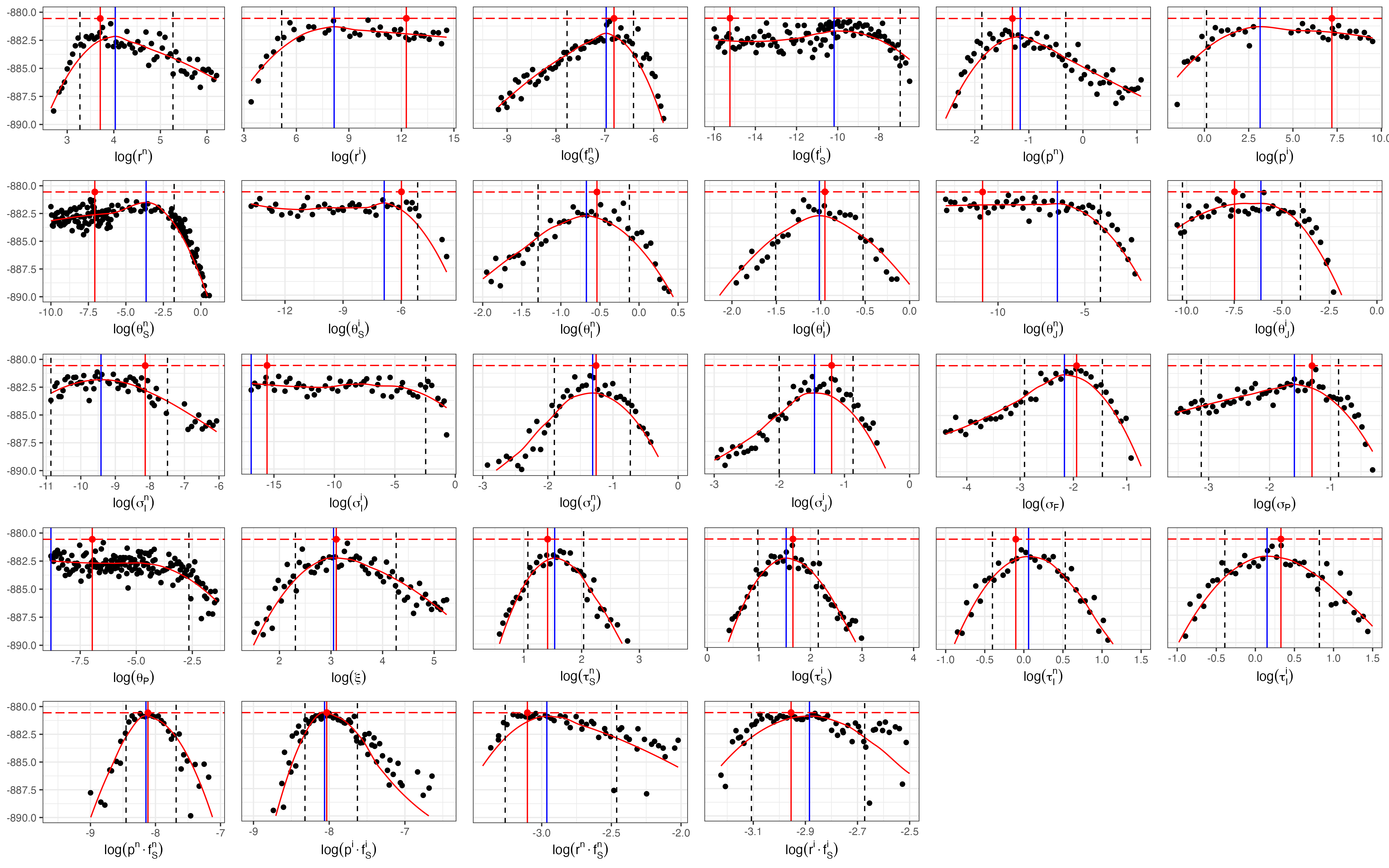}
    \caption{Monte Carlo Adjusted Profile results of SIRJPF2 model for mixed species Parasitized dynamics.
    The vertical dotted lines represent the 95\% confidence interval obtained by MCAP.
    The vertical blue lines show the MLE estimated using MCAP.
    The red vertical lines correspond to the model with the overall highest likelihood among all searches. }
    \label{fig:SIRJPF2_profile}
\end{figure}

\subsubsection{Model Comparison}
In the panelPOMP framework, one must decide whether or not each parameter should be unit specific, or shared across units.
These decisions greatly affect the degree of freedom for each model. For example, with ten units in the model, adding a shared parameter increases the model’s dimension by one, and adding a unit specific parameter will increase the dimension by ten.
An increase in parameter dimension increases the model’s ability to quantitatively describe the observed data, but leads to the possibility of over-fitting, resulting in a model with poor explanatory and predictive power.
In order to find a balance between fitting accuracy and simplicity when selecting models, we employed the Akaike information criterion (AIC) as a criteria to compare models while adjusting for dimension.
We estimate model parameters via maximum likelihood using the panel iterated filtering (PIF) \citep{breto20} and calculate AIC for all models.
The following table shows the comparison of different models under different settings of unit specific parameter.
Each row represents a different model configuration, with specific parameters designated as either unit-specific or shared across all units.
Within the table, the max log-likelihood and AIC columns report the maximum log-likelihood and AIC values for each model configuration, where a lower AIC indicates a more favorable balance between model fit and complexity. Similarly, the max log-likelihood (MPIF) and AIC (MPIF) columns provide analogous metrics under the marginalized method setting for the PIF procedure.
\begin{table}[h!]
\centering
\small
\begin{center}
\begin{tabular}{||c |c |c |c|c||}
 \hline
Specific parameters  & Max log-likelihood & AIC & \makecell[c]{Max log-likelihood\\(MPIF)} & \makecell[c]{AIC\\(MPIF)}\\ [0.5ex]
 \hline\hline
$\varnothing$ & -881.19& 1814.38& -881.19& 1814.38 \\
$\theta^{\species}_{I,\unit}$ & -868.51& 1817.03& -868.51& 1817.03 \\
$p^{\species}_{\unit}$ & -869.77& 1819.54& -867.25& 1814.50 \\
$\theta_{P,\unit}$ & -878.20& 1822.40& -875.03& 1816.06 \\
$\xi_{\unit}$ & -882.78& 1831.55& -879.69& 1825.38 \\
$\theta^{\species}_{S,\unit}$ & -877.00& 1834.00& -872.40& 1824.79 \\
$f^{\species}_{S,\unit}$ & -883.20& 1846.40& -872.92& 1825.84 \\
$r^{\species}_{\unit}$ & -885.42& 1850.85& -869.90& 1819.80 \\[1ex]
 \hline
\end{tabular}
\end{center}
\caption{Comparison of model fit and complexity across various configurations of unit-specific parameters within the panelPOMP framework for mixed species Parasitized dynamics. The unit-specific parameter setting assessed include ($\theta^{\species}_{I,\unit}$), ($p^{\species}_{\unit}$), $\xi_{\unit}$, ($f^{\species}_{S,\unit}$ ), ($\theta^{\species}_{S,\unit}$), $\theta_{P,\unit}$ and ($r^{\species}_{\unit}$) for $\species \in \{\native,\invasive\}$.}
\label{Table:SIRJPF2_model_comparison}
\end{table}

\subsection{Simulation}

\begin{figure}[H]
    \centering
    \includegraphics[width=\linewidth]{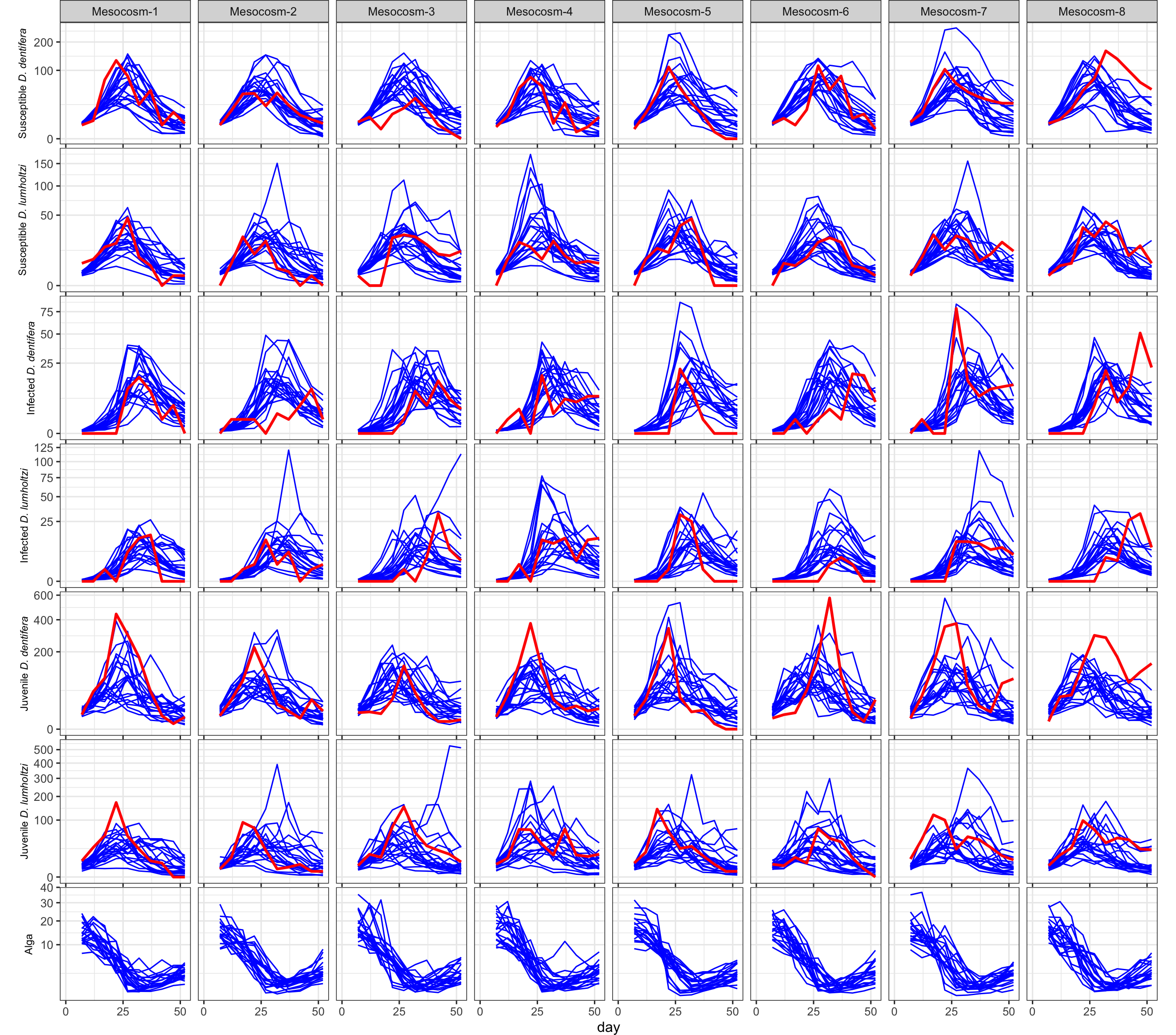}
   \caption{Simulated densities (Individuals$/$Liter) of susceptible, infected and juvenile \textit{D.~dentifera}, susceptible, infected and juvenile \textit{D.~lumholtzi} and alga density ($10^6\cdot$cells$/$Liter) over time for each experimental unit. The blue lines represent individual simulation runs, capturing the variability in susceptible density across replicates, while the red line represents the actual experiment data.}
    \label{fig:SIRJPF2_simulation_lines}
\end{figure}

\subsection{Residual Analysis}

\begin{figure}[H]
    \centering
    \begin{subfigure}{0.75\linewidth}
    \centering
    \includegraphics[width=\linewidth]{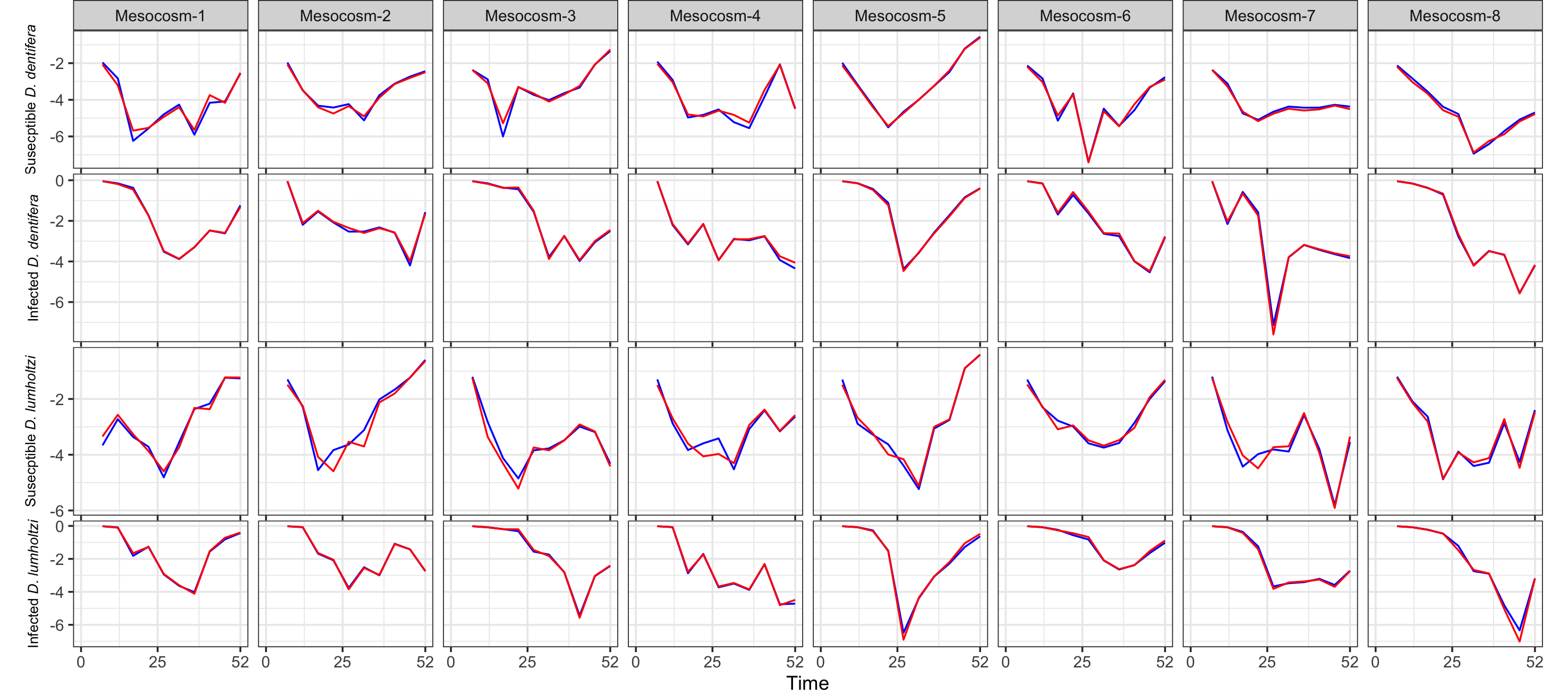}
   \caption{Comparison of conditional log likelihood of model fitting on susceptible, infected \textit{D.~dentifera} and \textit{D.~lumholtzi} at each observation for setting of $\xi_{J,\unit} = 1$ (blue) and $\xi_{J,\unit} = 0.001$ (red).}
    \label{fig:Res_loglik}
    \end{subfigure}
    \vspace{1em}
    
    \begin{subfigure}{0.75\linewidth}
    \centering
    \includegraphics[width=\linewidth]{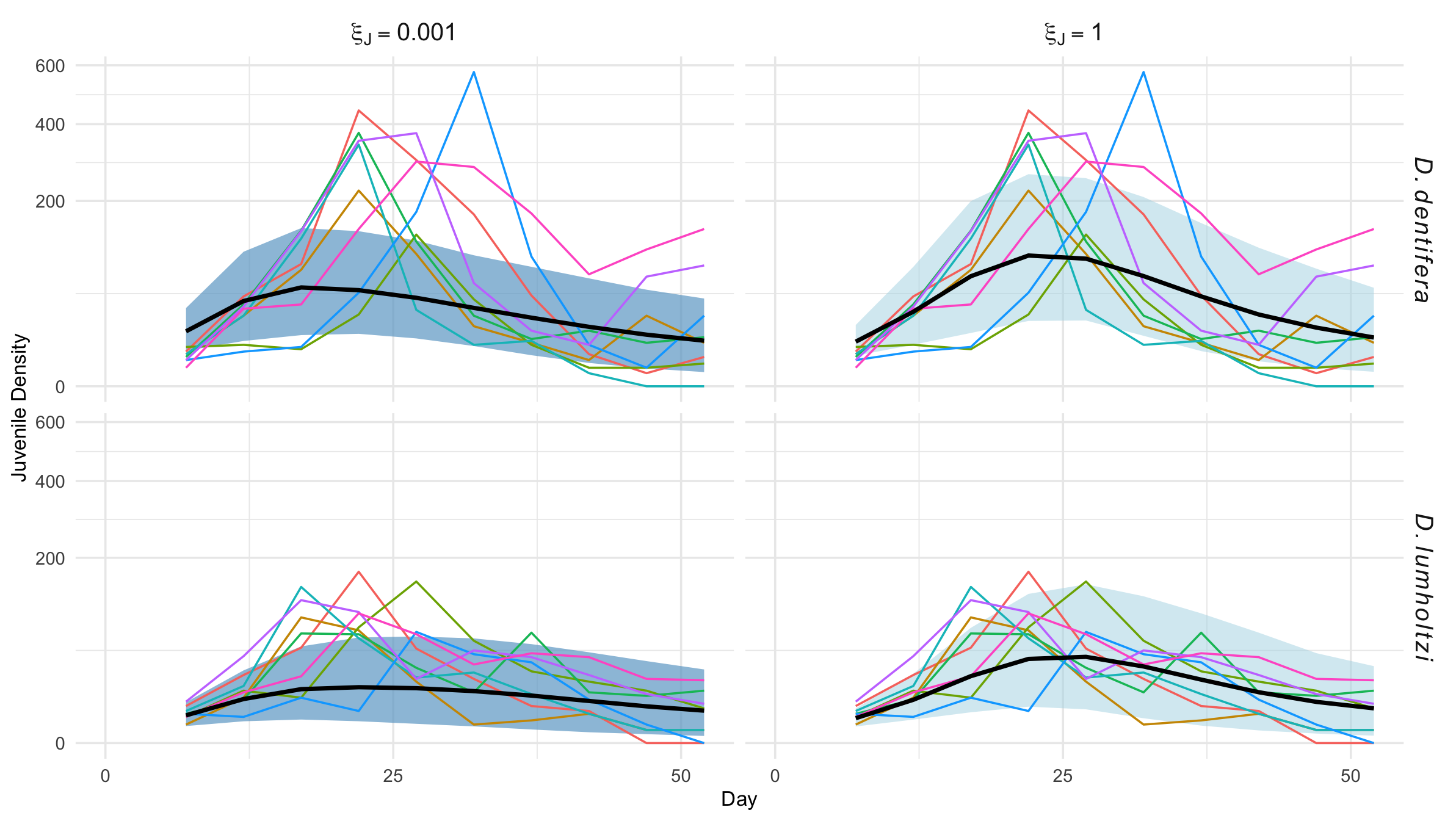}
     \caption{Simulated densities (Individuals$/$Liter) of juvenile \textit{D.~lumholtzi} and \textit{D.~dentifera} over time for in the mixed species parasitized dynamics based on maximized loglikelihood parameters, with simulated mean curves (solid black curves) and $95\%$ confidence band for setting $\xi_{J,\unit} = 0.001$ (purple) and  $\xi_{J,\unit} = 1$ (blue).
The observed densities are represented by the solid colorful curves.}
    \label{fig:Res_sim}
    \end{subfigure}
    \label{fig:Res}
    \caption{Residual analysis results}
\end{figure}

\section{SIRJPF Model}
The SIRJPF model presented here encapsulates the dynamics in each bucket($\unit$) among susceptible individuals ($S^{\species}$), infected individuals ($I^{\species}$), juvenile population ($J^{\species}$), algae as a food resource ($F$), the parasite population ($P$).
The superscript $\species$ denotes the species, which represents native or invasive species.
Each stochastic differential equation accounts for various biological and ecological processes, augmented with stochastic terms to capture the inherent randomness within the systems.
{\small
\begin{align}
    dS^{\species}_{\unit}(t) &= \lambda^{\species}_{J,{\unit}} J^{\species}_{\unit}(t) \, dt - \big\{\theta^{\species}_{S,{\unit}} + p^{\species}_{\unit} f^{\species}_{S,{\unit}} P_{\unit}(t) + \delta \big\} S^{\species}_{\unit}(t)\, dt  + S^{\species}_{\unit}(t)\, d\zeta^{\species}_{S,{\unit}}, \label{eq:SIRJPF_dS}\\
    dJ^{\species}_{\unit}(t) &= r^{\species}_{\unit}f^{\species}_{S,{\unit}}F_{\unit}(t) S^{\species}_{\unit}(t)\, dt - \theta^{\species}_{J,{\unit}} J^{\species}_{\unit}(t)\, dt - \delta J^{\species}_{\unit}(t)\, dt - \lambda^{\species}_{J,{\unit}} J^{\species}_{\unit}(t)\, dt + J^{\species}_{\unit}(t)\, d\zeta^{\species}_{J,{\unit}}, \label{eq:SIRJPF_dJ}\\
    dI^{\species}_{\unit}(t) &= p^{\species}_{\unit} f^{\species}_{S,{\unit}} S^{\species}_{\unit}(t) P_{\unit}(t)\, dt - \big\{\theta^{\species}_{I,{\unit}} + \delta \big\} I^{\species}_{\unit}(t)\, dt + I^{\species}_{\unit}(t) \, d\zeta^{\species}_{I,{\unit}}, \label{eq:SIRJPF_dI}\\
    dP_{\unit}(t) &= \beta^{\species}_{\unit} \theta^{\species}_{I,{\unit}} I^{\species}_{\unit}(t)dt - f^{\species}_{S,{\unit}} ( S^{\species}_{\unit}(t) + \xi_{\unit} I^{\species}_{\unit}(t))P_{\unit}(t) dt - \theta_{p,{\unit}} P_{\unit}(t)\, dt + P_{\unit}(t)\, d\zeta_{P,{\unit}}, \label{eq:SIRJPF_dP}\\
    dF_{\unit}(t) &= - f^{\species}_{S,{\unit}} F_{\unit}(t) \left(S^{\species}_{\unit}(t)+\xi_{J,{\unit}}J^{\species}_{\unit}(t) +\xi_{\unit} I^{\species}_{\unit}(t)\right)dt  + \mu \, dt + F_{\unit}(t)\, d\zeta_{F,{\unit}}, \label{eq:SIRJPF_dF}\\
d\zeta^{\species}_{S,{\unit}} &\sim \Normal \big[0, (\sigma^{\species}_{S,{\unit}})^{2}\, dt\big],
d\zeta^{\species}_{I,{\unit}} \sim \Normal \big[0, (\sigma^{\species}_{I,{\unit}})^{2}\, dt\big],
d\zeta^{\species}_{J,{\unit}} \sim \Normal \big[0, (\sigma^{\species}_{J,{\unit}})^{2}\, dt\big],\\
d\zeta_{F,{\unit}} &\sim \Normal \big[0, (\sigma_{F,{\unit}})^{2}\, dt\big],
d\zeta_{P,{\unit}} \sim \Normal \big[0, (\sigma_{P,{\unit}})^{2}\, dt\big].
\end{align}
}

Each equation represents specific biological interactions and processes.
Equation \eqref{eq:SIRJPF_dS} models the dynamics of the susceptible \textit{Daphnia} population across time, $S^{\species}_{\unit}(t)$, where growth occurs from the juvenile population $J^{\species}_{\unit}(t)$ at a maturation rate $\lambda^{\species}_{J,\unit}$, while losses include natural mortality ($\theta^{\species}_{S,\unit}$), infection by intaking parasite at rate $p^{\species}_{\unit} f^{\species}_{S,\unit} P^{\species}_{\unit}(t)$, and sampling rate ($\delta$).
\parskip 3mm

In Equation \eqref{eq:SIRJPF_dJ}, the dynamics of the juvenile population $J^{\species}_{\unit}(t)$ are captured, with growth occurring via feeding on algae, scaled by birth rate $r^{\species}_{\unit}$ and feeding rate $f^{\species}_{S,\unit} F^{\species}_{\unit}(t)$.
Losses in $J^{\species}_{\unit}(t)$ include natural mortality ($\theta^{\species}_{J,\unit}$), sampling rate ($\delta$), and maturation to the adult population at rate $\lambda^{\species}_{J,\unit}$.
Additionally, $J^{\species}_{\unit}(t) \, d\zeta^{\species}_{J,\unit}$ represents stochastic fluctuations affecting the juvenile population.
\parskip 3mm

Equation \eqref{eq:SIRJPF_dI} describes the infected population $I^{\species}_{\unit}(t)$, with new infections arising from interaction between susceptible individuals $S^{\species}_{\unit}(t)$ and alga at rate $p^{\species}_{\unit} f^{\species}_{S,\unit} S^{\species}_{\unit}(t) P^{\species}_{\unit}(t)$.
Infected individuals experience losses through natural mortality ($\theta^{\species}_{I,\unit}$) and sampling rate ($\delta$), while $I^{\species}_{\unit}(t) \, d\zeta^{\species}_{I,\unit}$ accounts for stochastic noise.
\parskip 3mm

Equation \eqref{eq:SIRJPF_dP} models the dynamics of \textit{M.} population $P_{\unit}(t)$, which grows through spore release from the death of infected individuals $I^{\species}_{\unit}(t)$ at a rate $\beta^{\species}_{\unit} \theta^{\species}_{I,\unit}$.
Losses are due to intaking by susceptible and infected individuals, scaled by filtering rate $f^{\species}_{S,\unit}$ and the ratio of filtering rate $\xi_{\unit}$ between subjects before and after infection, and natural mortality at rate $\theta^{\species}_{P,\unit}$.
Stochastic fluctuations are represented by the term $P_{\unit}(t) \, d\zeta_{P,\unit}$.
\parskip 3mm

Finally, Equation \eqref{eq:SIRJPF_dF} models the \textit{A.} alga population $F_{\unit}(t)$. The algae are consumed by susceptible individuals, juveniles, and infected individuals, scaled by consumption rate $f_S$ and ratio of filtering rate $\xi_{J,\unit}$ and $\xi_{\unit}$. Algae gets manually refilled at rate $\mu$, with $F_{\unit}(t) \, d\zeta_{F,\unit}$ representing stochastic noise.
\parskip 3mm

The stochastic terms $d\zeta^{\species}_{S,{\unit}}$, $d\zeta^{\species}_{J,{\unit}}$, $d\zeta^{\species}_{I,{\unit}}$, $d\zeta_{P,{\unit}}$, and $d\zeta_{F,{\unit}}$ represent Brownian motion affecting each population, modeled as Gaussian white noise with mean zero and variances $(\sigma^{\species}_{S,{\unit}})^{2}\, dt$, $(\sigma^{\species}_{J,{\unit}})^{2}\, dt$, $(\sigma^{\species}_{I,{\unit}})^{2}\, dt$, $(\sigma_{P,{\unit}})^{2}\, dt$, and $(\sigma_{F,{\unit}})^{2}\, dt$, respectively.
Similar to the SIRJPF2 model, we fixed the $\sigma^{\species}_{S}$ to be zero due to its flat profile plot.
\parskip 3mm

This model describes the interactions and feedback mechanisms among the populations under study.
The equations describe the growth or decline of each population, influenced by deterministic factors such as reproduction, filtering rates, predation, infection, and mortality, alongside stochastic factors captured through the $d\zeta_{\cdot,u}$ terms for each unit.
These stochastic components represent environmental variability and other unpredictable influences, adhering to the principles of stochastic differential equations.
By incorporating random fluctuations, the model offers a statistically motivated representation of population dynamics, extending beyond the deterministic framework used in previous analysis.
In the following sections, we will present the flow diagram and parameter estimation results to further illustrate the interaction between latent states.

\subsection{Flow Diagram}
The SIRJPF model illustrates the interdependent dynamics among the Susceptible ($S$), Infected ($I$), Juvenile ($J$), Food ($F$), and Parasite ($P$) populations, with transitions representing ecological and biological interactions.
The model integrates deterministic growth and mortality rates with stochastic components, providing a realistic framework for studying population fluctuations under variable environmental conditions.

\usetikzlibrary{positioning}
\usetikzlibrary {arrows.meta}
\usetikzlibrary{shapes.geometric}

\begin{figure}[H]
\begin{center}
\resizebox{8cm}{!}{
\begin{tikzpicture}[
  square/.style={rectangle, draw=black, minimum width=0.5cm, minimum height=0.5cm, rounded corners=.1cm, fill=blue!8},
  rhombus/.style={diamond, draw=black, minimum width=0.1cm, minimum height=0.1cm, fill=purple!8,aspect = 1},
  travel/.style={circle, draw=black, minimum width=0.5cm, minimum height=0.5cm, fill=green!8},
  report/.style={shape=regular polygon, regular polygon sides=8, draw, fill=red!8,minimum size=0.6cm,inner sep=0cm},
  bendy/.style={bend left=10},
  bendy2/.style={bend left=100},
  bendy3/.style={bend left=-100},
  >/.style={shorten >=0.25mm},
  >/.tip={Stealth[length=1.5mm,width=1.5mm]}
]
\tikzset{>={}};

\node (S) at (2.5,0) [square] {$S^{\species}$};
\node (I) at (5.5,0) [square] {$I^{\species}$};
\node (J) at (-0.5,0) [square] {$J^{\species}$};
\node (R) at (2.5,-2)  [rhombus] {$R$};
\node (P) at (5.5,-2)[travel] {$P$};
\node (F) at (-0.5,-2) [report] {$F$};

\draw [->, bendy] (S) to  (I);
\draw [->, bendy] (I) to  (S);
\draw [->, bendy] (J) to  (S);
\draw [->, bendy] (S) to  (J);

\draw [->] (P) --  (S);
\draw [->] (F) --  (R);
\draw [->] (P) --  (R);
\draw [->] (S) -- (R);
\draw [->] (I) -- (R);
\draw [->] (J) -- (R);
\draw [->] (F) -- (S);
\draw [->] (F) -- (J);
\draw [->] (F) -- (I);
\draw (F) edge[loop below] (F);

\draw [->, bendy] (P) to (I);
\draw [->, bendy] (I) to (P);

\end{tikzpicture}
}
\end{center}
\vspace{-5mm}
\caption{Flow diagram for the SIRJPF model illustrating population interactions. The model includes the $R$ state, representing mortality. Susceptible populations ($S^{\species}$) can reproduce into juvenile ($J^{\species}$), where the superscript $\species$ denotes the species of \textit{Daphnia}. And \textit{A.} Algae ($F$) will be refilled, as shown by the recycling arrows. Transitions from $S^{\species}$ to $I^{\species}$ denote infection, and transitions to the $R$ state occur upon death, where deceased individuals release parasites back into the environment. All $S^{\species}$, $I^{\species}$ and $J^{\species}$ consume resources from $F$, and over time, components in $F$ and $P$ also progress to $R$.}
\label{fig:flow_SIRJPF}
\end{figure}
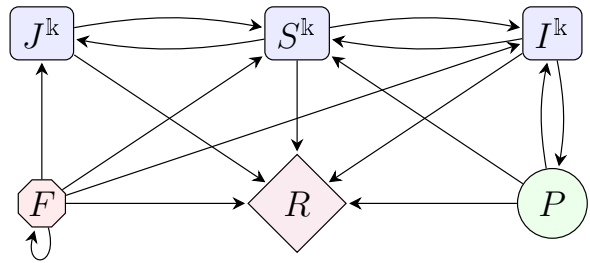

\subsection{Data Visualization}
\begin{figure}[H]
\centering
\begin{knitrout}
\definecolor{shadecolor}{rgb}{0.969, 0.969, 0.969}\color{fgcolor}
\includegraphics[width=\textwidth]{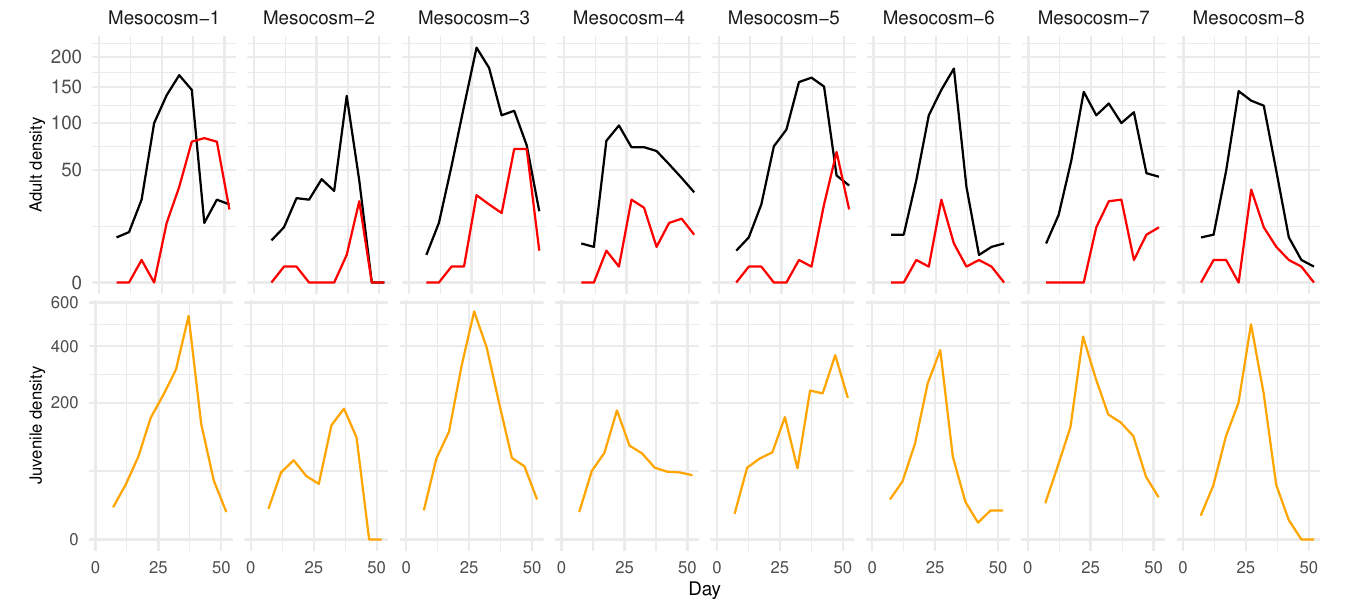} 
\end{knitrout}
\caption{\label{fig:data_vis_SIRPF_dent}
Density (Individuals$/$Liter) of \textit{D.~dentifera}.
The top panel shows adult susceptibles (\textit{D.~dentifera}, black) and infecteds (\textit{D.~dentifera}, red).
The bottom panel shows juvenile susceptibles (\textit{D.~dentifera}, orange).
There were negligible infected juveniles.
Columns are buckets corresponding to replications with same treatment setting.
}
\end{figure}

\begin{figure}[H]
\centering
\begin{knitrout}
\definecolor{shadecolor}{rgb}{0.969, 0.969, 0.969}\color{fgcolor}
\includegraphics[width=\textwidth]{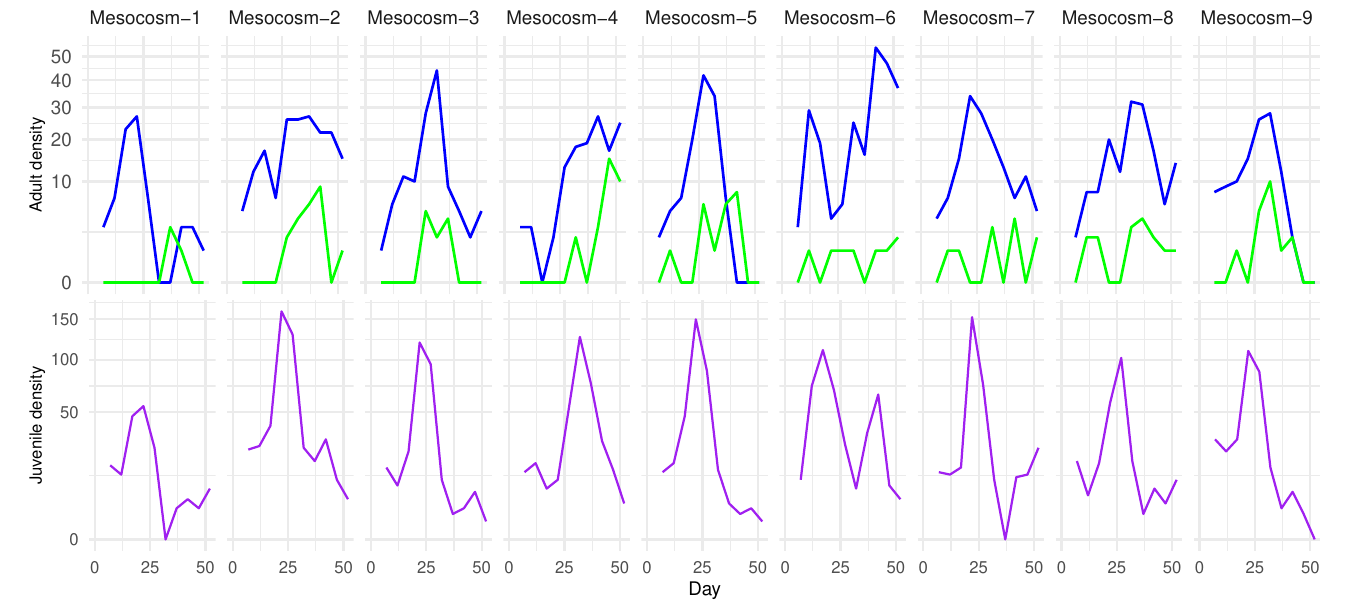} 
\end{knitrout}
\caption{\label{fig:data_vis_SIRPF_lum}
Density (Individuals$/$Liter) of \textit{D.~lumholtzi}.
The top panel shows adult susceptibles (\textit{D.~lumholtzi}, blue) and infecteds (\textit{D.~lumholtzi}, green).
The bottom panel shows juvenile susceptibles (\textit{D.~lumholtzi}, purple).
There were negligible infected juveniles.
Columns are buckets corresponding to replications with same treatment setting.
}
\end{figure}

\subsection{Results}

\subsubsection{Parameter Estimation}
\begin{table}[H]
\centering
\resizebox{\textwidth}{!}{
\begin{tabular}{||c|c |c|c|c||}
\hline
Parameter & Definition &Unit & Value & CI\\[0.5ex]
\hline
\hline
$S^{\native}$ & Susceptible host density   & $\mathrm{individual} \cdot L ^{-1}$                 & Variable                & \\
$I^{\native}$ & Infected host density     & $ \mathrm{individual} \cdot L ^{-1}$                & Variable                & \\
$J^{\native}$ & Juvenile host density    & $ \mathrm{individual} \cdot L ^{-1}$                & Variable                & \\
$F$   & Alga density                              & $10^6 \cdot \mathrm{cell} \cdot L ^{-1}$             & Variable                & \\
$P$   & Spore density                             & $10^3 \cdot \mathrm{spore} \cdot L ^{-1}$            & Variable                & \\
$r^{\native}$ &  Birth rate of juvenile                        & $\mathrm{individual} \cdot 10^{-6}\cdot \mathrm{cell}^{-1}$ & $5.52\cdot 10$  & ($3.67\cdot 10$,$1.68\cdot 10^{2}$)\\
$f^{\native}_{S}$ & Susceptible adult host filtering rate   &  $\mathrm{L} \cdot \mathrm{individual}^{-1} \cdot \mathrm{day}^{-1}$   & $7.64\cdot 10^{-4}$  & ($2.56\cdot 10^{-4}$,\,$1.31\cdot 10^{-3}$)\\
$p^{\native}$          & Number of infections per spore            &     $10^{-3} \cdot \mathrm{individual} \cdot \mathrm{spore}^{-1} $ & $3.06\cdot 10^{-1}$  & ($1.63\cdot 10^{-1}$,\,$8.67\cdot 10^{-1}$)\\
$p^{\native} \cdot f_{S^{\native}}$          & Effective infection rate of native adult hosts per‐spore      &     $10^{-3} \cdot \mathrm{L} \cdot \mathrm{spore}^{-1} \cdot \mathrm{day}^{-1}$  & $2.34\cdot 10^{-4}$  & ($1.77\cdot 10^{-4}$,$3.09\cdot 10^{-4}$)\\
$r^{\native} \cdot f_{S^{\native}}$          & Effective Birth rate of native juvenile            &     $10^{-6} \cdot \mathrm{L} \cdot \mathrm{cell}^{-1} \cdot \mathrm{day}^{-1} $ & $4.22\cdot 10^{-2}$  & ($3.81\cdot 10^{-2}$,$5.19\cdot 10^{-2}$)\\
$\theta^{\native}_{S}$ & Susceptible adult host mortality rate     &     $  \mathrm{day}^{-1}$     & $9.30\cdot 10^{-7}$  & (0,\,$1.83\cdot 10^{-4}$)\\
$\theta^{\native}_{I}$ & Infected adult host mortality rate        &     $  \mathrm{day}^{-1}$     & $4.54\cdot 10^{-1}$  & ($2.03\cdot 10^{-1}$,\,$7.57\cdot 10^{-1}$)\\
$\theta^{\native}_{J}$ & Juvenile mortality rate        &     $ \mathrm{day}^{-1}$     & $7.69\cdot 10^{-5}$  & (0,\,$4.43\cdot 10^{-2}$)\\
$\theta_{P}$   & Spore degradation rate                           &     $ \mathrm{day}^{-1}$     & $1.18\cdot 10^{-4}$   & (0,\,$2.15\cdot 10^{-2}$)\\
$\lambda^{\native}_{J}$ & Maturation rate of the juvenile                  &     $ \mathrm{day}^{-1} $                  & $1.00\cdot 10^{-1}$  & \\
$\xi$           & Ratio of infected host filtering rate       &     $\mathrm{Unitless}$                    & $1.13\cdot 10$         & ($2.48$,\,$4.37\cdot 10$)\\
$\xi_J$         & Ratio of juvenile individual filtering rate &     $\mathrm{Unitless}$                    & $1.00$ & \\
$\beta^{\native}$       & Spores produced by death per infected individual     &  $10^3 \cdot \mathrm{spore} \cdot \mathrm{individual}^{-1} \cdot \mathrm{day}^{-1}$  & $3.00\cdot 10$ & \\
$\mu$           & Algae refilling rate                                &  $10^6 \cdot \mathrm{cell} \cdot \mathrm{L} ^{-1} \cdot \mathrm{day}^{-1}$                 & $3.70\cdot 10^{-1}$ & \\
$\delta$        & Sampling rate                                      &  $day ^{-1}$                                                                         & $1.30\cdot 10^{-2}$ & \\
$\sigma^{\native}_{S}$ & Standard deviation of Brownian motion of susceptible adult         &   $\sqrt{\mathrm{individual} \cdot \mathrm{day}^{-1}}$  & 0  & \\
$\sigma^{\native}_{I}$ & Standard deviation of Brownian motion of infected adult         &   $\sqrt{\mathrm{individual} \cdot \mathrm{day}^{-1}}$  & $5.74\cdot 10^{-1}$  & (0,\,$6.19\cdot 10^{-1}$)\\
$\sigma^{\native}_{J}$ & Standard deviation of Brownian motion of juvenile               &   $\sqrt{\mathrm{individual} \cdot \mathrm{day}^{-1}}$  & $3.41\cdot 10^{-1}$  & ($1.52\cdot 10^{-1}$,\,$5.18\cdot 10^{-1}$)\\
$\sigma_{F}$   & Standard deviation of Brownian motion of alga                          &   $\sqrt{\mathrm{individual} \cdot \mathrm{day}^{-1}}$  & $6.97\cdot 10^{-2}$  & (0,\,$1.63\cdot 10^{-1}$)\\
$\sigma_{P}$   & Standard deviation of Brownian motion of parasite                      &   $\sqrt{\mathrm{individual} \cdot \mathrm{day}^{-1}}$  & $4.60\cdot 10^{-1}$  & ($2.10\cdot 10^{-1}$,\,$6.10\cdot 10^{-1}$)\\
$\tau^{\native}_{S}$      &  Measurement dispersion for susceptible adult                 &$\mathrm{Unitless}$           & $1.50\cdot 10$  & ($7.25$,\,$3.99\cdot 10$)\\
$\tau^{\native}_{I}$      &  Measurement dispersion for infected adult                &$\mathrm{Unitless}$            & $1.13$  & ($6.80\cdot 10^{-1}$,\,$2.61$)\\
\hline
\end{tabular}}
\caption{Variables and parameter definitions and estimates of SIRJPF model for \textit{D.~dentifera}-only dynamics. The confidence interval is generated by the Monte Carlo Adjusted Profile.}
\label{Table:SIRJPF_native}
\end{table}

\subsubsection{Parameter Estimation}
\begin{table}[H]
\centering
\resizebox{\textwidth}{!}{
\begin{tabular}{||c|c |c|c|c||}
\hline
Parameter & Definition &Unit & Value & CI\\[0.5ex]
\hline
\hline
$S^{\invasive}$ & Susceptible host density   & $\mathrm{individual} \cdot L ^{-1}$                 & Variable                & \\
$I^{\invasive}$ & Infected host density     & $ \mathrm{individual} \cdot L ^{-1}$                & Variable                & \\
$J^{\invasive}$ & Juvenile host density    & $ \mathrm{individual} \cdot L ^{-1}$                & Variable                & \\
$F$   & Alga density                              & $10^6 \cdot \mathrm{cell} \cdot L ^{-1}$             & Variable                & \\
$P$   & Spore density                             & $10^3 \cdot \mathrm{spore} \cdot L ^{-1}$            & Variable                & \\
$r^{\invasive}$ &  Birth rate of juvenile                        & $\mathrm{individual} \cdot 10^{-6}\cdot \mathrm{cell}^{-1}$ & $2.49\cdot 10^{2}$  & ($9.36\cdot 10$,$1.63\cdot 10^{3}$)\\
$f^{\invasive}_{S}$ & Susceptible adult host filtering rate   &  $\mathrm{L} \cdot \mathrm{individual}^{-1} \cdot \mathrm{day}^{-1}$   & $6.49\cdot 10^{-4}$  & ($6.57\cdot 10^{-5}$,$1.33\cdot 10^{-3}$)\\
$p^{\invasive}$          & Number of infections per spore            &     $10^{-3} \cdot \mathrm{individual} \cdot \mathrm{spore}^{-1} $ & $6.65\cdot 10^{-1}$  & ($3.42\cdot 10^{-1}$,$6.17$)\\
$p^{\invasive} \cdot f_{S^{\invasive}}$          & Effective infection rate of invasive adult hosts per‐spore      &     $10^{-3} \cdot \mathrm{L} \cdot \mathrm{spore}^{-1} \cdot \mathrm{day}^{-1}$ & $4.32\cdot 10^{-4}$  & ($3.78\cdot 10^{-4}$,$7.45\cdot 10^{-4}$)\\
$r^{\invasive} \cdot f_{S^{\invasive}}$          & Effective Birth rate of invasive juvenile            &     $10^{-6} \cdot \mathrm{L} \cdot \mathrm{cell}^{-1} \cdot \mathrm{day}^{-1} $ & $1.62\cdot 10^{-1}$  & ($1.08\cdot 10^{-1}$,$2.48\cdot 10^{-1}$)\\
$\theta^{\invasive}_{S}$ & Susceptible adult host mortality rate     &     $ \mathrm{day}^{-1}$     & $5.73\cdot 10^{-1}$  & ($3.01\cdot 10^{-1}$,$1.16$)\\
$\theta^{\invasive}_{I}$ & Infected adult host mortality rate        &     $ \mathrm{day}^{-1}$     & $1.74\cdot 10^{-1}$  & ($2.05\cdot 10^{-2}$,$4.16\cdot 10^{-1}$)\\
$\theta^{\invasive}_{J}$ & Juvenile mortality rate        &     $  \mathrm{day}^{-1}$     & $2.55\cdot 10^{-5}$  & (0,$1.28\cdot 10^{-1}$)\\
$\theta_{P}$   & Spore degradation rate                           &     $ \mathrm{day}^{-1}$     & $1.71\cdot 10^{-6}$   & (0,$8.61\cdot 10^{-2}$)\\
$\lambda^{\invasive}_{J}$ & Maturation rate of the juvenile                  &     $ \mathrm{day}^{-1} $                  & $1.00\cdot 10^{-1}$  & \\
$\xi$           & Ratio of infected host filtering rate       &     $\mathrm{Unitless}$                    & $7.33\cdot 10$         & ($9.54$,$8.02\cdot 10^{2}$)\\
$\xi_J$         & Ratio of juvenile individual filtering rate &     $\mathrm{Unitless}$                    & $1.00$ & \\
$\beta^{\invasive}$       & Spores produced by death per infected individual     &  $10^3 \cdot \mathrm{spore} \cdot \mathrm{individual}^{-1} \cdot \mathrm{day}^{-1}$  & $3.00\cdot 10$ & \\
$\mu$           & Algae refilling rate                                &  $10^6 \cdot \mathrm{cell} \cdot \mathrm{L} ^{-1} \cdot \mathrm{day}^{-1}$                 & $3.70\cdot 10^{-1}$ & \\
$\delta$        & Sampling rate                                      &  $day ^{-1}$                                                                         & $1.30\cdot 10^{-2}$ & \\
$\sigma^{\invasive}_{S}$ & Standard deviation of Brownian motion of susceptible adult         &   $\sqrt{\mathrm{individual} \cdot \mathrm{day}^{-1}}$  & 0  & \\
$\sigma^{\invasive}_{I}$ & Standard deviation of Brownian motion of infected adult         &   $\sqrt{\mathrm{individual} \cdot \mathrm{day}^{-1}}$  & $2.73\cdot 10^{-1}$  & (0, $5.75\cdot 10^{-1}$)\\
$\sigma^{\invasive}_{J}$ & Standard deviation of Brownian motion of juvenile               &   $\sqrt{\mathrm{individual} \cdot \mathrm{day}^{-1}}$  & $3.72\cdot 10^{-1}$  & ($2.75\cdot 10^{-1}$, $5.25\cdot 10^{-1}$)\\
$\sigma_{F}$   & Standard deviation of Brownian motion of alga                          &   $\sqrt{\mathrm{individual} \cdot \mathrm{day}^{-1}}$  & $5.22\cdot 10^{-8}$  & (0,$1.80\cdot 10^{-2}$)\\
$\sigma_{P}$   & Standard deviation of Brownian motion of parasite                      &   $\sqrt{\mathrm{individual} \cdot \mathrm{day}^{-1}}$  & $7.99\cdot 10^{-8}$  & (0,$3.55\cdot 10^{-1}$)\\
$\tau^{\invasive}_{S}$      &  Measurement dispersion for susceptible adult                  &$\mathrm{Unitless}$           & $7.00\cdot 10$  & (0,$4.40\cdot 10^{3}$)\\
$\tau^{\invasive}_{I}$      &  Measurement dispersion for infected adult                 &$\mathrm{Unitless}$            & $1.17$  & ($5.80\cdot 10^{-1}$,$1.08\cdot 10$)\\
\hline
\end{tabular}}
\caption{Variables and parameter definitions and estimates of SIRJPF model for \textit{D.~lumholtzi}-only dynamics. The confidence interval is generated by the Monte Carlo Adjusted Profile.}
\label{Table:SIRJPF_invasive}
\end{table}

\subsubsection{MCAP Results}
\begin{figure}[H]
    \centering
    \includegraphics[width=\linewidth]{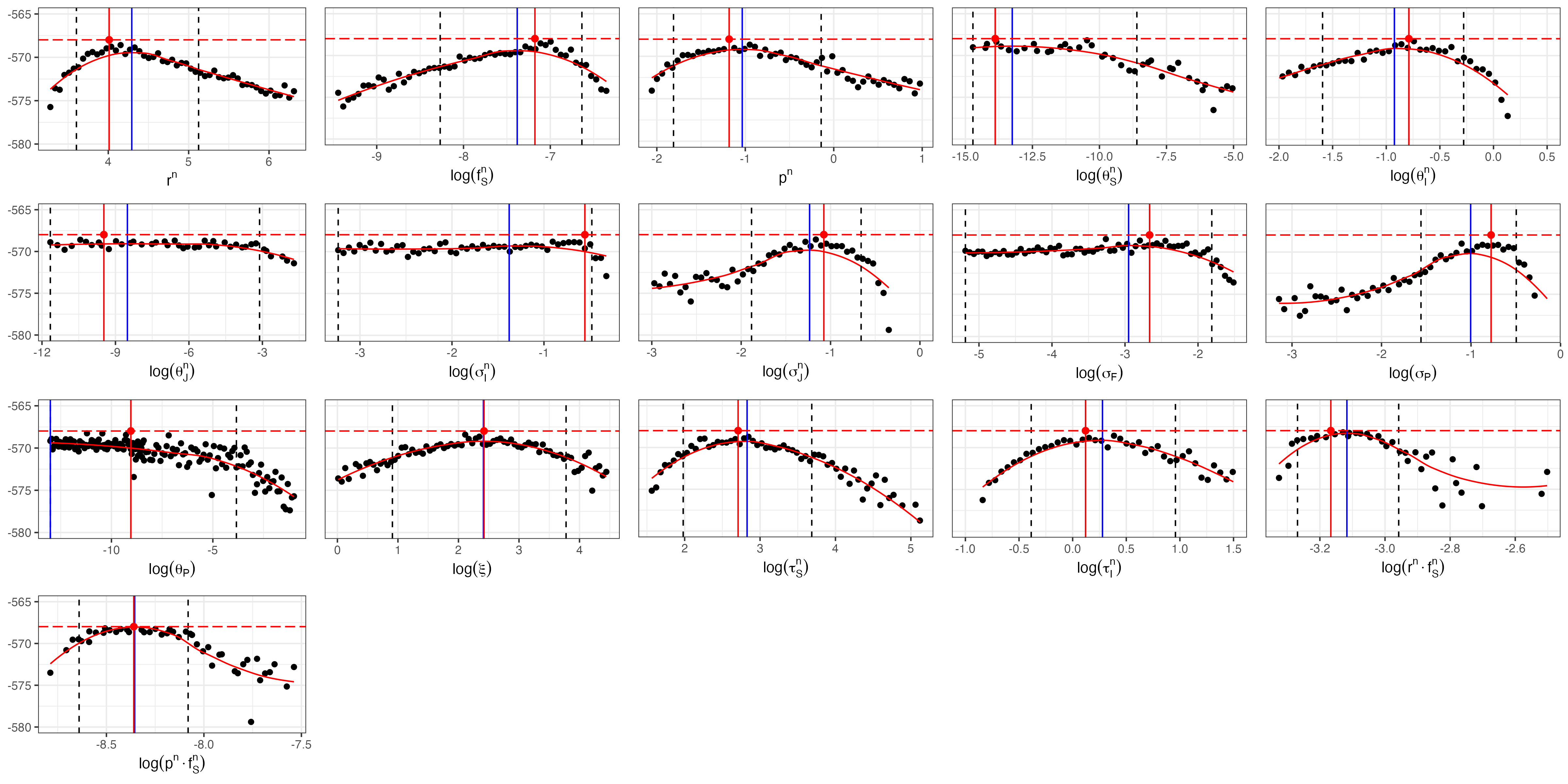}
    \caption{Monte Carlo Adjusted Profile results of SIRJPF model for \textit{D.~dentifera}-only dynamics.
    The vertical dotted lines represent the 95\% confidence interval obtained by MCAP.
    The vertical blue lines show the MLE estimated using MCAP.
    The red vertical lines correspond to the model with the overall highest likelihood among all searches.
    }
    \label{fig:SIRJPF_dent_profile}
\end{figure}

\begin{figure}[H]
    \centering
    \includegraphics[width=\linewidth]{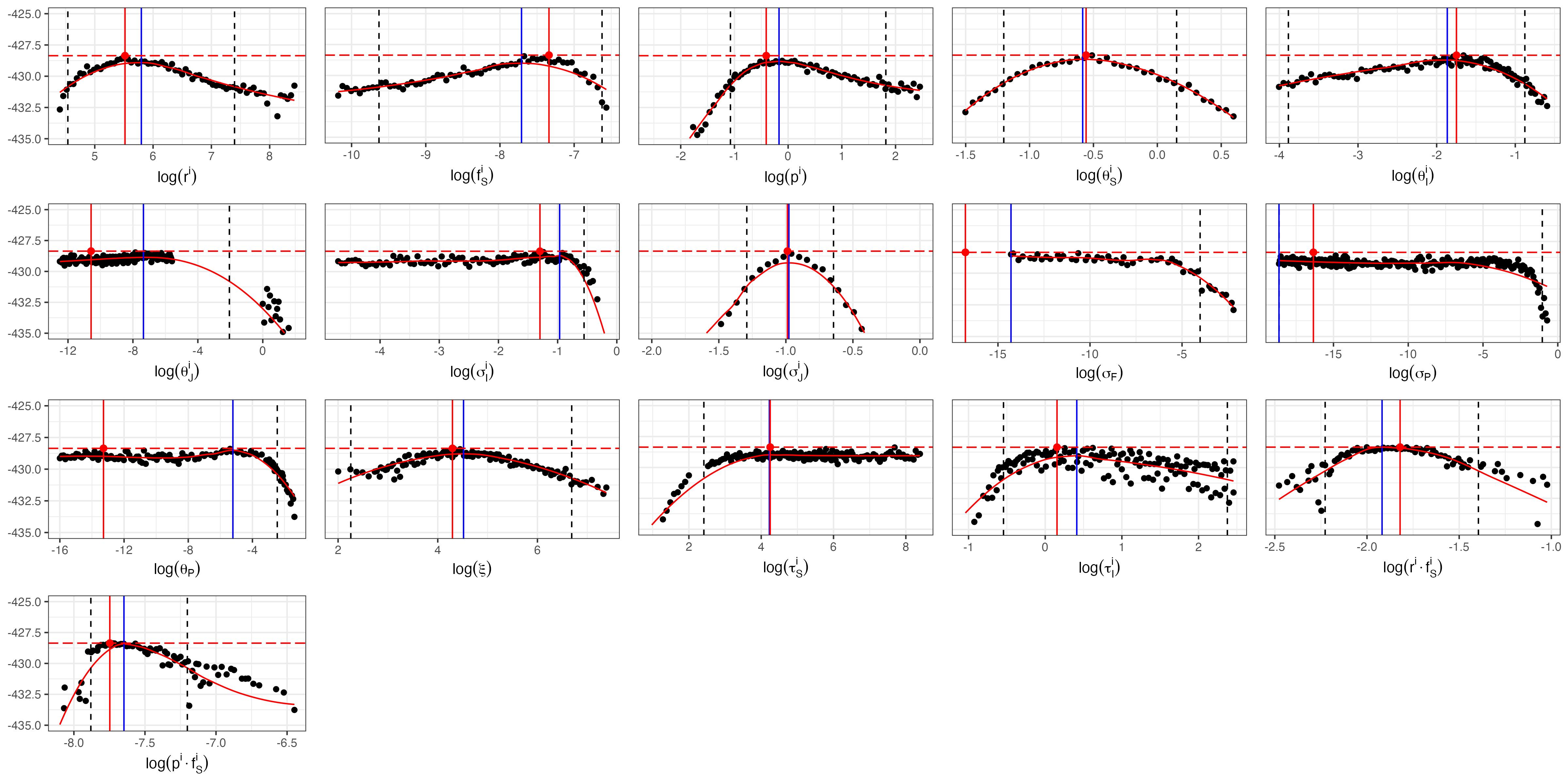}
    \caption{Monte Carlo Adjusted Profile results of SIRJPF model for \textit{D.~lumholtzi}-only dynamics.
    The vertical dotted lines represent the 95\% confidence interval obtained by MCAP.
    The vertical blue lines show the MLE estimated using MCAP.
    The red vertical lines correspond to the model with the overall highest likelihood among all searches.
}
    \label{fig:SIRJPF_native_profile}
\end{figure}

\subsubsection{Model Comparison}
\begin{table}[h!]
\centering
\begin{center}
\begin{tabular}{||c |c |c |c|c||}
 \hline
Specific parameters  & Max log-likelihood & AIC & \makecell[c]{Max log-likelihood\\(block)} & \makecell[c]{AIC\\(block)}\\ [0.5ex]
 \hline\hline
$\theta^{\native}_{I,\unit}$ & -549.38& 1140.77& -547.92& 1137.84 \\
$\varnothing$  & -567.98& 1163.97& -567.98& 1163.97 \\
 $r^{\native}_{\unit}$ & -563.38& 1168.76& -562.95& 1167.90 \\
$p^{\native}_{\unit}$ & -563.78& 1169.56& -562.75& 1167.49 \\
$\xi_{\unit}$ & -566.59& 1175.19& -565.80& 1173.60 \\
$\theta^{\native}_{S,\unit}$ & -567.31& 1176.61& -565.48& 1172.96 \\
$f^{\native}_{S,\unit}$ & -568.66& 1179.33& -566.32& 1174.63 \\
$\theta_{P,\unit}$& -568.21& 1178.42& -568.87& 1179.74 \\[1ex]
 \hline
\end{tabular}
\end{center}
\caption{Comparison of model fit and complexity across various configurations of unit-specific parameters within the panelPOMP framework for \textit{D.~dentifera}-only dynamics. The unit-specific parameter setting assessed include $\theta^{\native}_{I,\unit}$, $p^{\native}_{\unit}$, $r^{\native}_{\unit}$, $\xi_{\unit}$, $\theta^{\native}_{S,\unit}$, $\theta_{P,\unit}$.}
\label{Table:SIRJPF_dent_model_comparison}
\end{table}

\begin{table}[h!]
\centering
\begin{center}
\begin{tabular}{||c |c |c |c|c||}
 \hline
Specific parameters  & Max log-likelihood & AIC & \makecell[c]{Max log-likelihood\\(block)} & \makecell[c]{AIC\\(block)}\\ [0.5ex]
 \hline\hline
$\xi_{\unit}$ & -419.78& 883.560& -416.87& 877.747 \\
$\varnothing$  & -428.36& 884.715& -428.36& 884.715 \\
$p^{\invasive}_{\unit}$ & -421.21& 886.411& -420.66& 885.326 \\
$\theta^{\invasive}_{I,\unit}$ & -421.21& 886.427& -419.32& 882.636 \\
$f^{\invasive}_{S,\unit}$ & -422.85& 889.700& -423.26& 890.523 \\
$\theta_{P,\unit}$ & -429.20& 902.400& -428.99& 901.972 \\
$r^{\invasive}_{\unit}$ & -432.43& 908.868& -426.31& 896.612 \\
$\theta^{\invasive}_{S,\unit}$ & -432.87& 909.740& -430.01& 904.014 \\[1ex]
 \hline
\end{tabular}
\end{center}
\caption{Comparison of model fit and complexity across various configurations of unit-specific parameters within the panelPOMP framework for \textit{D.~lumholtzi}-only dynamics. The unit-specific parameter setting assessed include $\theta^{\invasive}_{I,\unit}$, $p^{\invasive}_{\unit}$, $r^{\invasive}_{\unit}$, $\xi_{\unit}$, $\theta^{\invasive}_{S,\unit}$, $\theta_{P,\unit}$.}
\label{Table:SIRJPF_lum_model_comparison}
\end{table}

Despite the model configuration with unit-specific $\theta^{\native}_{I,\unit}$ and $\xi_{\unit}$ achieving the lowest AIC value in Table \ref{Table:SIRJPF_dent_model_comparison} and \ref{Table:SIRJPF_lum_model_comparison}, the improvement over the all-shared-parameters model is marginal.
This slight advantage may be attributed to Monte Carlo error inherent in the parameter estimation process using the panelPOMP framework.
The stochastic nature of the iterated filtering algorithms employed in POMP relies on Monte Carlo simulations, which can introduce variability in the maximum likelihood estimates and, consequently, the AIC values.
\parskip 3mm

Given the potential for such computational noise and the relatively small difference in AIC, we prefer to adopt the model with all shared parameters.
This choice promotes parsimony and reduces the risk of overfitting, ensuring that the model remains both interpretable and generalizable.
By keeping $\theta_I$—the natural death rate of infected \textit{Daphnia}—shared across all units, we maintain a simpler model structure that is sufficient to describe the observed data without unnecessary complexity.

\subsection{Simulation}
\begin{figure}[H]
    \centering
    \includegraphics[width=0.75\linewidth]{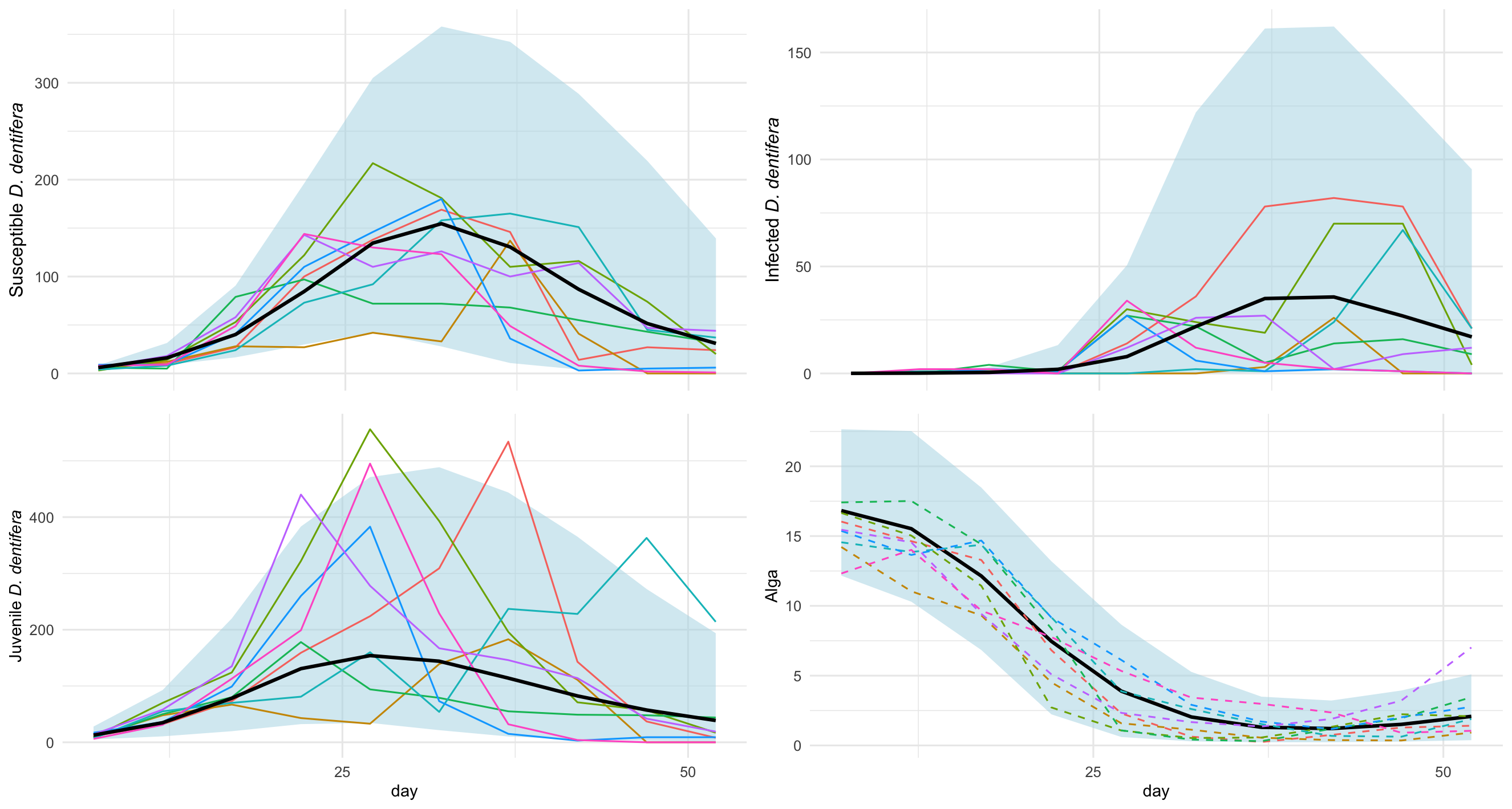}
    \caption{Simulation plots of densities (Individuals$/$Liter) for susceptible, infected, and juvenile \textit{D.~dentifera}, along with alga density ($10^6\cdot$cells$/$Liter) and parasite density ($10^3\cdot$spores$/$Liter), over time (days). The colorful solid lines represent observed experimental data, showing the variability across different experimental replicates. The black line represents the mean trajectory from the simulation model, capturing the general trend of each density. The light blue shaded area corresponds to the 95\% confidence interval of the simulations, indicating the range of model predictions. Dashed lines illustrate individual simulated trajectories, highlighting the model’s ability to capture fluctuations around the mean trajectory.}
    \label{fig:SIRJPF_native_simulation_band}
\end{figure}

\begin{figure}[H]
    \centering
    \includegraphics[width=0.75\linewidth]{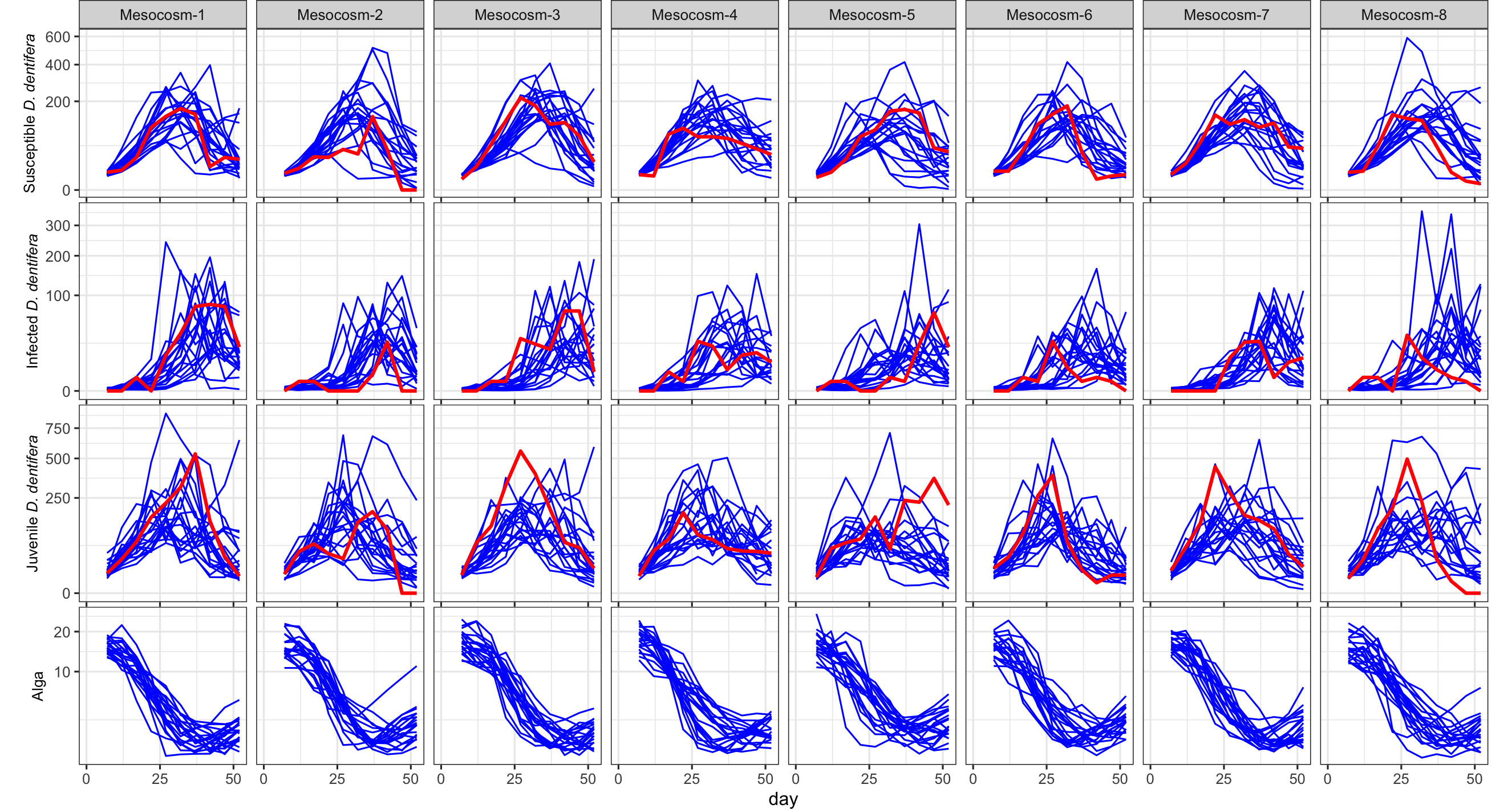}
   \caption{Simulated densities (Individuals$/$Liter) of susceptible \textit{D.~dentifera}, infected \textit{D.~dentifera} density, juvenile \textit{D.~dentifera} and alga density ($10^6\cdot$cells$/$Liter) over time (days) for each experimental unit. The blue lines represent individual simulation runs, capturing the variability in susceptible density across replicates, while the red line represents the actual experimental data.}
    \label{fig:SIRJPF_dent_simulation_lines}
\end{figure}

\begin{figure}[H]
    \centering
    \includegraphics[width=0.75\linewidth]{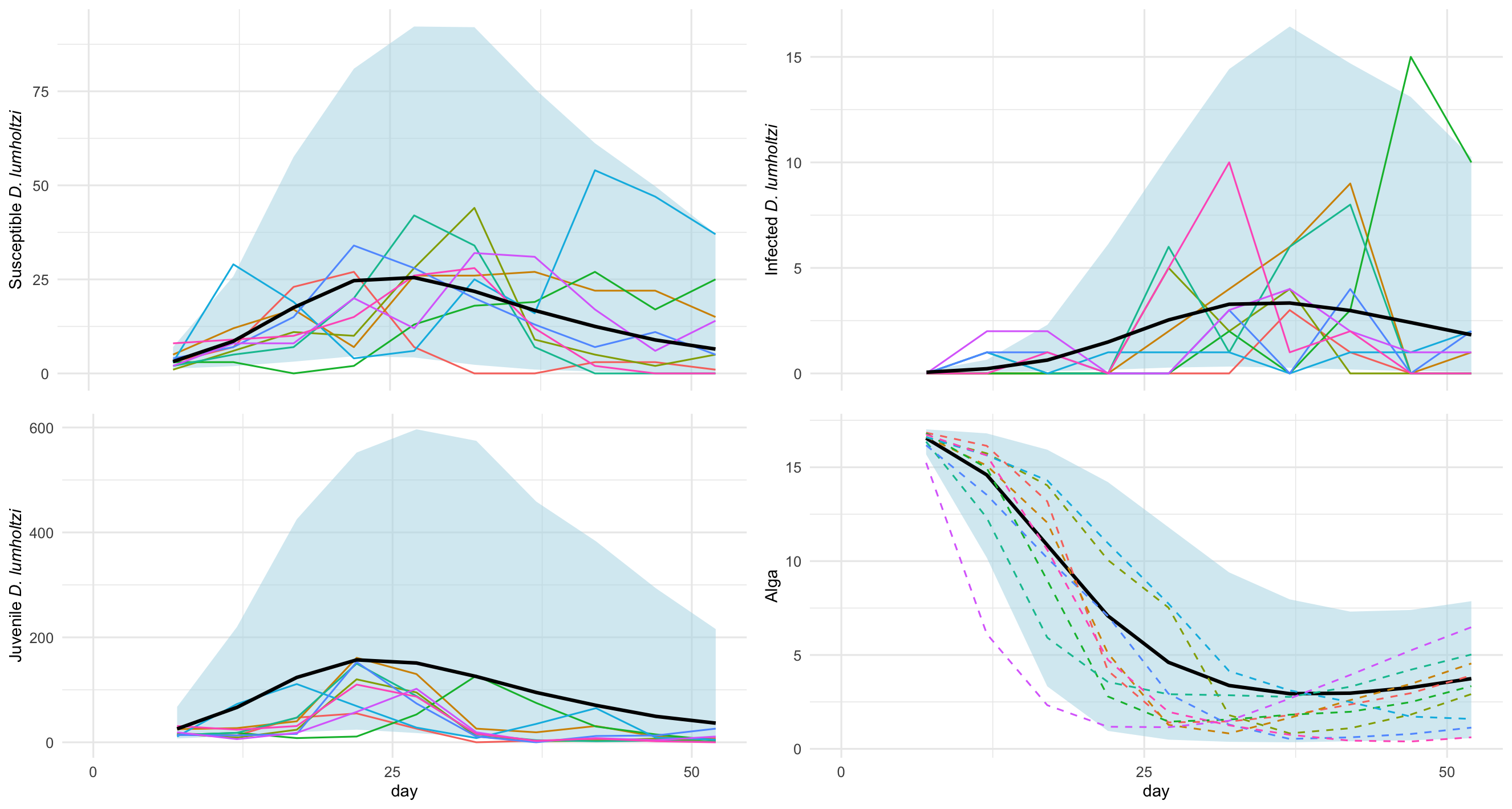}
    \caption{Simulation plots of densities (Individuals$/$Liter) for susceptible, infected, and juvenile \textit{D.~lumholtzi}, along with alga density ($10^6\cdot$cells$/$Liter) and parasite density ($10^3\cdot$spores$/$Liter), over time (days). The colorful solid lines represent observed experimental data, showing the variability across different experimental replicates. The black line represents the mean trajectory from the simulation model, capturing the general trend of each density. The light blue shaded area corresponds to the 95\% confidence interval of the simulations, indicating the range of model predictions. Dashed lines illustrate individual simulated trajectories, highlighting the model’s ability to capture fluctuations around the mean trajectory.}
    \label{fig:SIRJPF_invasive_simulation_band}
\end{figure}

\begin{figure}[H]
    \centering
    \includegraphics[width=0.75\linewidth]{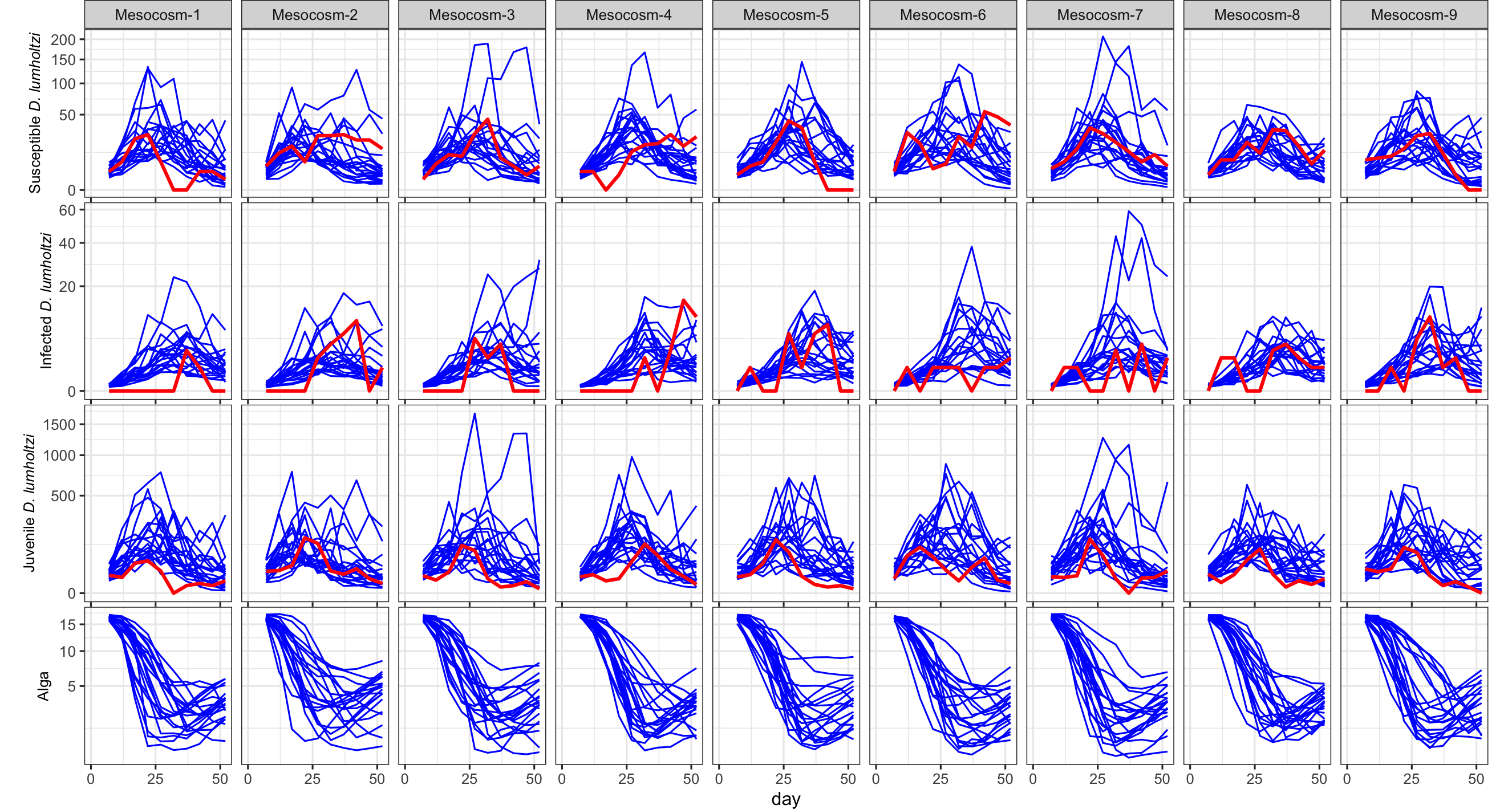}
   \caption{Simulated trajectories of susceptible \textit{D.~lumholtzi}, infected \textit{D.~lumholtzi}, juvenile \textit{D.~lumholtzi} and alga density ($10^6\cdot$cells$/$Liter) over time (days) for each experimental unit (u1 to u9). The blue lines represent individual simulation runs, capturing the variability in susceptible density across replicates, while the red line represents the actual experimental data.}
    \label{fig:SIRJPF_lum_simulation_lines}
\end{figure}

The simulation plots (Figures \ref{fig:SIRJPF_native_simulation_band} to \ref{fig:SIRJPF_lum_simulation_lines}) demonstrate our model's capability to capture the complex interactions among various states within the population dynamics of \textit{D.~dentifera} and \textit{D.~lumholtzi}.
Specifically, the model captures the latent pattern for susceptible and infected densities across multiple experimental units, as evidenced by the close alignment between the simulated mean trajectories (depicted by the black and light-blue bands) and the observed experimental data (colorful solid lines).
Each blue line represents an individual simulation case, capturing the stochastic variability across replicates, while the light blue shaded area (where present) denotes the 95\% confidence interval of the simulation results.
Remarkably, although juvenile density data were not included in the model fitting process, the model still effectively predicts juvenile density trends.
The promising performance of the model indicates the model's capacity to generalize underlying ecological processes, as the 95\% confidence intervals largely cover the observed juvenile data.

\parskip 3mm

At the same time, the depletion of  algal density over time, reflected in the simulation further explains the behavior of the \textit{Daphnia}.
The decline in algal density at later stages indicates that the system reaches a state of food limitation, which is a dynamic anticipated by our model.
The reduction in food availability provides a mechanistic explanation for the observed trends in \textit{Daphnia} densities, as resource scarcity constrains population growth, ultimately affecting the susceptible, infected, and juvenile densities.
Thus, the model captures not only the reproductive dynamics but also the resource-consumption interactions that shape these dynamics.
The model’s ability to reproduce these patterns accurately highlights its potential utility in explaining and predicting empirical trends in complex ecological systems.

\section{SRJF2 Model}
In this section, we introduce the SRJF2 model, which analyze the dynamics between native and invasive species competing for a shared food resource, algae ($F$) in each bucket($\unit$).
The stochastic differential equations governing the system are:
{\small
\begin{align}
    dS^{\species}_{\unit}(t) &= \lambda^{\species}_{J,{\unit}} J^{\species}_{\unit}(t) \, dt - (\theta^{\species}_{S,{\unit}} + \delta) S^{\species}_{\unit}(t)\, dt + S^{\species}_{\unit}(t)\, d\zeta^{\species}_{S,{\unit}}, \label{eq:SRJF2_dS}\\
    dJ^{\species}_{\unit}(t) &= r^{\species}_{\unit}f^{\species}_{S,{\unit}}F_{\unit}(t) S^{\species}_{\unit}(t)\, dt - \theta^{\species}_{J,{\unit}} J^{\species}_{\unit}(t)\, dt - \delta J^{\species}_{\unit}(t)\, dt - \lambda^{\species}_{J,{\unit}} J^{\species}_{\unit}(t)\, dt + J^{\species}_{\unit}(t)\, d\zeta^{\species}_{J,{\unit}}, \label{eq:SRJF2_dJ}\\
    dF_{\unit}(t) &= - \sum_{\species \in \{\native, \invasive\}} f^{\species}_{S,{\unit}} F_{\unit}(t) \left(S^{\species}_{\unit}(t)+\xi_{J,{\unit}}J^{\species}_{\unit}(t)\right)dt  + \mu \, dt + F_{\unit}(t)\, d\zeta_{F,{\unit}}, \label{eq:SRJF2_dF}\\
d\zeta^{\species}_{S,{\unit}} &\sim \Normal \big[0, (\sigma^{\species}_{S,{\unit}})^{2}\, dt\big]\,
d\zeta^{\species}_{J,{\unit}} \sim \Normal \big[0, (\sigma^{\species}_{J,{\unit}})^{2}\, dt\big]\,
d\zeta_{F,{\unit}} \sim \Normal \big[0, \sigma_{F,{\unit}}^{2}\, dt\big]\label{eq:SRJF2_dzetaS}.
\end{align}
}
Equation~\eqref{eq:SRJF2_dS} models the rate of change of the specific species' susceptible population $S^{\species}_{\unit}(t)$.
The term $\lambda^{\species}_{J,\unit} J^{\species}_{\unit}(t) \, dt$ represents the maturation of juveniles into susceptible adults at rate $\lambda^{\species}_{J,\unit}$.
The mortality term $-\left(\theta^{\species}_{S,\unit} + \delta\right) S^{\species}_{\unit}(t) \, dt$ accounts for natural death at rate $\theta^{\species}_{S,\unit}$ and the sampling rate $\delta$.
For consistency with the SIRJPF2 and SIRJPF model, we further fix $\sigma^{\species}_{S,{\unit}}$ to be zero for specific analysis of this dynamics.
\parskip 3mm

Equation~\eqref{eq:SRJF2_dJ} describes the dynamics of the juvenile population $J^{\species}_{\unit}(t)$.
The growth term $r^{\species}_{\unit} f^{\species}_{S,\unit} F_{\unit}(t) S^{\species}_{\unit}(t) \, dt$ signifies the production of juveniles through reproduction, where $r^{\species}_{\unit}$ is the growth factor and $f^{\species}_{S,\unit}$ is the filtering rate of susceptible individuals on algae.
The loss term $-\left(\theta^{\species}_{J,\unit} + \delta + \lambda^{\species}_{J,\unit}\right) J^{\species}(t) \, dt$ includes natural juvenile mortality at rate $\theta^{\species}_{J,\unit}$, sampling rate $\delta$, and maturation into the susceptible adult class at rate $\lambda^{\species}_{J,\unit}$.
The stochastic term $J^{\species}_{\unit}(t) \, d\zeta^{\species}_{J,\unit}$ captures random fluctuations affecting the juvenile population, with $d\zeta^{\species}_{J,\unit}$ being the Brownian motion characterized by variance $(\sigma^{\species}_{J,{\unit}})^{2}\, dt$.
\parskip 3mm

Equation~\eqref{eq:SRJF2_dF} captures the dynamics of the algae population $F_{\unit}(t)$, which serves as a shared food resource for both native and invasive species.
The consumption terms $\sum_{\species \in \{\native, \invasive\}} f^{\species}_{S,{\unit}} F_{\unit}(t) \left(S^{\species}_{\unit}(t)+\xi_{J,{\unit}}J^{\species}_{\unit}(t)\right)dt$ represent the reduction of algae due to feeding by native and invasive susceptible individuals and juveniles, respectively.
The parameter $\xi_{J,\unit}$ denotes the relative ratio of juveniles on algae consumption compared to susceptible adults.
The term $\mu \, dt$ represents the refill rate of algae, contributing to its replenishment.
The stochastic fluctuation $F_{\unit}(t) \, d\zeta_{F,\unit}$ introduces randomness into the algae dynamics, where $d\zeta_{F,\unit}$ is the Brownian motion with variance $\sigma_{F,{\unit}}^{2}\, dt$.
\parskip 3mm

\subsection{Flow Diagram}
The following Figure~\ref{fig:SRJF2} presents the flow diagram of the SRJF2 model, illustrating the interactions among susceptible individuals ($S^{\native}$, $S^{\invasive}$), juveniles ($J^{\native}$,$J^{\invasive}$), algae as a food resource ($F$), and mortality ($R$).
The bidirectional arrows between $J^{\species}$ and $S^{\species}$, represent the maturation of juveniles into susceptible individuals and the reproduction of susceptible individuals into juveniles.
Both susceptible individuals and juveniles, consume resources from the alga population $F$, as indicated by the arrows from $F$ to each population node.
The algal food resource is regularly replenished, depicted by the recycling loop, and contributes to the growth of both susceptible individuals and juveniles.
All populations have pathways leading to the mortality state $R$, representing natural death and disease-induced mortality.
Two species of \textit{Daphnia} are competing through the food resources in this model.

\begin{figure}[H]
\begin{center}
\resizebox{8cm}{!}{
\begin{tikzpicture}[
  square/.style={rectangle, draw=black, minimum width=0.5cm, minimum height=0.5cm, rounded corners=.1cm, fill=blue!8},
  rhombus/.style={diamond, draw=black, minimum width=0.1cm, minimum height=0.1cm, fill=purple!8,aspect = 1},
  travel/.style={circle, draw=black, minimum width=0.5cm, minimum height=0.5cm, fill=green!8},
  report/.style={shape=regular polygon, regular polygon sides=8, draw, fill=red!8,minimum size=0.6cm,inner sep=0cm},
  bendy/.style={bend left=10},
  bendy2/.style={bend left=100},
  bendy3/.style={bend left=-100},
  >/.style={shorten >=0.25mm},
  >/.tip={Stealth[length=1.5mm,width=1.5mm]}
]
\tikzset{>={}};

\node (Sn) at (5.5,0) [square] {$S^{\native}$};
\node (Si) at (-0.5,0) [square] {$S^{\invasive}$};
\node (Jn) at (5.5,2) [square] {$J^{\native}$};
\node (Ji) at (-0.5,2) [square] {$J^{\invasive}$};
\node (R) at (2.5,0)  [rhombus] {$R$};
\node (F) at (2.5,2) [report] {$F$};

\draw [->, bendy] (Jn) to  (Sn);
\draw [->, bendy] (Ji) to  (Si);
\draw [->, bendy] (Sn) to  (Jn);
\draw [->, bendy] (Si) to  (Ji);

\draw [->] (F) --  (R);
\draw [->] (Sn) -- (R);
\draw [->] (Si) -- (R);
\draw [->] (Jn) -- (R);
\draw [->] (Ji) -- (R);
\draw [->] (F) -- (Sn);
\draw [->] (F) -- (Si);
\draw [->] (F) -- (Jn);
\draw [->] (F) -- (Ji);
\draw (F) edge[loop above] (F);

\end{tikzpicture}
}
\end{center}
\vspace{-5mm}
\caption{Flow diagram for the SRJF2 model illustrating population interactions. The model includes the $R$ state, representing mortality. For $\species \in \{\native,\invasive\}$, Susceptible populations ($S^{\species}$) can reproduce into juvenile ($J^{\species}$). And \textit{A.} Algae ($F$) will be refilled, as shown by the recycling arrows. Both $S^{\species}$ and $J^{\species}$ consume resources from $F$, and over time, components in $F$ also progress to $R$.}
\label{fig:SRJF2}
\end{figure}
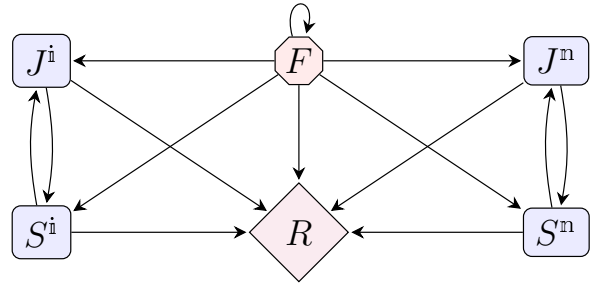

\subsection{Data Visualization}
\begin{figure}[H]
\centering
\begin{knitrout}
\definecolor{shadecolor}{rgb}{0.969, 0.969, 0.969}\color{fgcolor}
\includegraphics[width=\textwidth]{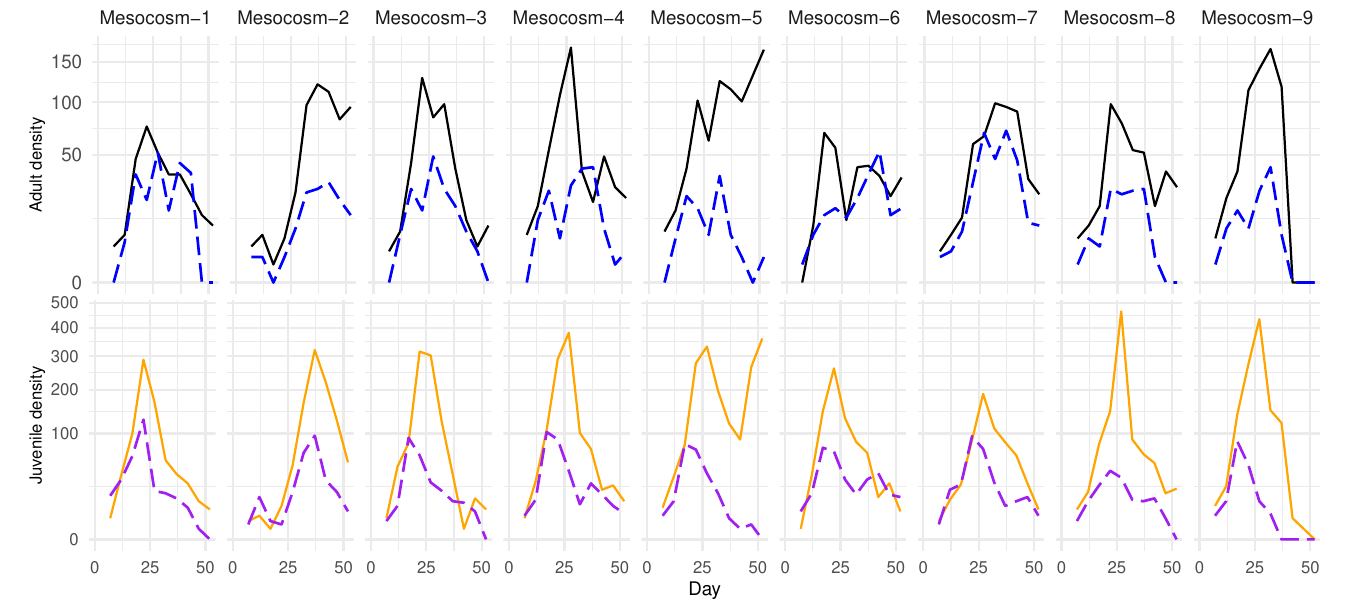} 
\end{knitrout}
\caption{\label{fig:data_vis_SRJF2}
Density (Individuals$/$Liter) of \textit{D.~dentifera} (solid lines) and \textit{D.~lumholtzi} (dashed lines).
The top panel shows adult susceptibles (\textit{D.~dentifera}, black and \textit{D.~lumholtzi}, blue).
The bottom panel shows juvenile susceptibles (\textit{D.~dentifera}, orange; \textit{D.~lumholtzi}, purple).
There were negligible infected juveniles.
Columns are buckets corresponding to replications with same treatment setting.
}
\end{figure}

\subsection{Results}
\subsubsection{Parameter Estimation}

\begin{table}[H]
\centering
\resizebox{\textwidth}{!}{
\begin{tabular}{||c|c |c|c|c||}
\hline
Parameter & Definition &Unit & Value & CI\\[0.5ex]
\hline
\hline
$S^{\species}$ & Susceptible host density for species $\species$  & $\mathrm{individual} \cdot L ^{-1}$                 & Variable                & \\
$J^{\species}$ & Juvenile host density for species $\species$     & $ \mathrm{individual} \cdot L ^{-1}$                & Variable                & \\
$F$   & Alga density                              & $10^6 \cdot \mathrm{cell} \cdot L ^{-1}$             & Variable                & \\
$P$   & Spore density                             & $10^3 \cdot \mathrm{spore} \cdot L ^{-1}$            & Variable                & \\
$r^{\native}$ & Birth rate of native juvenile                        & $\mathrm{individual} \cdot 10^{-6}\cdot \mathrm{cell}^{-1}$ & $6.29\cdot 10^{3}$  & ($2.42\cdot 10^{3}$,$3.76\cdot 10^{4}$)\\
$r^{\invasive}$ & Birth rate of invasive juvenile                      & $\mathrm{individual} \cdot 10^{-6}\cdot \mathrm{cell}^{-1}$ & $4.75\cdot 10^{3}$  & ($1.78\cdot 10^{3}$,$6.79\cdot 10^{4}$)\\
$f^{\native}_{S}$ & Native susceptible adult host filtering rate   &  $\mathrm{L} \cdot \mathrm{individual}^{-1} \cdot \mathrm{day}^{-1}$   & $7.10\cdot 10^{-5}$  & ($1.54\cdot 10^{-5}$,$1.17\cdot 10^{-4}$)\\
$f^{\invasive}_{S}$ & Invasive susceptible adult host filtering rate &  $\mathrm{L} \cdot \mathrm{individual}^{-1} \cdot \mathrm{day}^{-1}$   & $8.90\cdot 10^{-5}$  & (0,$2.04\cdot 10^{-4}$)\\
$r^{\native} \cdot f_{S^{\native}}$          & Effective Birth rate of native juvenile            &     $10^{-6} \cdot \mathrm{L} \cdot \mathrm{cell}^{-1} \cdot \mathrm{day}^{-1} $ & $4.47\cdot 10^{-1}$  & ($3.18\cdot 10^{-1}$,$1.16$)\\
$r^{\invasive} \cdot f_{S^{\invasive}}$          & Effective Birth rate of invasive juvenile            &     $10^{-6} \cdot \mathrm{L} \cdot \mathrm{cell}^{-1} \cdot \mathrm{day}^{-1} $ & $4.23\cdot 10^{-1}$  & ($2.67\cdot 10^{-1}$,$3.55$)\\
$\theta^{\native}_{S}$ & Native susceptible adult host mortality rate     &     $  \mathrm{day}^{-1}$     & $1.03$  & ($4.63\cdot 10^{-1}$,$1.86$)\\
$\theta^{\invasive}_{S}$ & Invasive susceptible adult host mortality rate   &     $  \mathrm{day}^{-1}$     & $4.49\cdot 10^{-1}$  & ($3.12\cdot 10^{-1}$,$1.40$)\\
$\theta^{\native}_{J}$ & Native juvenile mortality rate        &     $   \mathrm{day}^{-1}$     & $2.13\cdot 10^{-1}$  & ($7.94\cdot 10^{-2}$,$9.35\cdot 10^{-1}$)\\
$\theta^{\invasive}_{J}$ & Invasive juvenile mortality rate      &     $  \mathrm{day}^{-1}$     & $7.35\cdot 10^{-1}$  & ($2.91\cdot 10^{-1}$,$4.08$)\\
$\lambda^{\native}_{J}$ & Maturation rate of native juvenile                  &     $ \mathrm{day}^{-1} $                  & $1.00\cdot 10^{-1}$  & \\
$\lambda^{\invasive}_{J}$ & Maturation rate of invasive juvenile                &     $ \mathrm{day}^{-1} $                  & $1.00\cdot 10^{-1}$  & \\
$\xi_J$         & Ratio of juvenile individual filtering rate &     $\mathrm{Unitless}$                    & $1.00$ & \\
$\beta^{\native}$       & Spores produced per infected native individual     &  $10^3 \cdot \mathrm{spore} \cdot \mathrm{individual}^{-1} \cdot \mathrm{day}^{-1}$  & $3.00\cdot 10$ & \\
$\beta^{\invasive}$       & Spores produced per infected invasive individual   &  $10^3 \cdot \mathrm{spore} \cdot \mathrm{individual}^{-1} \cdot \mathrm{day}^{-1}$  & $3.00\cdot 10$ & \\
$\mu$           & Alga refilling rate                                &  $10^6 \cdot \mathrm{cell} \cdot \mathrm{L} ^{-1} \cdot \mathrm{day}^{-1}$                 & $3.70\cdot 10^{-1}$ & \\
$\delta$        & Sampling rate                                      &  $day ^{-1}$                                                                         & $1.30\cdot 10^{-2}$ & \\
$\sigma^{\native}_{S}$ & Standard deviation of Brownian motion of native susceptible adult             &   $\sqrt{\mathrm{individual} \cdot \mathrm{day}^{-1}}$  & 0  & \\
$\sigma^{\invasive}_{S}$ & Standard deviation of Brownian motion of invasive susceptible adult              &   $\sqrt{\mathrm{individual} \cdot \mathrm{day}^{-1}}$  & 0  & \\
$\sigma^{\native}_{J}$ & Standard deviation of Brownian motion of native juvenile               &   $\sqrt{\mathrm{individual} \cdot \mathrm{day}^{-1}}$  & $2.79\cdot 10^{-1}$  & ($1.38\cdot 10^{-1}$,$4.45\cdot 10^{-1}$)\\
$\sigma^{\invasive}_{J}$ & Standard deviation of Brownian motion of invasive juvenile             &   $\sqrt{\mathrm{individual} \cdot \mathrm{day}^{-1}}$  & $4.99\cdot 10^{-4}$  & ($5.87\cdot 10^{-5}$,$7.10\cdot 10^{-2}$)\\
$\sigma_{F}$   & Standard deviation of Brownian motion of alga                          &   $\sqrt{\mathrm{individual} \cdot \mathrm{day}^{-1}}$  & $5.37\cdot 10^{-2}$  & ($2.72\cdot 10^{-2}$,$9.19\cdot 10^{-2}$)\\
$\tau^{\native}_{S}$      &  Measurement dispersion for susceptible native adult             &$\mathrm{Unitless}$           & $1.12\cdot 10$  & ($4.26$,$1.10\cdot 10^{2}$)\\
$\tau^{\invasive}_{S}$      &  Measurement dispersion for susceptible invasive adult              &$\mathrm{Unitless}$            & $2.43$  & ($1.55$,$4.44$)\\
\hline
\end{tabular}}
\caption{Variables and parameter definitions and estimates of SRJF2 model for dynamics with both species of \textit{Daphnia}. The confidence interval is generated by the Monte Carlo Adjusted Profile.}
\label{Table:SRJF2}
\end{table}

\subsubsection{MCAP Results}
\begin{figure}[H]
    \centering
    \includegraphics[width=\linewidth]{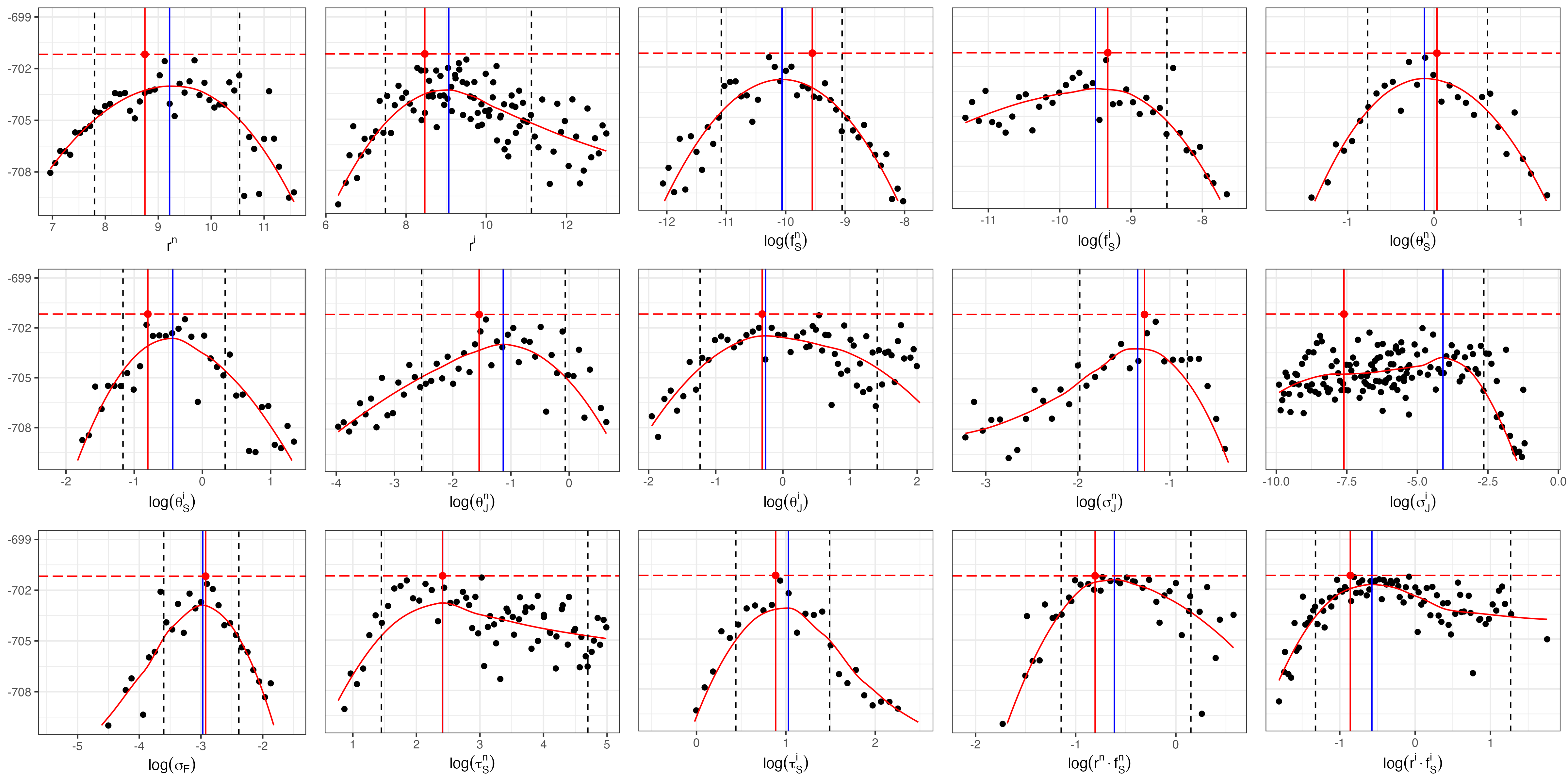}
    \caption{Monte Carlo Adjusted Profile results of SRJF2 model for dynamics with both species of \textit{Daphnia}.
    The vertical dotted lines represent the 95\% confidence interval obtained by MCAP.
    The vertical blue lines show the MLE estimated using MCAP.
    The red vertical lines correspond to the model with the overall highest likelihood among all searches.
}
    \label{fig:SRJF2_profile}
\end{figure}

\subsubsection{Model Comparison}

\begin{table}[h!]
\centering
\begin{center}
\begin{tabular}{||c |c |c |c|c||}
 \hline
Specific parameters  & Max log-likelihood & AIC & \makecell[c]{Max log-likelihood\\(block)} & \makecell[c]{AIC\\(block)}\\ [0.5ex]
 \hline\hline
$\varnothing$  & -701.17& 1428.34& -701.17& 1428.34 \\
$\theta^{\species}_{S,\unit}$ & -705.50& 1468.99& -691.20& 1440.39 \\
$f^{\species}_{S,\unit}$ & -720.72& 1499.44& -692.71& 1443.43 \\
$r^{\species}_{\unit}$  & -725.74& 1509.49& -696.35& 1450.70 \\[1ex]
 \hline
\end{tabular}
\end{center}
\caption{Comparison of model fit and complexity across various configurations of unit-specific parameters within the panelPOMP framework with SRJF2 model for dynamics with both species of \textit{Daphnia}. The unit-specific parameter setting assessed include (1)$\theta^{\species}_{S,\unit}$, (2)$f^{\species}_{S,\unit}$ , and (3)$r^{\species}_{\unit}$ for $\species \in \{\native,\invasive\}$.}
\label{Table:SRJF2_model_comparison}
\end{table}

Table~\ref{Table:SRJF2_model_comparison} presents a comparison of model fit and complexity across various configurations of unit-specific parameters within the panelPOMP framework for the SRJF2 model, which captures the dynamics of both native and invasive \textit{Daphnia} species.
We employed MCAP methods to estimate the maximum log-likelihood and AIC for each model configuration.
The models assessed include those with unit-specific parameters for the mortality rates ($\theta^{\species}_{S,\unit}$), filter rates ($f^{\species}_{S,\unit}$), and growth efficiency factors ($r^{\species}_{\unit}$) for $\species \in \{\native,\invasive\}$.
The model with all parameters shared (no unit-specific parameters) achieves the minimum AIC, indicating the best balance between model fit and complexity.
This outcome suggests that introducing unit-specific parameters does not significantly enhance the model's performance and may unnecessarily increase complexity.
Consequently, we choose to retain the model with all shared parameters, as it offers a more parsimonious representation of the system dynamics without compromising the explanatory power.

\subsection{Simulation}
\begin{figure}[H]
    \centering
    \includegraphics[width=0.75\linewidth]{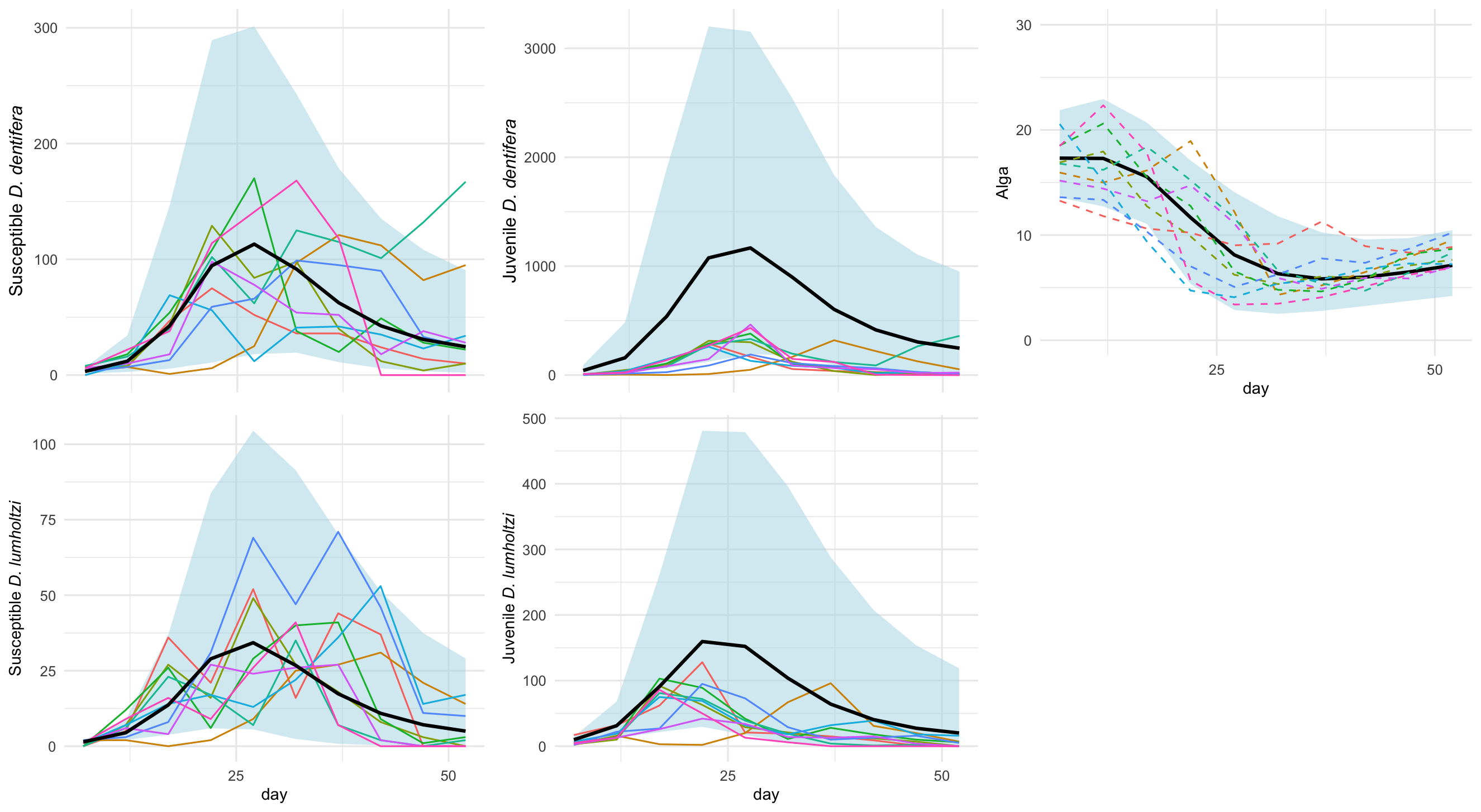}
    \caption{Simulation plots of densities (Individuals$/$Liter) for susceptible and juvenile populations of both native \textit{D.~dentifera} and invasive \textit{D.~lumholtzi}, along with alga density ($10^6\,$cells$/$Liter) over time (days). Colorful solid lines represent observed experimental data for each experimental replicate, showcasing variability across replicates. The black line represents the mean simulation trajectory, while the light blue shaded area corresponds to the 95\% confidence interval of the simulation. Dashed lines indicate individual simulation cases, illustrating the model's capacity to capture fluctuations around the mean trajectory.}
    \label{fig:SRJF2_simulation_band}
\end{figure}

\begin{figure}[H]
    \centering
    \includegraphics[width=0.75\linewidth]{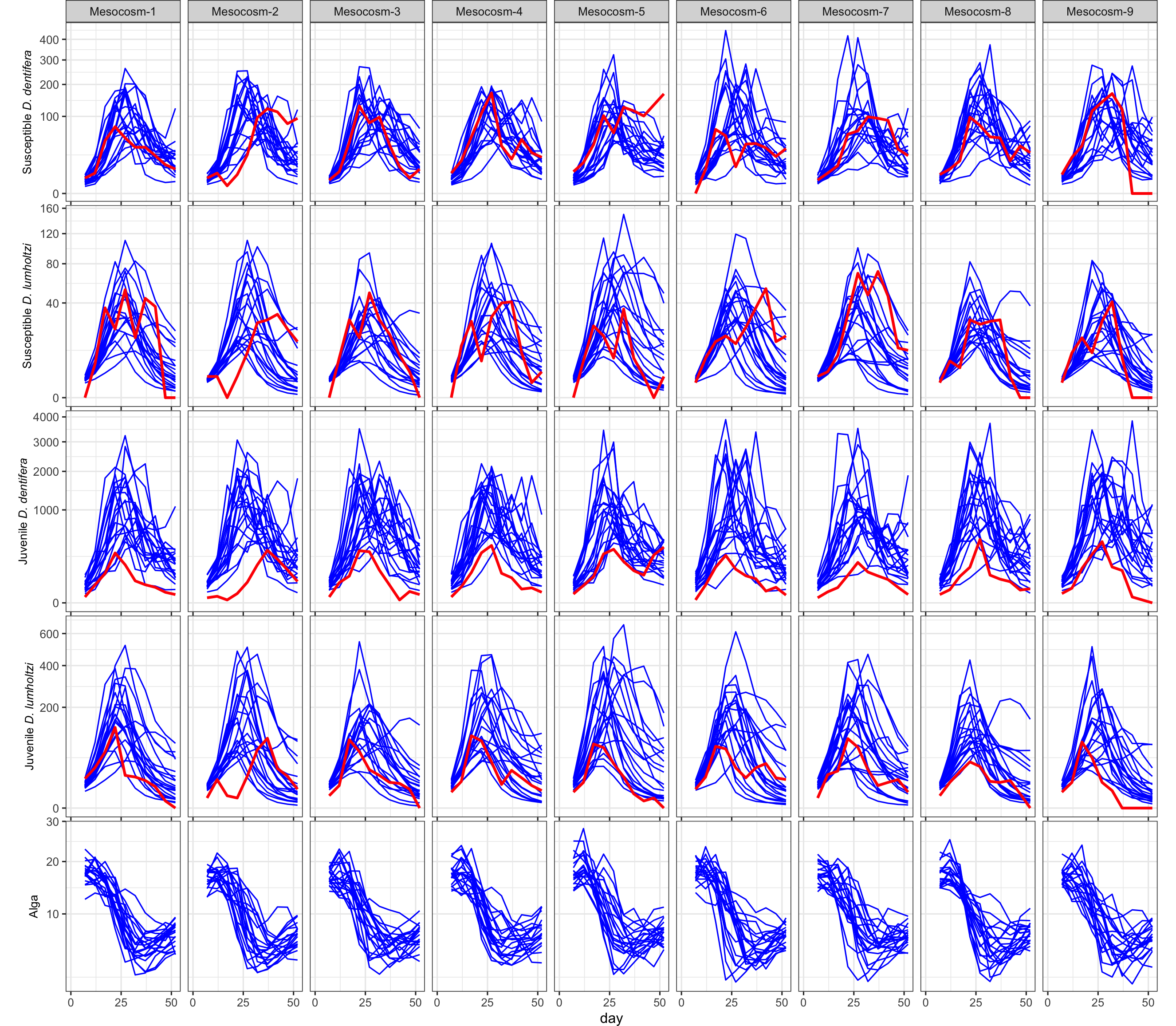}
    \caption{Simulated density ($10^6\,$cells$/$Liter) of susceptible native \textit{D.~dentifera}, susceptible invasive \textit{D.~lumholtzi}, juvenile native \textit{D.~dentifera}, invasive \textit{D.~lumholtzi}, and alga density ($F$) over time (days) for each experimental unit. The blue lines represent individual simulation cases, while the red line denotes observed experimental data, showing how simulated variability compares to actual data.}
    \label{fig:SRJF2_simulation_lines}
\end{figure}

The simulation results presented in Figures \ref{fig:SRJF2_simulation_band} to \ref{fig:SRJF2_simulation_lines} showcase our model's ability to represent the complex population interactions between \textit{D.~dentifera} and \textit{D.~lumholtzi}.
The model captures the patterns of susceptible and juvenile densities across multiple experimental units, as evidenced by the close alignment of the simulated mean trajectories (illustrated by the black line and light blue confidence bands) with the experimental data (shown by the colorful solid lines).
Despite the fact that juvenile density data were not used in calibrating the model, the model accurately predicts juvenile density trends, indicating its capability to generalize ecological processes effectively.
The 95\% confidence intervals largely encompass the observed juvenile data, supporting the construction of the model's latent process.

\parskip 3mm

Figure \ref{fig:SRJF2_simulation_lines} provides further insight into the dynamics of algal depletion over time, an aspect that is integral to the model’s predictions for \textit{Daphnia} populations.
The observed decrease in algal density at later stages, captured in the simulations, suggests a shift to a resource-limited environment—a behavior that aligns well with our model's assumptions.
As algae availability declines, population growth of both susceptible and juvenile \textit{Daphnia} becomes constrained, providing a mechanistic explanation for the observed patterns in their densities.
By incorporating resource-consumption feedback into the population dynamics, the model demonstrates not only an ability to simulate individual population trends but also to capture this ecological feedback loop.

\section{SRJF Model}
In this section, we introduce the SRJF model, which formalizes the dynamics among susceptible individuals ($S^{\species}$), the juvenile population ($J^{\species}$), and algae serving as a food resource ($F$) as follows:
{\small
\begin{align}
    dS^{\species}_{\unit}(t) &= \lambda^{\species}_{J,{\unit}} J^{\species}_{\unit}(t) \, dt - (\theta^{\species}_{S,{\unit}} + \delta) S^{\species}_{\unit}(t)\, dt + S^{\species}_{\unit}(t)\, d\zeta^{\species}_{S,{\unit}} , \label{eq:SRJF_dS}\\
    dJ^{\species}_{\unit}(t) &= r^{\species}_{\unit}f^{\species}_{S,{\unit}}F_{\unit}(t) S^{\species}_{\unit}(t)\, dt - \theta^{\species}_{J,{\unit}} J^{\species}_{\unit}(t)\, dt - \delta J^{\species}_{\unit}(t)\, dt - \lambda^{\species}_{J,{\unit}} J^{\species}_{\unit}(t)\, dt + J^{\species}_{\unit}(t)\, d\zeta^{\species}_{J,{\unit}}, \label{eq:SRJF_dJ}\\
    dF_{\unit}(t) &= - f^{\species}_{S,{\unit}} F_{\unit}(t) \left(S^{\species}_{\unit}(t)+\xi_{J,{\unit}}J^{\species}_{\unit}(t)\right)dt  + \mu \, dt + F_{\unit}(t)\, d\zeta_{F,{\unit}}, \label{eq:SRJF_dF}\\
d\zeta^{\species}_{S,{\unit}} &\sim \Normal \big[0, (\sigma^{\species}_{S,{\unit}})^{2}\, dt\big]\,
d\zeta^{\species}_{J,{\unit}} \sim \Normal \big[0, (\sigma^{\species}_{J,{\unit}})^{2}\, dt\big]\,
d\zeta_{F,{\unit}} \sim \Normal \big[0, \sigma_{F,{\unit}}^{2}\, dt\big].\label{eq:SRJF_dzetaS}
\end{align}
}
Equation~\eqref{eq:SRJF_dS} describes the rate of change of the susceptible population $S^{\species}_{\unit}(t)$. The term $\lambda^{\species}_{J,\unit} J^{\species}(t) \, dt$ represents the growth of susceptible individuals from the maturation of juvenile population at a rate $\lambda^{\species}_{J,\unit}$.
The loss term $-\left(\theta^{\species}_{S,{\unit}}+ \delta\right) S^{\species}_{\unit}(t) \, dt$ accounts for the mortality of susceptible individuals due to natural causes at rate $\theta^{\species}_{S,\unit}$ and sampling at rate $\delta$.
The stochastic component $S^{\species}_{\unit}(t) \, d\zeta^{\species}_{S,\unit}$ introduces random fluctuations into the susceptible population dynamics, with $d\zeta^{\species}_{S,\unit}$ being a Brownian motion characterized by variance $(\sigma^{\species}_{S,{\unit}})^{2}\, dt$.
For consistency with the SIRJPF2 and SIRJPF model, we further fix $\sigma^{\species}_{S,{\unit}}$ to be zero for specific analysis of this dynamics.
\parskip 3mm

In Equation~\eqref{eq:SRJF_dJ}, the dynamics of the juvenile population $J^{\species}_{\unit}(t)$ are modeled. The growth term $r^{\species}_{\unit} f^{\species}_{S,\unit} F^{\species}_{\unit}(t) S^{\species}_{\unit}(t) \, dt$ signifies the increase in juveniles resulting from feeding on alga, where $r^{\species}_{\unit}$ is the growth efficiency factor and $f^{\species}_{S,\unit}$ is the filtering rate of susceptible individuals.
The interaction $F^{\species}_{\unit}(t) S^{\species}_{\unit}(t)$ captures how the availability of algae influences juvenile growth.
The loss term $-\left(\theta^{\species}_{J,\unit} + \delta + \lambda^{\species}_{J,\unit}\right) J^{\species}_{\unit}(t) \, dt$ reflects natural juvenile mortality at rate $\theta^{\species}_{J,\unit}$, sampling at rate $\delta$, and the maturing of juveniles to susceptible individuals at rate $\lambda^{\species}_{J,\unit}$.
The stochastic term $J^{\species}_{\unit}(t) \, d\zeta^{\species}_{J,\unit}$ reflects random environmental and demographic fluctuations affecting juveniles, with $d\zeta^{\species}_{J,\unit}$ being Brownian motion with variance $(\sigma^{\species}_{J,{\unit}})^{2}\, dt$.
However, introducing several stochastic terms when describing \textit{Daphnia} behavior allows too much freedom for model, which results in weakly identified parameters and overfitting.
For the purpose of better generalization, we allow the stochastic term in juvenile states to affect the density of adult to show the fluctuations.
\parskip 3mm

Equation~\eqref{eq:SRJF_dF} captures the dynamics of the \textit{A.} algae population $F^{\species}_{\unit}(t)$.
The consumption term $- f^{\species}_{S,\unit} F^{\species}_{\unit}(t) \left(S^{\species}_{\unit}(t) + \xi_{J,\unit} J^{\species}_{\unit}(t) \right) \, dt$ represents the reduction of algae due to feeding by both susceptible individuals and juveniles, with $\xi_{J,\unit}$ indicating the ratio of juveniles on algae consumption compared to susceptible individuals.
The term $\mu \, dt$ denotes the refill rate of alga.
The stochastic fluctuation $F^{\species}_{\unit}(t) \, d\zeta^{\species}_{F,\unit}$ introduces randomness into the algae dynamics, where $d\zeta^{\species}_{F,\unit}$ shows the Brownian motion with variance $\sigma_{F,{\unit}}^{2}\, dt$.
These terms model the latent randomness in the system due to environmental variability and other stochastic factors influencing the susceptible individuals, juveniles, and alga populations.

\subsection{Flow Diagram}
\begin{figure}[H]
\begin{center}
\resizebox{8cm}{!}{
\begin{tikzpicture}[
  square/.style={rectangle, draw=black, minimum width=0.5cm, minimum height=0.5cm, rounded corners=.1cm, fill=blue!8},
  rhombus/.style={diamond, draw=black, minimum width=0.1cm, minimum height=0.1cm, fill=purple!8,aspect = 1},
  travel/.style={circle, draw=black, minimum width=0.5cm, minimum height=0.5cm, fill=green!8},
  report/.style={shape=regular polygon, regular polygon sides=8, draw, fill=red!8,minimum size=0.6cm,inner sep=0cm},
  bendy/.style={bend left=10},
  bendy2/.style={bend left=100},
  bendy3/.style={bend left=-100},
  >/.style={shorten >=0.25mm},
  >/.tip={Stealth[length=1.5mm,width=1.5mm]}
]
\tikzset{>={}};

\node (Sn) at (5.5,0) [square] {$S$};
\node (Jn) at (5.5,2) [square] {$J$};
\node (R) at (2.5,0)  [rhombus] {$R$};
\node (F) at (2.5,2) [report] {$F$};

\draw [->, bendy] (Jn) to  (Sn);
\draw [->, bendy] (Sn) to  (Jn);

\draw [->] (F) --  (R);
\draw [->] (Sn) -- (R);
\draw [->] (Jn) -- (R);
\draw [->] (F) -- (Sn);
\draw [->] (F) -- (Jn);
\draw (F) edge[loop above] (F);

\end{tikzpicture}
}
\end{center}
\vspace{-5mm}
\caption{Flow diagram for the SRJF model illustrating population interactions. The model includes the $R$ state, representing mortality. For $\species \in \{\native,\invasive\}$, susceptible populations ($S^{\species}$) can reproduce into juvenile ($J^{\species}$) and \textit{A.} Algae ($F$) will be refilled, as shown by the recycling arrows. Both $S^{\species}$ and $J^{\species}$ consume resources from $F$, and over time, components in $F$ also progress to $R$.}
\label{fig:SRJF}
\end{figure}
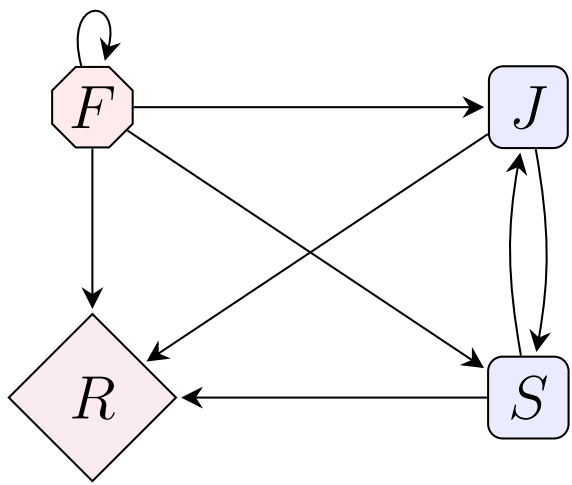

\subsection{Data Visualization}
\begin{figure}[H]
\centering
\begin{knitrout}
\definecolor{shadecolor}{rgb}{0.969, 0.969, 0.969}\color{fgcolor}
\includegraphics[width=\textwidth]{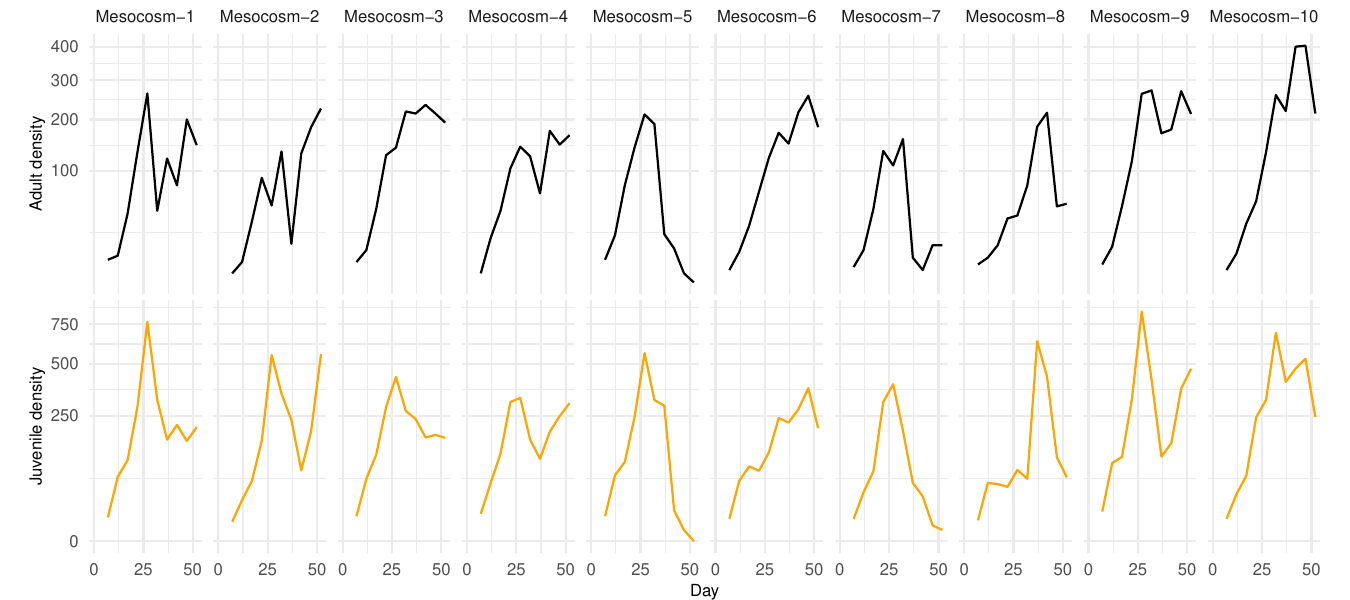} 
\end{knitrout}
\caption{\label{fig:data_vis_SRJF_dent}
Density (Individuals$/$Liter) of \textit{D.~dentifera}.
The top panel shows adult susceptibles (\textit{D.~dentifera}, black).
The bottom panel shows juvenile susceptibles (\textit{D.~dentifera}, orange).
There were negligible infected juveniles.
Columns are buckets corresponding to replications with same treatment setting.
}
\end{figure}

\begin{figure}[H]
\centering
\begin{knitrout}
\definecolor{shadecolor}{rgb}{0.969, 0.969, 0.969}\color{fgcolor}
\includegraphics[width=\textwidth]{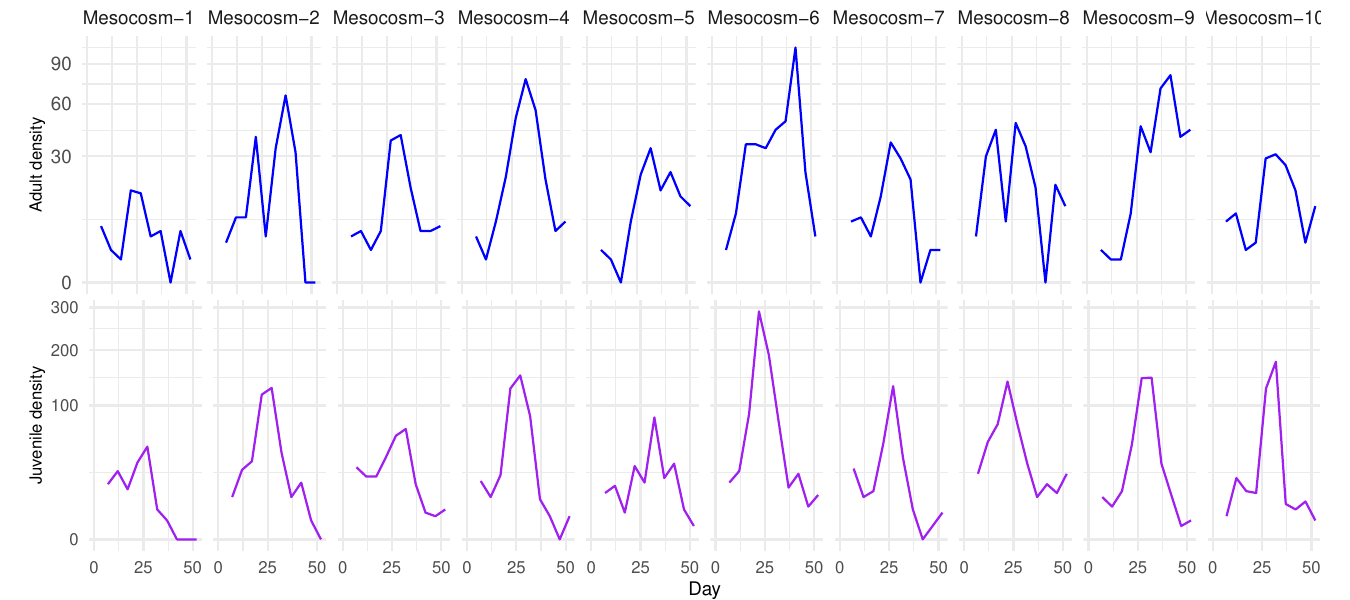} 
\end{knitrout}
\caption{\label{fig:data_vis_SIPJF_lum}
Density (Individuals$/$Liter) of \textit{D.~lumholtzi}.
The top panel shows adult susceptibles (\textit{D.~lumholtzi}, blue).
The bottom panel shows juvenile susceptibles (\textit{D.~lumholtzi}, purple).
There were negligible infected juveniles.
Columns are buckets corresponding to replications with same treatment setting.
}
\end{figure}

\subsection{Result}

\subsubsection{Parameter Estimation}
\begin{table}[H]
\centering
\resizebox{\textwidth}{!}{
\begin{tabular}{||c|c |c|c|c||}
\hline
Parameter & Definition &Unit & Value & CI\\[0.5ex]
\hline
\hline
$S^{\native}$ & Susceptible host density   & $\mathrm{individual} \cdot L ^{-1}$                 & Variable                & \\
$J^{\native}$ & Juvenile host density    & $ \mathrm{individual} \cdot L ^{-1}$                & Variable                & \\
$F$   & Alga density                              & $10^6 \cdot \mathrm{cell} \cdot L ^{-1}$             & Variable                & \\
$r^{\native}$ &  Birth rate of juvenile                        & $\mathrm{individual} \cdot 10^{-6}\cdot \mathrm{cell}^{-1}$ & $1.69\cdot 10^{3}$  & ($6.78\cdot 10^{2}$,$4.53\cdot 10^{3}$)\\
$f^{\native}_{S}$ & Susceptible adult host filtering rate   &  $\mathrm{L} \cdot \mathrm{individual}^{-1} \cdot \mathrm{day}^{-1}$   & $1.32\cdot 10^{-4}$  & ($7.29\cdot 10^{-5}$,$2.39\cdot 10^{-4}$)\\
$r^{\native} \cdot f_{S^{\native}}$          & Effective Birth rate of native juvenile            &     $10^{-6} \cdot \mathrm{L} \cdot \mathrm{cell}^{-1} \cdot \mathrm{day}^{-1} $ & $2.23\cdot 10^{-1}$  & ($1.53\cdot 10^{-1}$,$3.59\cdot 10^{-1}$)\\
$\theta^{\native}_{S}$ & Susceptible adult host mortality rate     &     $  \mathrm{day}^{-1}$     & $6.92\cdot 10^{-1}$  & ($4.27\cdot 10^{-1}$,$1.27$)\\
$\theta^{\native}_{J}$ & Juvenile mortality rate        &     $ \mathrm{day}^{-1}$     & $2.13\cdot 10^{-8}$  & (0,$4.56\cdot 10^{-2}$)\\
$\lambda^{\native}_{J}$ & Maturation rate of the juvenile                  &     $ \mathrm{day}^{-1} $                  & $1.00\cdot 10^{-1}$  & \\
$\xi_J$         & Ratio of juvenile individual filtering rate &     $\mathrm{Unitless}$                    & $1.00$ & \\
$\beta^{\native}$       & Spores produced by death per infected individual     &  $10^3 \cdot \mathrm{spore} \cdot \mathrm{individual}^{-1} \cdot \mathrm{day}^{-1}$  & $3.00\cdot 10$ & \\
$\mu$           & Algae refilling rate                                &  $10^6 \cdot \mathrm{cell} \cdot \mathrm{L} ^{-1} \cdot \mathrm{day}^{-1}$                 & $3.70\cdot 10^{-1}$ & \\
$\delta$        & Sampling rate                                      &  $day ^{-1}$                                                                         & $1.30\cdot 10^{-2}$ & \\
$\sigma^{\native}_{S}$ & Standard deviation of Brownian motion of susceptible adult               &   $\sqrt{\mathrm{individual} \cdot \mathrm{day}^{-1}}$  & 0  & \\
$\sigma^{\native}_{J}$ & Standard deviation of Brownian motion of juvenile               &   $\sqrt{\mathrm{individual} \cdot \mathrm{day}^{-1}}$  & $3.07\cdot 10^{-1}$  & ($2.31\cdot 10^{-1}$,$3.81\cdot 10^{-1}$)\\
$\sigma_{F}$   & Standard deviation of Brownian motion of alga                          &   $\sqrt{\mathrm{individual} \cdot \mathrm{day}^{-1}}$  & $9.31\cdot 10^{-7}$  & (0,$4.60\cdot 10^{-2}$)\\
$\tau^{\native}_{S}$      &  Measurement dispersion for susceptible adult          &$\mathrm{Unitless}$           & $1.34\cdot 10$  & ($6.77$,$9.81\cdot 10$)\\
\hline
\end{tabular}}
\caption{Variables and parameter definitions and estimates of SRJF model for \textit{D.~dentifera}-only dynamics. The confidence interval is generated by the Monte Carlo Adjusted Profile.}
\label{Table:SRJF_native}
\end{table}

\subsubsection{Parameter Estimation}
\begin{table}[H]
\centering
\resizebox{\textwidth}{!}{
\begin{tabular}{||c|c |c|c|c||}
\hline
Parameter & Definition &Unit & Value & CI\\[0.5ex]
\hline
\hline
$S^{\invasive}$ & Susceptible host density   & $\mathrm{individual} \cdot L ^{-1}$                 & Variable                & \\
$J^{\invasive}$ & Juvenile host density    & $ \mathrm{individual} \cdot L ^{-1}$                & Variable                & \\
$F$   & Alga density                              & $10^6 \cdot \mathrm{cell} \cdot L ^{-1}$             & Variable                & \\
$r^{\invasive}$ &  Birth rate of juvenile                        & $\mathrm{individual} \cdot 10^{-6}\cdot \mathrm{cell}^{-1}$ & $9.77\cdot 10^{2}$  & ($8.86\cdot 10$,$\infty$)\\
$f^{\invasive}_{S}$ & Susceptible adult host filtering rate   &  $\mathrm{L} \cdot \mathrm{individual}^{-1} \cdot \mathrm{day}^{-1}$   & $2.67\cdot 10^{-4}$  & (0,$1.34\cdot 10^{-3}$)\\
$r^{\invasive} \cdot f_{S^{\invasive}}$          & Effective Birth rate of invasive juvenile            &     $10^{-6} \cdot \mathrm{L} \cdot \mathrm{cell}^{-1} \cdot \mathrm{day}^{-1} $ & $2.61\cdot 10^{-1}$  & ($8.60\cdot 10^{-2}$,$4.22$)\\
$\theta^{\invasive}_{S}$ & Susceptible adult host mortality rate     &     $ \mathrm{day}^{-1}$     & $5.99\cdot 10^{-1}$  & ($2.26\cdot 10^{-1}$,$1.40$)\\
$\theta^{\invasive}_{J}$ & Juvenile mortality rate        &     $ \mathrm{day}^{-1}$     & $2.98\cdot 10^{-1}$  & (0,$6.41$)\\
$\lambda^{\invasive}_{J}$ & Maturation rate of the juvenile                  &     $ \mathrm{day}^{-1} $                  & $1.00\cdot 10^{-1}$  & \\
$\xi_J$         & Ratio of juvenile individual filtering rate &     $\mathrm{Unitless}$                    & $1.00$ & \\
$\beta^{\invasive}$       & Spores produced by death per infected individual     &  $10^3 \cdot \mathrm{spore} \cdot \mathrm{individual}^{-1} \cdot \mathrm{day}^{-1}$  & $3.00\cdot 10$ & \\
$\mu$           & Algae refilling rate                                &  $10^6 \cdot \mathrm{cell} \cdot \mathrm{L} ^{-1} \cdot \mathrm{day}^{-1}$                 & $3.70\cdot 10^{-1}$ & \\
$\delta$        & Sampling rate                                      &  $day ^{-1}$                                                                         & $1.30\cdot 10^{-2}$ & \\
$\sigma^{\invasive}_{S}$ & Standard deviation of Brownian motion of susceptible adult               &   $\sqrt{\mathrm{individual} \cdot \mathrm{day}^{-1}}$  & 0  & \\
$\sigma^{\invasive}_{J}$ & Standard deviation of Brownian motion of juvenile               &   $\sqrt{\mathrm{individual} \cdot \mathrm{day}^{-1}}$  & $3.78\cdot 10^{-1}$  & (0,$5.07\cdot 10^{-1}$)\\
$\sigma^{\invasive}_{F}$   & Standard deviation of Brownian motion of alga                          &   $\sqrt{\mathrm{individual} \cdot \mathrm{day}^{-1}}$  & $4.73\cdot 10^{-2}$  & (0,$1.38\cdot 10^{-1}$)\\
$\tau^{\invasive}_{S}$      &  Measurement dispersion for susceptible adult           &$\mathrm{Unitless}$           & $2.35$  & ($1.32$,$5.72$)\\
\hline
\end{tabular}}
\caption{Variables and parameter definitions and estimates of SRJF model for \textit{D.~lumholtzi}-only dynamics. The confidence interval is generated by the Monte Carlo Adjusted Profile.}
\label{Table:SRJF_invasive}
\end{table}

\subsubsection{MCAP Results}
\begin{figure}[H]
    \centering
    \includegraphics[width=\linewidth]{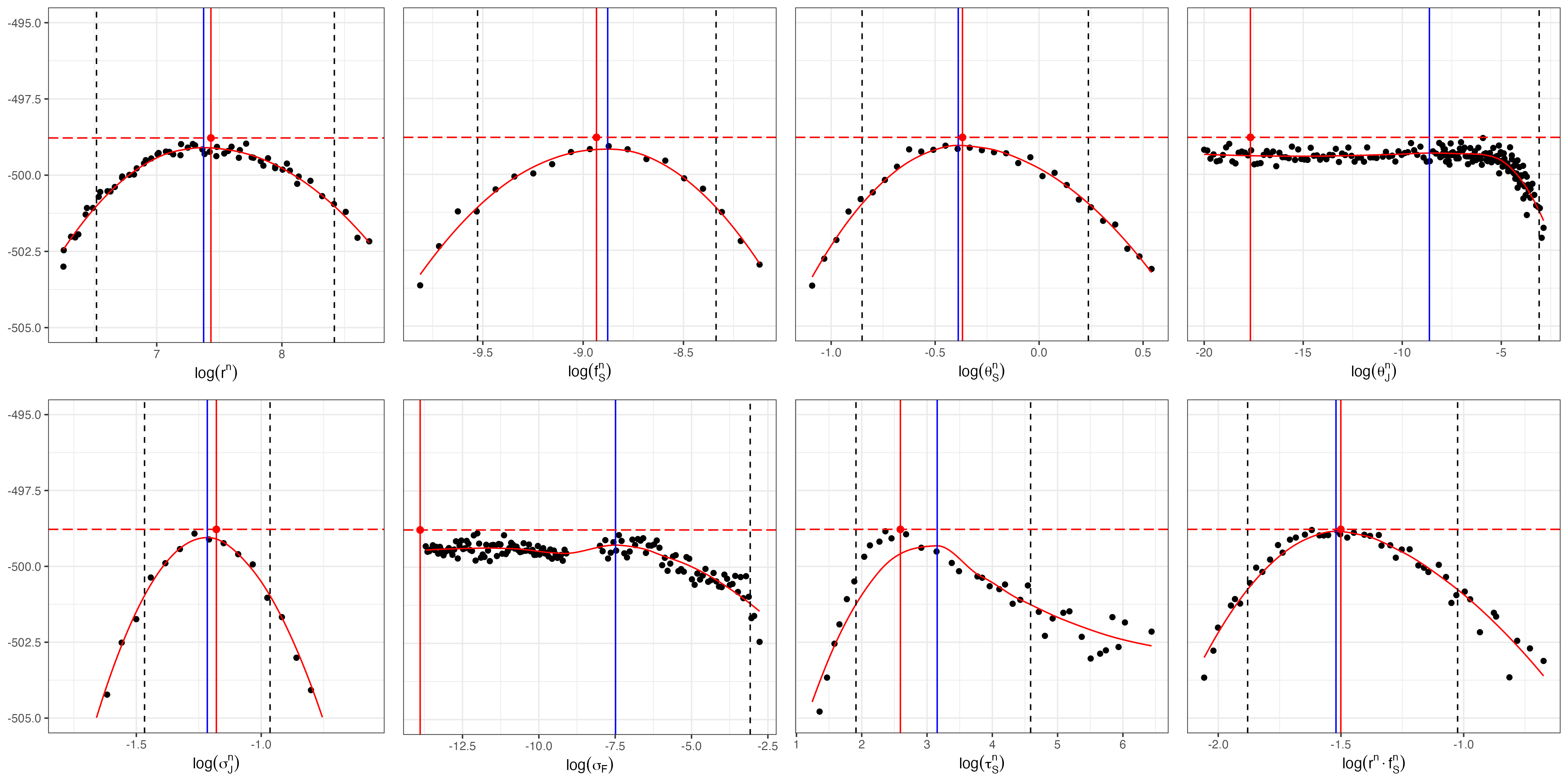}
    \caption{Monte Carlo Adjusted Profile results of SRJF model for \textit{D.~dentifera}-only dynamics.
    The vertical dotted lines represent the 95\% confidence interval obtained by MCAP.
    The vertical blue lines show the MLE estimated using MCAP.
    The red vertical lines correspond to the model with the overall highest likelihood among all searches.
}
    \label{fig:SRJF_dent_profile}
\end{figure}

\begin{figure}[H]
    \centering
    \includegraphics[width=\linewidth]{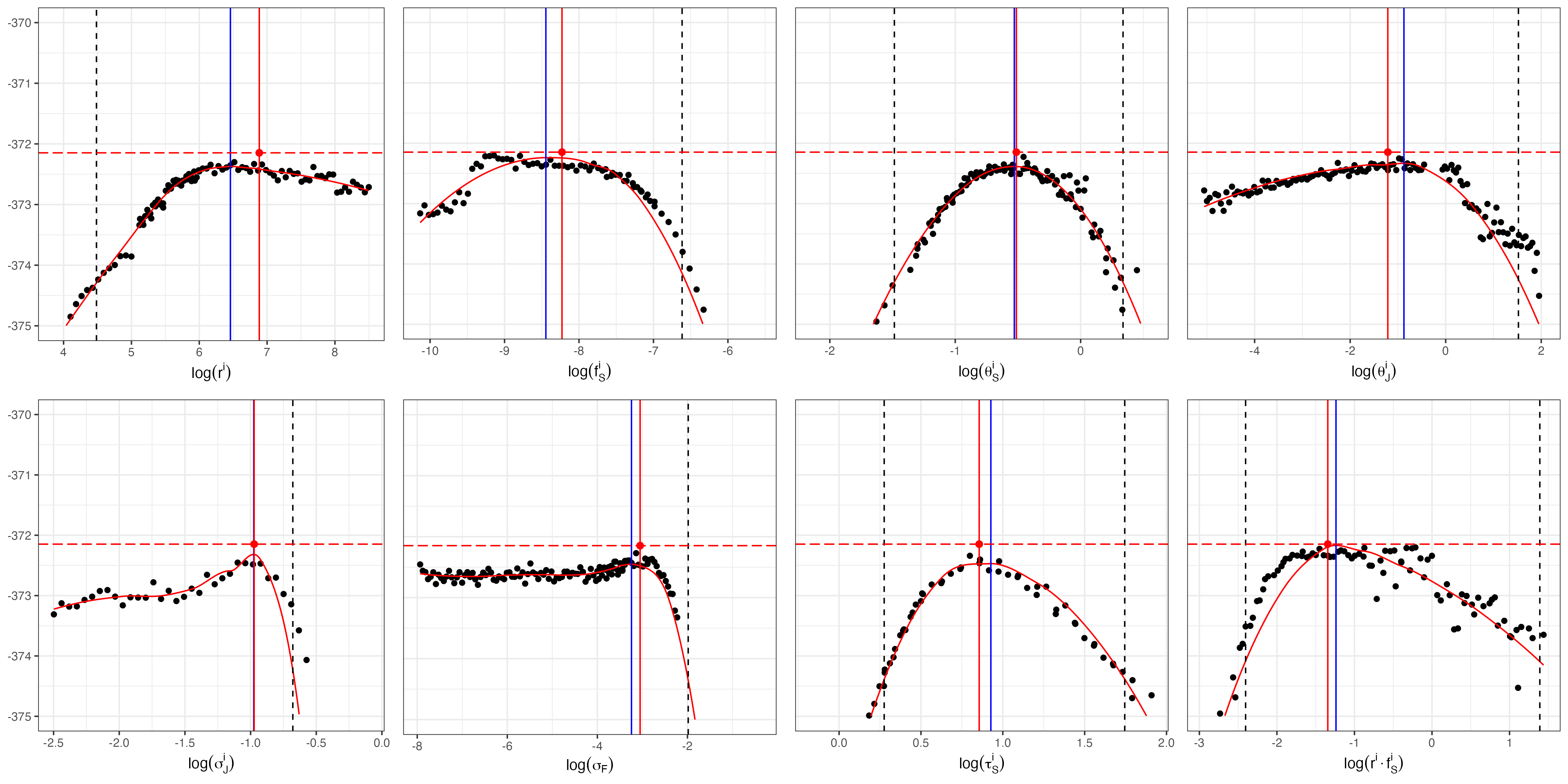}
    \caption{Monte Carlo Adjusted Profile results of SRJF model for \textit{D.~lumholtzi}-only dynamics.
    The vertical dotted lines represent the 95\% confidence interval obtained by MCAP.
    The vertical blue lines show the MLE estimated using MCAP.
    The red vertical lines correspond to the model with the overall highest likelihood among all searches.
}
    \label{fig:SRJF_lum_profile}
\end{figure}

\subsubsection{Model Comparison}
\begin{table}[h!]
\centering
\begin{center}
\begin{tabular}{||c |c |c |c|c||}
 \hline
Specific parameters  & Max log-likelihood & AIC & \makecell[c]{Max log-likelihood\\(block)} & \makecell[c]{AIC\\(block)}\\ [0.5ex]
 \hline\hline
$\varnothing$ & -498.78& 1011.56& -498.78& 1011.56 \\
$\theta^{\native}_{S,\unit}$  & -490.57& 1013.14& -489.85& 1011.71 \\
$r^{\native}_{\unit}$ & -497.43& 1026.85& -496.44& 1024.89 \\
$f^{\native}_{S,\unit}$ & -497.43& 1026.86& -496.16& 1024.32\\[1ex]
 \hline
\end{tabular}
\end{center}
\caption{Comparison of model fit and complexity across various configurations of unit-specific parameters within the panelPOMP framework for \textit{D.~dentifera}-only dynamics. The unit-specific parameter setting assessed include $f^{\native}_{S,\unit}$, $r^{\native}_{\unit}$, $\theta^{\native}_{S,\unit}$.}
\label{Table:SRJF_dent_model_comparison}
\end{table}

\begin{table}[h!]
\centering
\begin{center}
\begin{tabular}{||c |c |c |c|c||}
 \hline
Specific parameters  & Max log-likelihood & AIC & \makecell[c]{Max log-likelihood\\(block)} & \makecell[c]{AIC\\(block)}\\ [0.5ex]
 \hline\hline
$\varnothing$ & -372.15& 758.294& -372.15& 758.294 \\
$\theta^{\invasive}_{S,\unit}$  & -365.33& 762.658& -365.37& 762.733 \\
$r^{\invasive}_{\unit}$ & -366.84& 765.676& -366.29& 764.571 \\
$f^{\invasive}_{S,\unit}$ & -368.15& 768.300& -368.07& 768.147 \\[1ex]
 \hline
\end{tabular}
\end{center}
\caption{Comparison of model fit and complexity across various configurations of unit-specific parameters within the panelPOMP framework for \textit{D.~lumholtzi}-only dynamics. The unit-specific parameter setting assessed include $f^{\invasive}_{S,\unit}$, $r^{\invasive}_{\unit}$, $\theta^{\invasive}_{S,\unit}$.}
\label{Table:SRJF_lum_model_comparison}
\end{table}

The results in the tables indicate that the model configuration with all shared parameters achieves the minimum AIC score for both dynamics, suggesting an optimal balance between model fit and complexity.
These results align with our previous analysis.

\subsection{Simulation}
\begin{figure}[H]
    \centering
    \includegraphics[width=0.75\linewidth]{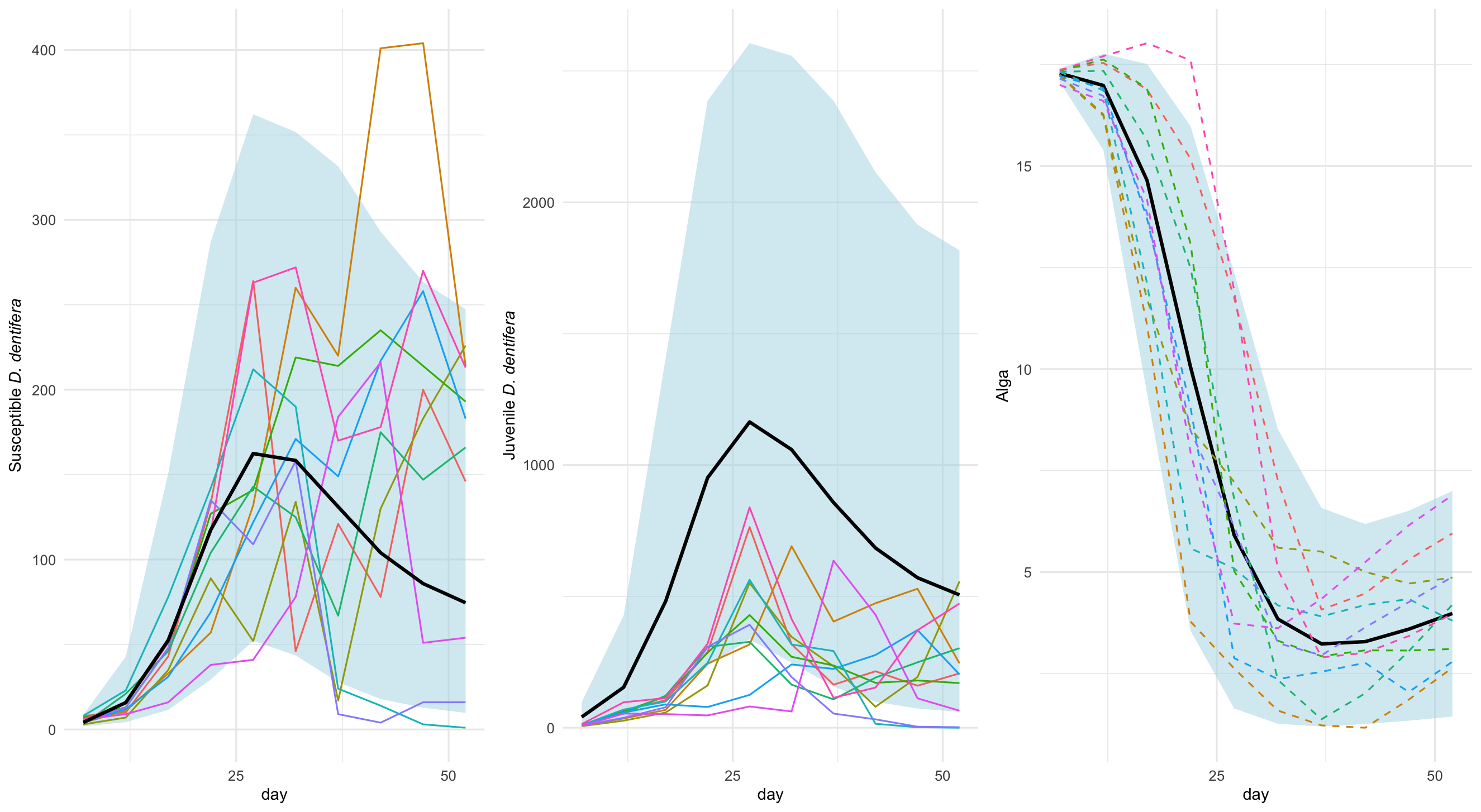}
    \caption{Simulation plots of densities (Individuals$/$Liter) for susceptible, and juvenile \textit{D.~dentifera}, along with alga density ($10^6\,$cells$/$Liter) and parasite density  ($10^3\,$spores$/$Liter), over time (days). The colorful solid lines represent observed experimental data, showing the variability across different experimental replicates. The black line represents the mean trajectory from the simulation model, capturing the general trend of each density. The light blue shaded area corresponds to the 95\% confidence interval of the simulations, indicating the range of model predictions. Dashed lines illustrate individual simulated trajectories, highlighting the model’s ability to capture fluctuations around the mean trajectory.}
    \label{fig:SRJF_native_simulation_band}
\end{figure}

\begin{figure}[H]
    \centering
    \includegraphics[width=0.75\linewidth]{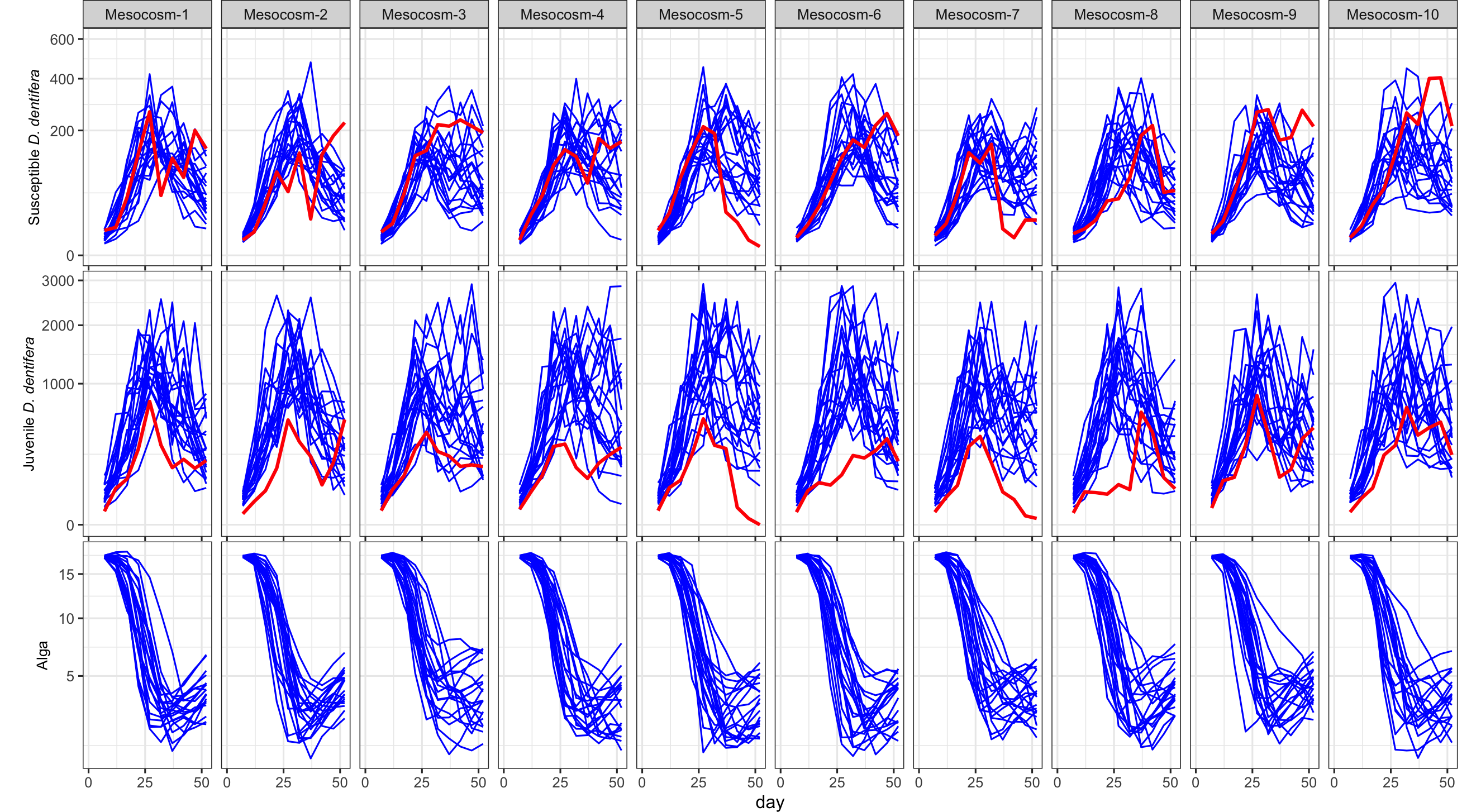}
   \caption{Simulated densities (Individuals$/$Liter) of susceptible \textit{D.~dentifera}, juvenile \textit{D.~dentifera} and alga density ($10^6\,$cells$/$Liter) over time (days) for each experimental unit. The blue lines represent individual simulation runs, capturing the variability in susceptible density across replicates, while the red line represents the actual experiment data.}
    \label{fig:SRJF_dent_simulation_lines}
\end{figure}

\begin{figure}[H]
    \centering
    \includegraphics[width=0.75\linewidth]{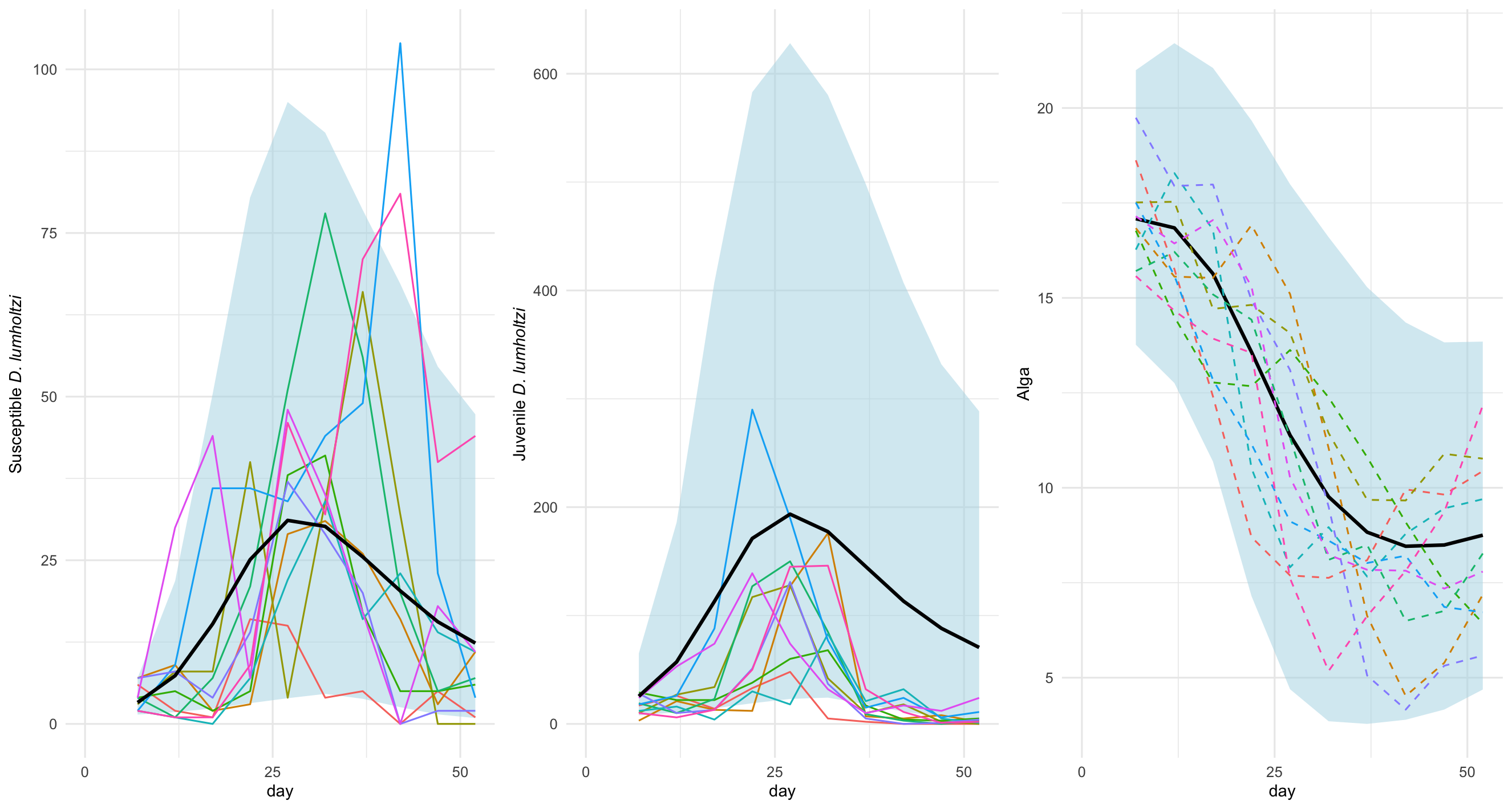}
    \caption{Simulation plots of densities (Individuals$/$Liter) for susceptible, and juvenile \textit{D.~lumholtzi}, along with alga density ($10^6\cdot$cells$/$Liter) and parasite density  ($10^3\cdot$spores$/$Liter), over time (days). The colorful solid lines represent observed experimental data, showing the variability across different experimental replicates. The black line represents the mean trajectory from the simulation model, capturing the general trend of each density. The light blue shaded area corresponds to the 95\% confidence interval of the simulations, indicating the range of model predictions. Dashed lines illustrate individual simulated trajectories, highlighting the model’s ability to capture fluctuations around the mean trajectory.}
    \label{fig:SRJF_invasive_simulation_band}
\end{figure}

\begin{figure}[H]
    \centering
    \includegraphics[width=0.75\linewidth]{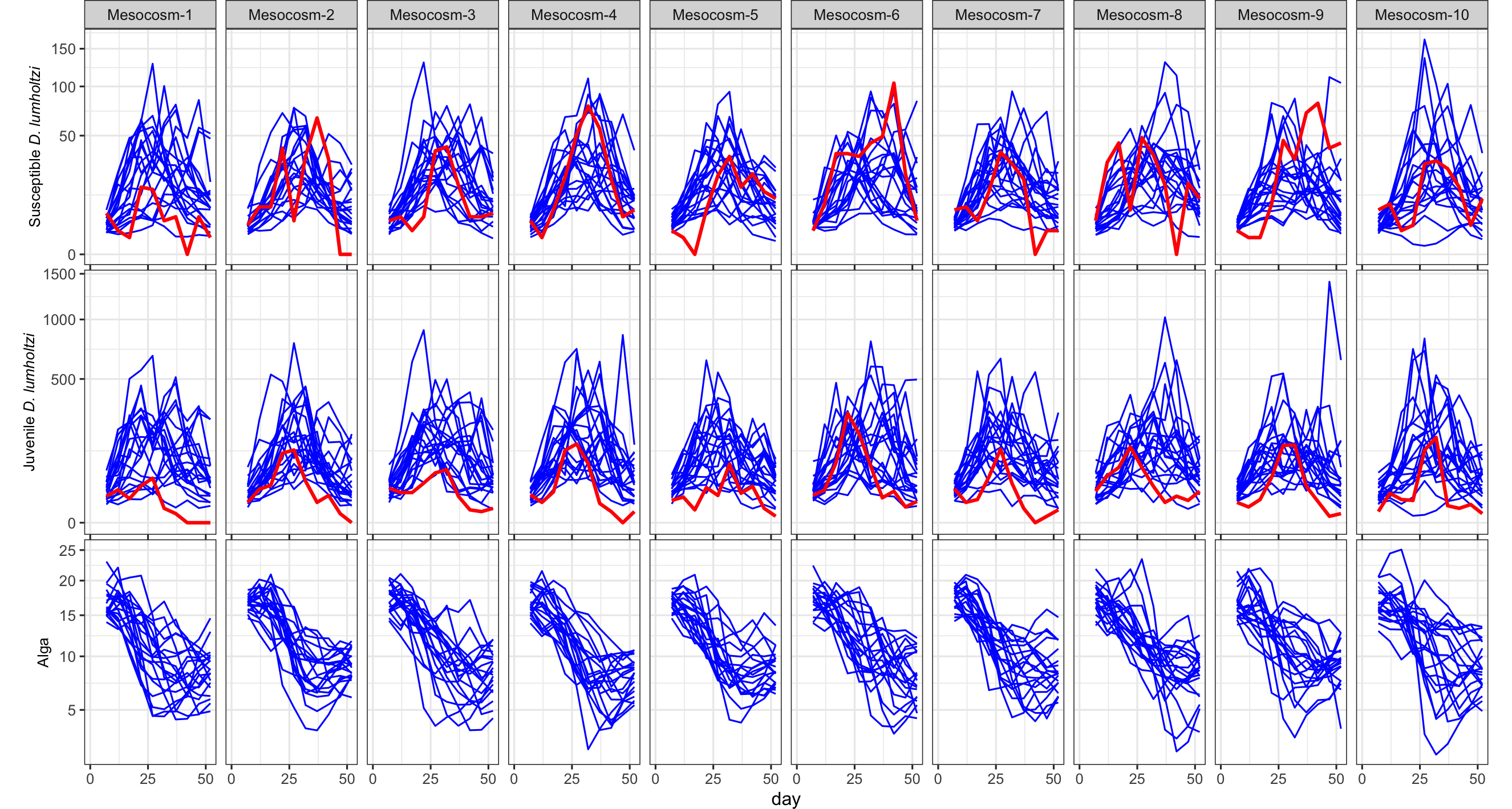}
   \caption{Simulated densities (Individuals$/$Liter) of susceptible \textit{D.~lumholtzi}, juvenile \textit{D.~lumholtzi} and  alga density ($10^6\cdot$cells$/$Liter) ays) for each experimental unit. The blue lines represent individual simulation runs, capturing the variability in susceptible density across replicates, while the red line represents the actual experiment data.}
    \label{fig:SRJF_lum_simulation_lines}
\end{figure}

The simulation plots (Figures \ref{fig:SRJF_native_simulation_band} to \ref{fig:SRJF_lum_simulation_lines}) collectively demonstrate the SRPF model's capability to capture the complex interactions among various states within the population dynamics of \textit{D.~dentifera} and \textit{D.~lumholtzi}.
Specifically, the model effectively reproduces the latent patterns for susceptible and infected densities across multiple experimental units, as evidenced by the close alignment between the simulated mean trajectories (depicted by the black and light-blue bands) and the observed experimental data (represented by colorful solid lines).
Each blue line illustrates an individual simulation case, capturing the stochastic variability across replicates, while the light blue shaded area (where present) denotes the 95\% confidence interval of the simulation results.
Although juvenile density data were not included in the model fitting process, the model still accurately predicts juvenile density trends (Figures \ref{fig:SRJF_lum_simulation_lines}).
This performance indicates the model's capacity to generalize underlying ecological processes, as the 95\% confidence intervals largely encompass the observed juvenile data.
The model also predicts a decline in algal density as the experiment progresses.
This predicted reduction in food availability provides a mechanistic explanation for the observed trends in \textit{Daphnia} densities, as resource scarcity constrains population growth, ultimately affecting the susceptible, infected, and juvenile densities.

\section{Benchmark Models}
\subsection{Searle et al.’s Model}

Using the mechanistic model introduced by \citet{searle16}, we conduct predictions of population $\tilde{y}_{\unit,\benchmarktime}$ for susceptible, infected \textit{D.~dentifera} and \textit{D.~lumholtzi} for all time $\benchmarktime \in \{1,2,\dots,N_{\unit}\}$ and units $\unit \in \{1,2,\dots,U\}$.
In our case, each unit is observed at the same number of time points, so $N_{\unit} = N$ for all $\unit$.
Then, we obtain the overall log-likelihood by summing the log-probabilities of the observed density of \textit{Daphnia}.
Specifically, we use the observed densities $y^*_{\unit,\benchmarktime}$ and the predicted means $\tilde{y}_{\unit,\benchmarktime}$ derived from the mechanistic model, along with the dispersion parameter $\tau$ to specify a NBinom distribution and calculate the likelihood:
\begin{align}
\ell(\beta, \theta) &= \sum_{\unit=1}^{\Unit}\sum_{\benchmarktime=1}^{N}
\log P(Y_{\unit,\benchmarktime} = y^*_{\unit,\benchmarktime} \mid \mu_{\unit,\benchmarktime}, \tau),
\label{eq:total_loglik}
\end{align}

To obtain the maximum likelihood estimate, we conduct the linear optimization on $\tau$ and choose the $\hat{\tau}$ that yields the highest overall likelihood. 
With this $\hat{\tau}$, we obtain the likelihood by \ref{eq:total_loglik} for the model on the data. And the result of the model fitting can be found in the Table~2 in the main article.

\subsection{Negative Binomial Regression}

Count data from ecological studies often exhibit overdispersion and unit-specific variation.
The NBinom distribution addresses overdispersion, while random intercepts allow each unit to have its own baseline level.
As a benchmark model, we employ NBinom generalized linear mixed models (GLMMs) to capture overdispersed count data and correlation across repeated measures within each unit $\unit$.
The NBinom distribution is parameterized by a mean $\mu_{\unit,\benchmarktime}$ and a dispersion parameter $\theta$.

We present three GLMMs with increasing of order of time factor to capture the trend of these \textit{Daphnia} species in the experiment.
While a linear model is easy to interpret and straightforward to implementing, observed \textit{Daphnia} data exhibit a non-linear pattern.
Introducing quadratic or cubic terms to the model enhance the capability of model to capture these non-linear trends in the data.
By starting with a linear model and then considering quadratic and cubic cases, we can evaluate the complexity of temporal patterns and better approximate the latent process of in the dynamics.

\subsubsection{Model 1 (Linear Trend Model)}
We start with a simple GLMM having only with linear terms:
\begin{align}
Y_{\unit,\benchmarktime} &\sim \text{NBinom}(\mu_{\unit,\benchmarktime}, \tau), \\[6pt]
\log(\mu_{\unit,\benchmarktime}) &= \beta_0 + \beta_1\, t_{\unit,\benchmarktime} + b_{\unit},
\label{model:GLMM_linear}
\end{align}
where $\mu_{\unit,\benchmarktime}$ is the expected count for unit $\unit$ at time $\benchmarktime$.
We fit this model independently for each {\it Daphnia} species, life stage, and infection status, and so the total log-likelihood is the sum over these categories.
The vector coefficient $\beta_1$ measures the effect of time on the log-scale of \textit{Daphnia} density.
The term $b_\unit$ accounts for the random intercept drawn from a normal distribution with variance $\sigma^2$.
Conditional on the random intercept $b_\unit$, observations within a unit are assumed independent.
The dispersion parameter $\tau$ from the NBinom distribution accommodates overdispersion in the count data, which is density in our case.

\subsubsection{Model 2 (Quadratic Trend Model)}
To accommodate the non-linear pattern, we include a quadratic term in time in model~\ref{model:GLMM_quad}:
\begin{align}
Y_{\unit,\benchmarktime} &\sim \text{NBinom}(\mu_{\unit,\benchmarktime}, \tau), \\[6pt]
\log(\mu_{\unit,\benchmarktime}) &= \beta_0 + \beta_1\, t_{\unit,\benchmarktime} + \beta_2\, t_{\unit,\benchmarktime}^2 + b_{\unit}.
\label{model:GLMM_quad}
\end{align}
Here, $\beta_2$ captures curvature in the trend, allowing for a turning point as time progresses.

\subsubsection{Model 3 (Cubic Trend Model)}
To capture more complex temporal pattern, we further introduce the cubic term in the model~\ref{model:GLMM_cubic}:
\begin{align}
Y_{\unit,\benchmarktime} &\sim \text{NBinom}(\mu_{\unit,\benchmarktime}, \tau), \\[6pt]
\log(\mu_{\unit,\benchmarktime}) &= \beta_0 + \beta_1\, t_{\unit,\benchmarktime} + \beta_2\, t_{\unit,\benchmarktime}^2 + \beta_3\, t_{\unit,\benchmarktime}^3 + b_{\unit}.
\label{model:GLMM_cubic}
\end{align}
The cubic term $\beta_3$ provides additional capability for capturing multiple fluctuation patterns in the \textit{Daphnia} dynamics.

The likelihood for these models is constructed by considering the joint distribution of all observations $\{Y_{u,n}\}$, integrating over the random effects, $b_{1:\Unit}$. Letting $\beta = (\beta_0, \beta_1, \dots)$ denote the fixed-effect parameters, the conditional likelihood of the observations for unit $\unit$ given $b_\unit$ is
\begin{align}
L_\unit(\beta, \tau, \sigma^2) &= \prod_{\benchmarktime=1}^{N} f_{\text{NB}}\bigl(Y_{\unit,\benchmarktime}; \mu_{\unit,\benchmarktime}(\beta, \sigma^2), \tau\bigr),
\label{eq:conditional_likelihood}
\end{align}
where $f_{\text{NB}}(\cdot;\mu,\tau)$ is the NBinom probability mass function and $\mu_{\unit,\benchmarktime}(\beta, b_\unit) = \exp(\beta_0 + \beta_1 t_{\unit,\benchmarktime} + \dots + b_\unit)$. 
The total likelihood is then the product over all units due to the independence:
\begin{align}
L(\beta, \tau, \sigma^2) &= \prod_{\unit=1}^{\Unit} L_\unit(\beta, \tau, \sigma^2)\\
                         &= \prod_{\unit=1}^{\Unit} \prod_{\benchmarktime=1}^{N} f_{\text{NB}}\bigl(Y_{\unit,\benchmarktime}; \mu_{\unit,\benchmarktime}(\beta, \sigma^2), \tau\bigr)
\label{eq:total_likelihood}
\end{align}
and the log-likelihood is:
\begin{align}
\ell(\beta, \tau, \sigma^2) &= \sum_{\unit=1}^{\Unit} \sum_{\benchmarktime=1}^{N} \log \bigl(f_{\text{NB}}(Y_{\unit,\benchmarktime}; \mu_{i,j}(\beta, \sigma^2), \tau)\bigl).
\label{eq:log_likelihood}
\end{align}
The calculated log-likelihood used to compute the AIC for model comparison. And the result of the model fitting of model~\ref{model:GLMM_linear},~\ref{model:GLMM_quad} and \ref{model:GLMM_cubic} can be found in the Table~2 in the main article.

\subsection{SIRPF2 Model}

The SIRPF2 model presented here ignores the state of juvenile \textit{Daphnia} and describes the dynamics in each bucket ($\unit$) among susceptible individuals ($S^{\species}$), infected individuals ($I^{\species}$), alga as a food resource ($F$), the parasite population ($P$).
The superscript $\species$ denotes the species, which represents native or invasive species.
Each stochastic differential equation accounts for various biological and ecological processes, augmented with stochastic terms to capture the inherent randomness within the systems.
{\small
\begin{align}
    dS^{\species}_{\unit}(t) &= r^{\species}_{\unit} \, f^{\species}_{S,{\unit}} \, F_{\unit}(t) \, S^{\species}_{\unit}(t)\, dt -
    \big\{
    \theta^{\species}_{S,{\unit}} + p^{\species}_{\unit} \, f^{\species}_{S,{\unit}} \, P_{\unit}(t) + \delta
    \big\}
    \, S^{\species}_{\unit}(t)\, dt,\\
    dI^{\species}_{\unit}(t) &= p^{\species}_{\unit} \, f^{\species}_{S,{\unit}}\,  S^{\species}_{\unit}(t) \, P_{\unit}(t)\, dt -
    \big\{
    \theta^{\species}_{I,{\unit}} + \delta
    \big\}
    \, I^{\species}_{\unit}(t)\, dt + I^{\species}_{\unit}(t) \, d\zeta^{\species}_{I,{\unit}},\label{eq:SIRPF2_dS}\\
    dP_{\unit}(t) &= \sum_{\species \in \{\native, \invasive\}} \left(\beta^{\species}_{\unit} \, \theta^{\species}_{I,{\unit}} \,  I^{\species}_{\unit}(t) - f^{\species}_{S,{\unit}} \big\{ S^{\species}_{\unit}(t) + \xi_{\unit} I^{\species}_{\unit}(t) \big\} P_{\unit}(t)\right) dt - \theta_{P,{\unit}} \, P_{\unit}(t)\, dt + P_{\unit}(t)\, d\zeta_{P,{\unit}},\\
    dF_{\unit}(t) &= - \sum_{\species \in \{\native, \invasive\}} f^{\species}_{S,{\unit}} \, F_{\unit}(t) \left(S^{\species}_{\unit}(t)+\xi_{\unit} \, I^{\species}_{\unit}(t)\right)dt  + \mu \, dt + F_{\unit}(t)\, d\zeta_{F,{\unit}},\label{eq:SIRPF2_dF}\\
d\zeta^{\species}_{I,{\unit}} &\sim \Normal \big[0, (\sigma^{\species}_{I,{\unit}})^{2}\, dt\big],
\hspace{5mm}
d\zeta_{F,{\unit}} \sim \Normal \big[0, \sigma_{F,{\unit}}^{2}\, dt\big],
\hspace{5mm}
d\zeta_{P,{\unit}} \sim \Normal \big[0, \sigma_{P,{\unit}}^{2}\, dt\big].
\end{align}
}

Unlike the SIRJPF2 model, we ignore the maturation process from juvenile to adult in equation \eqref{eq:SIRPF2_dS}.
Instead, this model uses $r^{\species}_{\unit}$ to represent the average rate of growth in density of \textit{Daphnia} adult.

\subsubsection{Flow Diagram}
The SIRPF2 model illustrates the interdependent dynamics among the Susceptible ($S$), Infected ($I$), Food ($F$), and Parasite ($P$) populations, with transitions representing ecological and biological interactions.

\usetikzlibrary{positioning}
\usetikzlibrary {arrows.meta}
\usetikzlibrary{shapes.geometric}

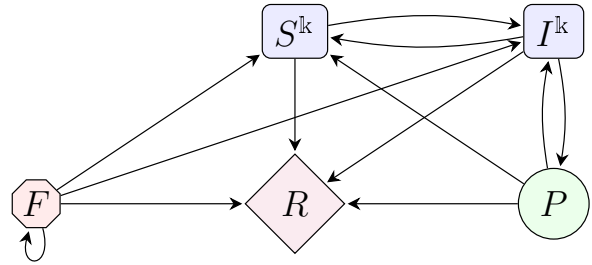
\begin{figure}[H]
\begin{center}
\resizebox{8cm}{!}{
\begin{tikzpicture}[
  square/.style={rectangle, draw=black, minimum width=0.5cm, minimum height=0.5cm, rounded corners=.1cm, fill=blue!8},
  rhombus/.style={diamond, draw=black, minimum width=0.1cm, minimum height=0.1cm, fill=purple!8,aspect = 1},
  travel/.style={circle, draw=black, minimum width=0.5cm, minimum height=0.5cm, fill=green!8},
  report/.style={shape=regular polygon, regular polygon sides=8, draw, fill=red!8,minimum size=0.6cm,inner sep=0cm},
  bendy/.style={bend left=10},
  bendy2/.style={bend left=100},
  bendy3/.style={bend left=-100},
  >/.style={shorten >=0.25mm},
  >/.tip={Stealth[length=1.5mm,width=1.5mm]}
]
\tikzset{>={}};

\node (S) at (2.5,0) [square] {$S^{\species}$};
\node (I) at (5.5,0) [square] {$I^{\species}$};
\node (R) at (2.5,-2)  [rhombus] {$R$};
\node (P) at (5.5,-2)[travel] {$P$};
\node (F) at (-0.5,-2) [report] {$F$};

\draw [->, bendy] (S) to  (I);
\draw [->, bendy] (I) to  (S);

\draw [->] (P) --  (S);
\draw [->] (F) --  (R);
\draw [->] (P) --  (R);
\draw [->] (S) -- (R);
\draw [->] (I) -- (R);
\draw [->] (F) -- (S);
\draw [->] (F) -- (I);
\draw (F) edge[loop below] (F);

\draw [->, bendy] (P) to (I);
\draw [->, bendy] (I) to (P);

\end{tikzpicture}
}
\end{center}
\vspace{-5mm}
\caption{Flow diagram for the SIRPF2 model illustrating population interactions.}
\label{fig:flow_SIRPF2}
\end{figure}

\section{Experimental Data on Peak Density}
\begin{table}[H]
\centering
\begin{tabular}{lccccc}
\toprule
Mesocosm
& \shortstack{\textit{D.~dentifera} \\ juvenile} 
& \shortstack{\textit{D.~dentifera} \\ female adult} 
& \shortstack{\textit{D.~lumholtzi} \\ juvenile} 
& \shortstack{\textit{D.~lumholtzi} \\ female adult} 
& \shortstack{\textit{D.~lumholtzi} \\ male adult} \\
\midrule
1 & 22 & 22 & 22 & 27 & 27 \\
2 & 22 & 32 & 17 & 17 & 22 \\
3 & 27 & 32 & 27 & 22 & 37 \\
4 & 22 & 22 & 37 & 17 & 27 \\
5 & 22 & 22 & 17 & 32 & 27 \\
6 & 32 & 27 & 27 & 32 & 42 \\
7 & 27 & 22 & 17 & 17 & 27 \\
8 & 27 & 32 & 22 & 22 & 42 \\
\bottomrule
\end{tabular}
\caption{This table shows the day on which each species reached its peak density in mesocosms containing mixed \textit{Daphnia} species and a parasite. \textit{D.~dentifera} male adults are excluded because they were almost absent.}
\end{table}

\section{SIRJPF2-Gamma Model}
The SIRJPF2-Gamma model differs from the original SIRJPF2 model by its incorporation of gamma noise in transition processes between states, in place of the Gaussian noise used previously. 
Specifically, in the SIRJPF2-Gamma model, random fluctuations for transitions between states are explained by gamma distributions to represent ecological events occur through one-way transitions which are nonnegative. 
The use of gamma noise in modeling transitions explicitly enforces that individuals move from one class to another without the artificial creation or destruction of hosts that do not transit to or from another compartment.
This aligns with the underlying dynamics that the jump-like transitions observed in ecosystem reflect one-directional movements rather than symmetrical fluctuations.

{\small
\begin{align}
    dS^{\species}_{\unit}(t) &= \lambda^{\species}_{J,{\unit}} \,  J^{\species}_{\unit}(t)dt -
    \big\{
    \theta^{\species}_{S,{\unit}} + \delta
    \big\}
    \, S^{\species}_{\unit}(t)\, dt - p^{\species}_{\unit} \, f^{\species}_{S,{\unit}} \, P_{\unit}(t) S^{\species}_{\unit}(t) \, d\Gamma^{\species}_{SI,{\unit}},\\
    dI^{\species}_{\unit}(t) &= p^{\species}_{\unit} \, f^{\species}_{S,{\unit}}\,  S^{\species}_{\unit}(t) \, P_{\unit}(t)\, d\Gamma^{\species}_{SI,{\unit}} -
    \theta^{\species}_{I,{\unit}} I^{\species}_{\unit}(t)d\Gamma^{\species}_{IP,{\unit}}- \delta\, I^{\species}_{\unit}(t)\, dt,\\
    dJ^{\species}_{\unit}(t) &= r^{\species}_{\unit} \, f^{\species}_{S,{\unit}} \, F_{\unit}(t) \, S^{\species}_{\unit}(t)\, d\Gamma^{\species}_{SJ,{\unit}} -
    \big\{
    \theta^{\species}_{J,{\unit}}  + \delta
    \big\} 
    \, J^{\species}_{\unit}(t)\, dt - \lambda^{\species}_{J,{\unit}} \,  J^{\species}_{\unit}(t)dt, \\
    dP_{\unit}(t) &= \sum_{\species \in \{\native, \invasive\}} \left(\beta^{\species}_{\unit} \, \theta^{\species}_{I,{\unit}} \,  I^{\species}_{\unit}(t)d\Gamma^{\species}_{IP,{\unit}} - f^{\species}_{S,{\unit}} \big\{ S^{\species}_{\unit}(t) + \xi_{\unit} I^{\species}_{\unit}(t) \big\} P_{\unit}(t)dt\right) - \theta_{P,{\unit}} \, P_{\unit}(t)\, dt,\\
    dF_{\unit}(t) &= - \sum_{\species \in \{\native, \invasive\}} f^{\species}_{S,{\unit}} \, F_{\unit}(t) \left(S^{\species}_{\unit}(t)+\xi_{J,{\unit}} \, J^{\species}_{\unit}(t) +\xi_{\unit} \, I^{\species}_{\unit}(t)\right)dt  + \mu \, dt + F_{\unit}(t)\, d\Gamma_{F,{\unit}},\\
d\Gamma^{\species}_{JS,{\unit}} &\sim \Gamma
  \bigg[\frac{dt}{(\sigma^{\species}_{JS,{\unit}})^{2}}, (\sigma^{\species}_{JS,{\unit}})^{2}\bigg],
\quad
d\Gamma^{\species}_{SI,{\unit}} \sim \Gamma
  \bigg[\frac{dt}{(\sigma^{\species}_{SI,{\unit}})^{2}}, (\sigma^{\species}_{SI,{\unit}})^{2}\bigg],\\
d\Gamma^{\species}_{IP,{\unit}} &\sim \Gamma
  \bigg[\frac{dt}{(\sigma^{\species}_{IP,{\unit}})^{2}}, (\sigma^{\species}_{IP,{\unit}})^{2}\bigg],
\quad
d\Gamma_{F,{\unit}} \sim \Normal \big[0, \sigma_{F,{\unit}}^{2}\, dt\big].
\end{align}
}